\documentclass[peerreview]{IEEEtran}
\usepackage{amsthm}
\usepackage{amsmath}
\usepackage{mathrsfs}
\usepackage{amssymb} 
\usepackage{bbm,dsfont} 
\usepackage{multirow} 
\usepackage{units}
\usepackage{stmaryrd}   
\usepackage{verbatim}
\usepackage{enumerate} 
\usepackage[table]{xcolor}
\usepackage{slashbox} 
\usepackage{ wasysym }
\usepackage{tikz}
\usetikzlibrary{calc}
\usepackage{tabularx}

\usepackage{lipsum}
\usepackage{multicol}
\usepackage{float}

\usepackage{ifthen}

\usepackage{xcolor}
\usepackage{hyperref} 
\usepackage
{hyperref} 
\hypersetup{
    colorlinks,
    linkcolor={blue!80!black},
    citecolor={green!50!black},
    urlcolor={blue!80!black}
}

\usepackage{graphicx}
\graphicspath{{images/}}

\title{Quantum Channel State Masking} 
\author{
		\vspace{0.1cm}
    \IEEEauthorblockN{Uzi Pereg\IEEEauthorrefmark{1}, Christian Deppe\IEEEauthorrefmark{1}, and Holger Boche\IEEEauthorrefmark{2}} \\
		\vspace{0.25cm}
    \IEEEauthorblockA{\normalsize \IEEEauthorrefmark{1}Institute of Communication Engineering, Technical University of Munich \\
    \IEEEauthorrefmark{2}Theoretical Information Technology, Technical University of Munich\\
    Email: {\tt $\{$uzi.pereg,christian.deppe,boche$\}$@tum.de}}
}

\usepackage[square,numbers,sort&compress]{natbib}
\usepackage{accents}
\newlength{\dhatheight}

\newcommand{\bieee}{\begin{IEEEeqnarray}{rCl}}
\newcommand{\eieee}{\end{IEEEeqnarray}}
\newcommand{\prob}[1]{\Pr\left(#1\right)}

\renewcommand{\mathbbm}[1]{\text{\usefont{U}{bbm}{m}{n}#1}} 
\newcommand{\eps}{\varepsilon}

\newcommand{\norm}[1]{\left\lVert#1\right\rVert}
\newcommand{\trace}{\mathrm{Tr}}

\newcommand{\identity}{\mathbbm{1}}
\newcommand{\kb}[1]{ | #1 \rangle\langle #1 | } 

\newcommand{\ie}{\emph{i.e.} }
\newcommand{\eg}{\emph{e.g.} }
\newcommand{\etal}{\emph{et al.} }

\newcommand{\hM}{\hat{M}}

\newcommand{\Dset}{\mathcal{D}}
\newcommand{\Fset}{\mathcal{F}}

\newcommand{\Hset}{\mathcal{H}}

\newcommand{\Lset}{\mathcal{L}}

\newcommand{\Pset}{\mathcal{P}}
\newcommand{\Uset}{\mathcal{U}}
\newcommand{\Vset}{\mathcal{V}}

\newcommand{\Sset}{\mathcal{S}}

\newcommand{\Xset}{\mathcal{X}}
\newcommand{\Yset}{\mathcal{Y}}
\newcommand{\Zset}{\mathcal{Z}}

\theoremstyle{remark}	\newtheorem{theorem}{Theorem}
\theoremstyle{remark}	\newtheorem{lemma}[theorem]{Lemma}
\theoremstyle{remark}	\newtheorem{corollary}[theorem]{Corollary}
\theoremstyle{remark}	
\theoremstyle{remark} \newtheorem{definition}{Definition}
\theoremstyle{remark} \newtheorem{remark}{Remark}
\theoremstyle{remark}

\newcommand{\Tset}{\mathcal{T}}												

\newcommand{\channel}{\mathcal{N}}

\newcommand{\inC}{\mathsf{C}}
\newcommand{\inR}{\mathsf{R}}

\newcommand{\opR}{\mathbb{R}}
\newcommand{\opC}{\mathbb{C}}

\begin{document}
\maketitle

{}

\begin{abstract} 
Communication over a quantum channel that depends on a quantum state is considered when the encoder has channel side information (CSI) and is required to mask information on the quantum channel state  from the decoder.
A full characterization is established for the entanglement-assisted masking equivocation region with a maximally correlated channel state, and a regularized formula is given for the quantum capacity-leakage function without assistance.
For Hadamard channels without assistance, we derive single-letter inner and outer bounds, which coincide 
in the standard case of a channel that does not depend on a state. 
%
\end{abstract}


\section{Introduction}

Security and privacy are critical aspects in modern communication systems 
\cite{BHCPDA:13p,LRBW:17p,PiquerasJoverMarojevic:19p,WYDP:19p}. 
The classical wiretap channel was first introduced by Wyner \cite{Wyner:75p,LiangPoorShamai:09n} to model communication in the presence of a passive eavesdropper, and further studied in various scenarios, as in
\cite{LiangKramerPoorShamai:05p,BarrosRodrigues:06c,
BellareTessaroVardy:12a,BocheSchaeferPoor:15p,GoldfeldCuffPermuter:16p,
XingLiuZhang:16p,BGPSCP:18p,BocheCaiNotzelDeppe:19p,LiLiangPoorShamai:19p}. 
%
On the other hand, Merhav and Shamai \cite{MerhavShamai:07p} introduced a different communication system with the privacy requirement of masking.
In this setting, the sender transmits a sequence $X^n$ over a memoryless state-dependent channel $p_{Y|X,S}$, where the state sequence $S^n$ has a fixed memoryless distribution and is not affected by the transmission.  The transmitter of $X^n$ is informed of $S^n$ and is required to send information to the receiver while limiting the amount of information that the receiver can learn about $S^n$.  
It was shown in \cite{MerhavShamai:07p} that the achievable masking equivocation region  consists of rate-leakage pairs $(R,L)$ such that
\begin{align}
R\leq& I(U;Y)-I(U;S) \\
L\geq& I(S;U,Y)
\end{align}
for $(S,U,X,Y)\sim p_S \times p_{U|S} \times p_{X|U,S} \times p_{Y|X,S} $,
where $U$ is an auxiliary random variable, with cardinality $|\mathcal{U}|\leq |\Xset| |\Sset|$.
Related settings and extensions are also considered in \cite{LeTreustBloch:20p,KoyluogluSoundararajanVishwanath:16p,KoyluogluSoundararajanVishwanath:11c,DikshteinSomekhBaruchShamai:19c,AsoodehDizaAlajajiLinder:16p,LeTreustBloch:16c,TutunchuogluOzelYenerUlukus:14c,Courtade:12c}.


The 
field of quantum information is rapidly evolving in both practice and theory 
\cite{NCLMSYH:20p,DowlingMilburn:03p,JKLGD:13p,BennettBrassard:14p,BecerraFanMigdall:15p,YCLZRC:17p,ZDSZSG:17p,PERLHPCVV:20p}.
As technology 
approaches the atomic scale, 
we seem to be on the verge of the  ``Quantum Age" \cite{BouwmeesterZeilinger:00b,ImreGyongyosi:12b}.
Dynamics can be modeled by a noisy quantum channel, describing
 physical evolutions, density transformation,
discarding of sub-systems, quantum measurements, etc. \cite{Kitaev:97b} \cite[Section 4.6]{Wilde:17b}.
%
Quantum information theory is the natural extension of classical information theory. Nevertheless, 
this generalization reveals astonishing phenomena with no parallel in classical communication \cite{GyongyosiImreNguyen:18p}. For example, 
 two memoryless quantum channels, each with zero quantum capacity, can have a nonzero quantum capacity
when used together \cite{SmithYard:08p}. This property is known as super-activation.

Communication through quantum channels can be separated into different categories.
 For classical communication, the Holevo-Schumacher-Westmoreland (HSW) Theorem provides a regularized (``multi-letter")  formula for the capacity of a quantum channel without assistance \cite{Holevo:98p,SchumacherWestmoreland:97p}. 
Although calculation of such a formula is intractable in general, it provides computable lower bounds, and there are special cases where the capacity can be computed exactly. 
The reason for this difficulty is that the Holevo information is not necessarily additive \cite{Holevo:12b}. 
A similar difficulty occurs with transmission of quantum information. 
A regularized formula for the quantum capacity is given in \cite{BarnumNielsenSchumacher:98p,Loyd:97p,Shor:02l,Devetak:05p}, in terms of the coherent information. 
A computable 
formula is obtained in the special case where the channel is  degradable or less noisy 
\cite{DevetakShor:05p}.

Another scenario of interest is when Alice and Bob are provided with entanglement resources \cite{NielsenChuang:02b,BocheJanssenKaltenstadler:17p}.
While entanglement can be used to produce shared randomness, it is a much more powerful aid \cite{Wilde:17b,ChitambarGour:19p}. 
In particular, super-dense coding \cite{BennetWiesner:92p} is a well known communication protocol where two classical bits are transmitted using a single use of a noiseless qubit channel and a maximally entangled pair. Thereby, entanglement assistance doubles the transmission rate of classical messages over a noiseless qubit channel. 
 The entanglement-assisted capacity of 
a noisy quantum channel was fully characterized by Bennett \etal 
\cite{BennettShorSmolin:99p,BennettShorSmolin:02p} in terms of the quantum mutual information. 
Entanglement resources are thus instrumental for the performance analysis of quantum communication systems,  as the characterization with entanglement assistance 
provides a computable upper bound for unassisted communication as well.
In the other direction, i.e. using information measures to understand quantum physics, the quantum mutual information plays a role in investigating the entanglement structure of quantum field theories 
\cite{Swingle:10a,PanJing:08p,CasiniHuertaMyersYale:15p,AgonGaulkner:16p}.

The entanglement-assisted capacity theorem  can be viewed as the quantum generalization of Shannon's classical capacity theorem \cite{Shannon:48p} (see page 2640 in \cite{BennettShorSmolin:02p}).
Nonetheless, there are communication settings where entanglement can increase the capacity of a \emph{classical} channel, such as
the zero-error capacity problem \cite{LMMOR:12p} and the multiple access channel with entangled encoders \cite{LeditzkyAlhejjiLevinSmith:20p}. 
Entanglement assistance also has striking effects in 
 communication  games and their security applications \cite{CHSH:69p,PappaChaillouxWehnerDiamantiKerenidis:12p,VaziraniVidick:14p,JaiWeiWuGuo:20a,JiNatarajanVidickWrightYuen:20a,LeditzkyAlhejjiLevinSmith:20p}.
Furthermore, entanglement can assist  
the transmission of quantum information. By employing the teleportation protocol \cite{BennettBrassardJozsaPeres:93p}, qubits can be sent 
at half the rate of classical bits given entanglement resources. Thus, for a given quantum channel,  the entanglement-assisted quantum capacity has half the value of the entanglement-assisted classical capacity in units of qubits per channel use.  

From a practical standpoint, it is also important to determine the amount of entanglement supply that is consumed in the process of sending information. 
The tradeoff between communication and resource rates is considered in \cite{Shor:04p,DevetakHarrowWinter:08p,HsiehWilde:10p,HsiehWilde:10p1,WildeHsieh:12p1,WangHayashi:20c}.
Furthermore, the study of such tradeoffs led to the development of general ``father" and ``mother" protocols 
\cite{DevetakHarrowWinter:04p,Devetak:06p,HorodeckiOppenheim:07p,AbeyesingheDevetakHaydenWinter:09p,DupuisHaydenLi:10p}, which produce achievability schemes for various settings including those mentioned above. Many of those protocols can be presented as a consequence of the decoupling theorem \cite{HaydenHorodeckiWinterYard:08p,Dupuis:08a,Dupuis:10z}. Roughly speaking, the decoupling approach  shows that quantum information can be reliably communicated when Bob's environment is decoupled from Alice's purifying reference system.
Further work on entanglement-assisted communication
can be found in \cite{Holevo:02p,HsiehDevetakWinter:08p,
Shirokov:12p,DattaHsieh:13p,WildeHsiehBabar:14p,QianZhan:18p,AnshuJainWarsi:17a,BertaGharibyanWalter:17p,
CCVH:19a,AnshuJainWarsi:19p} and references therein.

Boche, Cai, and N\"{o}tzel \cite{BocheCaiNotzel:16p} addressed the classical-quantum channel  with channel state information (CSI) at the encoder. 
The 
capacity 
was determined given causal CSI, and a regularized formula was provided given 
non-causal CSI \cite{BocheCaiNotzel:16p} (see also \cite{Pereg:20c1,Pereg:19a3}). Warsi and Coon \cite{WarsiCoon:17p} used an information-spectrum approach to derive  multi-letter bounds for a similar setting, where the side information has a limited rate. The entanglement-assisted capacity of a quantum channel with non-causal CSI was determined by    Dupuis in \cite{Dupuis:08a,Dupuis:09c}, and with causal CSI  in \cite{Pereg:19c3,Pereg:19a}. One-shot communication with 
CSI is considered in \cite{AnshuJainWarsi:19p} as well.
 Luo and Devetak \cite{LuoDevetak:09p} considered channel simulation with source side information (SSI) at the 
decoder, and also solved the quantum generalization of the Wyner-Ziv problem \cite{WynerZiv:76p}.  Quantum data  compression with SSI  is also studied in \cite{DevetakWinter:03p,YardDevetak:09p,HsiehWatanabe:16p,DattaHircheWinter:19c,DattaHircheWinter:18a,
CHDH:19c, CHDH:18a}. 
Compression with SSI given entanglement assistance was recently considered by 
Khanian and Winter  \cite{KhanianWinter:19c2,KhanianWinter:18a,KhanianWinter:19c,KhanianWinter:19a}. 

Considering secure communication over the quantum wiretap channel,
 Devetak \cite{Devetak:05p} and Cai \etal \cite{CaiWinterYeung:04p}  established a regularized characterization of the secrecy capacity   without assistance. Connections to the coherent information of 
a quantum point to point channel were drawn in \cite{DevetakWinter:05p}. Related models appear in 
\cite{HsiehLuoBrun:08p,LiWinterZouGuo:09,Wilde:11p,Watanabe:12p,ElkoussStrelchuk:15p,AnshuHayashiWarsi:18c} as well.
The entanglement-assisted secrecy capacity was determined by Qi \etal \cite{QiSharmaWilde:18p} (see also \cite{HsiehWilde:10p,SharmaWakauwaWilde:17a}).
%
Boche \etal \cite{BocheCaiNotzelDeppe:19p,BochCaiDeppeNotzel:17p} studied the quantum wiretap channel with an active jammer.
Furthermore, the capacity-equivocation region was established, characterizing the tradeoff between secret key consumption and private classical communication \cite{HsiehLuoBrun:08p,Wilde:11p} (see also \cite{WildeHsieh:12p}\cite[Section 23.5.3]{Wilde:17b}).
In \cite{Devetak:05p}, Devetak considered entanglement generation using a secret-key-assisted quantum channel. 
The quantum 
Gel'fand-Pinsker wiretap channel is considered in \cite{AnshuHayashiWarsi:18c} and
other related scenarios can be found in \cite{KonigRennerBariskaMaurer:07p,GHKLLSTW:14p,LupoWildeLloyd:16p}.
%
The quantum broadcast and multiple access channels with confidential messages were recently considered in  \cite{SalekHsiehFonollosa:19a,SalekHsiehFonollosa:19c} and  \cite{AghaeeAkhbari:19c,BochJanssenSaeedianaeeni:20p}, respectively. 

\begin{center}
\begin{figure}[ht!]
\caption{Coding for a quantum state-dependent channel $\channel_{EA'\rightarrow B}$ given state information at the encoder and masking from the decoder, with and without entanglement assistance. The quantum systems of Alice and Bob are marked in red and blue, respectively. The channel state systems $E^n$ and $C^n$ are marked in brown.
}

\includegraphics[scale=0.75,trim={0 9.5cm 0 8.5cm},clip]{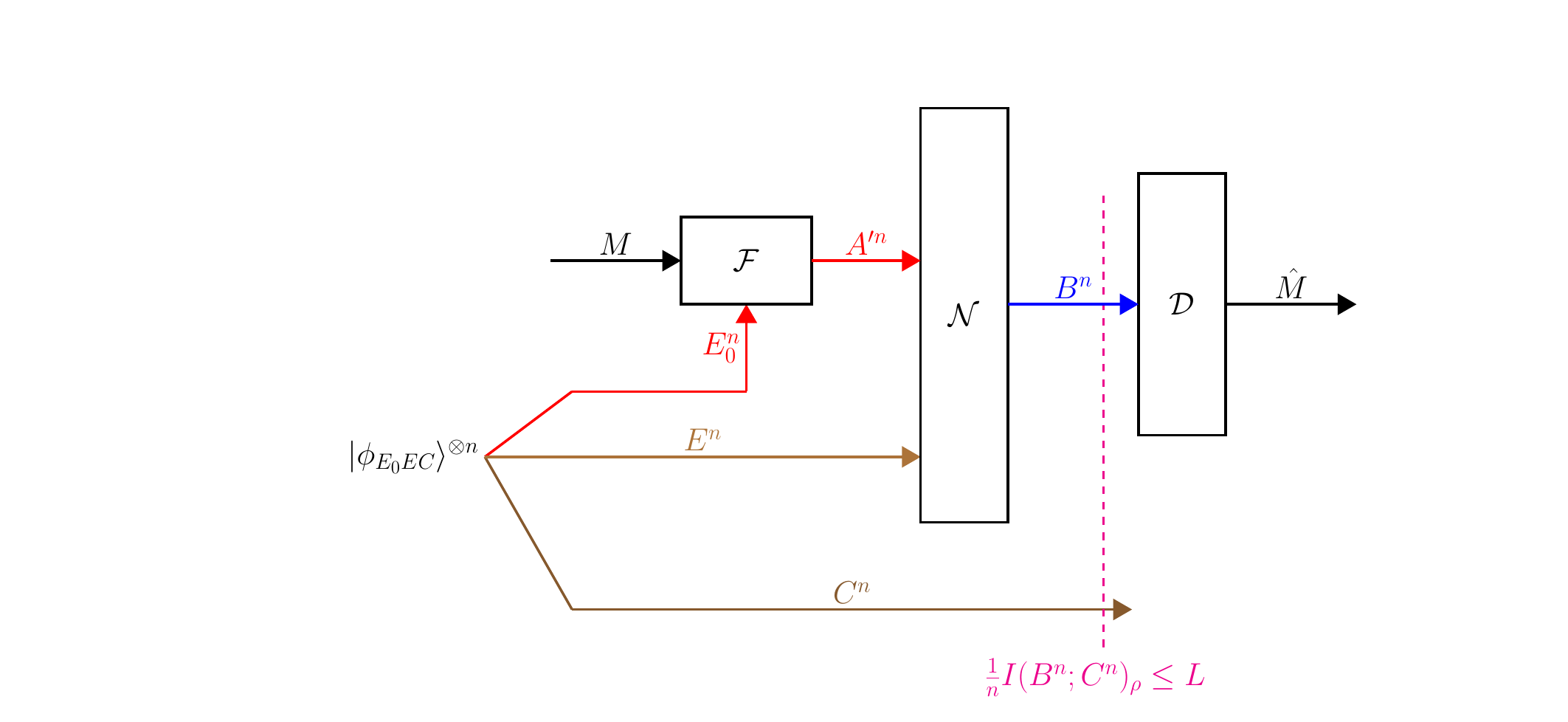} 
\\
{\footnotesize
(a) Unassisted coding: The quantum message is stored in $M$.   
 Alice encodes the quantum message using her access to the side information systems $E_0^n$, which are entangled with the channel state systems $E^n$. To this end, she applies the encoding map $\Fset_{  M E_0^n \rightarrow A'^n}$,
 and transmits the systems $A'^n$ over the channel. 
 Bob receives the channel output systems $B^n$ and applies the decoding map $\Dset_{B^n\rightarrow \hM}$.
A leakage rate $L$ is achieved if 
$
 \frac{1}{n} I(B^n;C^n)_\rho \leq L 
$.
}
\\ 
\includegraphics[scale=0.75,trim={0 8.3cm 0 8cm},clip]{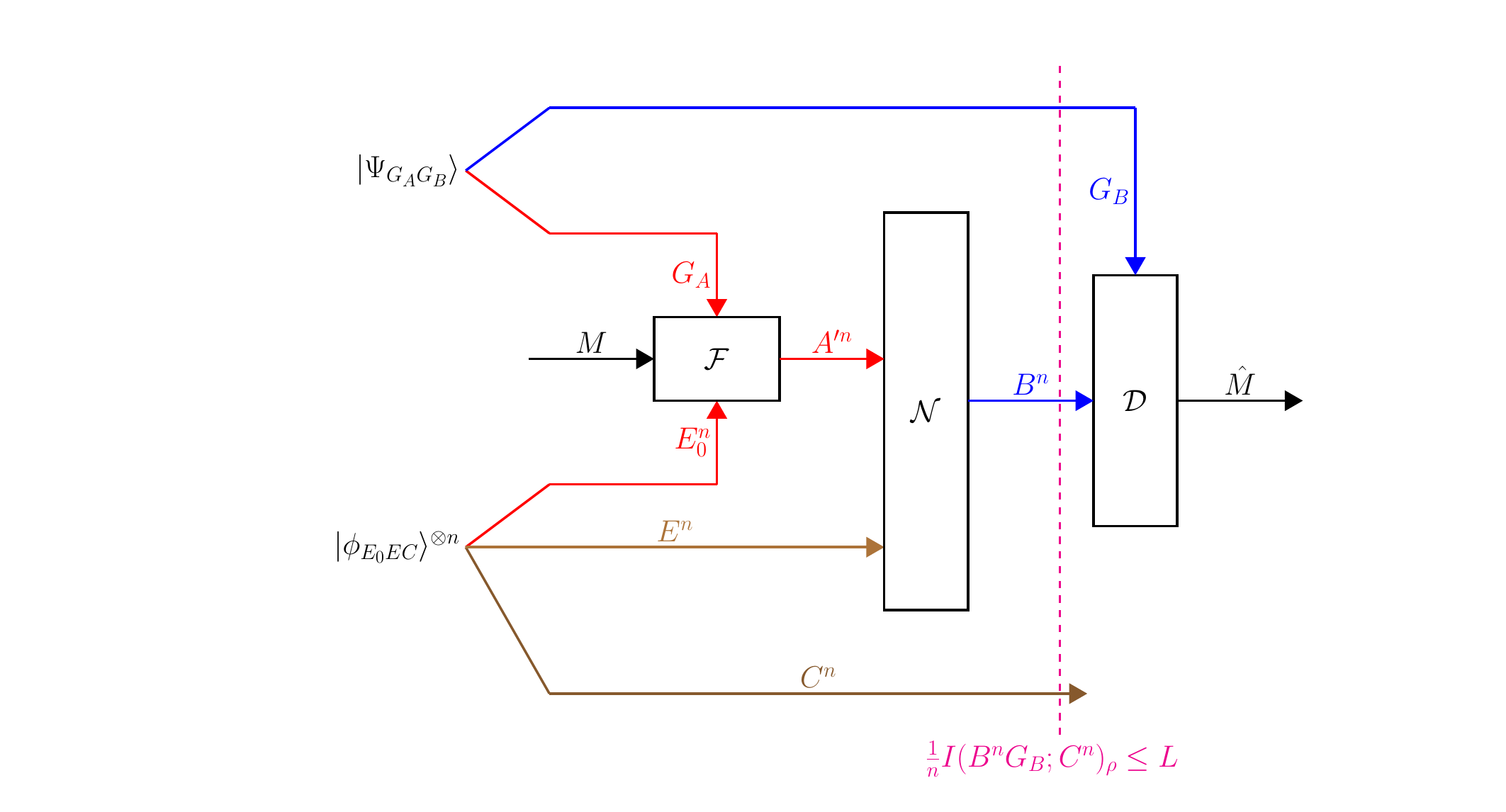} 
\\
{\footnotesize
(b) Entanglement-assisted coding: The quantum message is stored in $M$, while Alice and Bob's entanglement resources are in the quantum systems $G_A$ and $G_B$, respectively.  
 Alice encodes the quantum message using $G_A$ along with her access to the side information systems $E_0^n$, which are entangled with the channel state systems $E^n$. To this end, she applies the encoding map $\Fset_{  M G_A E_0^n \rightarrow A'^n}$,
 and transmits the systems $A'^n$ over the channel. 
 Bob receives the channel output systems $B^n$ and applies the decoding map $\Dset_{B^n G_B\rightarrow \hM}$ to $B^n$ and $G_B$.
A leakage rate $L$ is achieved if 
$
 \frac{1}{n} I(B^n G_B;C^n)_\rho \leq L 
$.
}
\label{fig:MSKsiCode}
\end{figure}

\end{center}

In this paper, we consider a quantum state-dependent channel $\channel_{EA'\rightarrow B}$, when the encoder has CSI and is required to mask information on the quantum channel state from the decoder.
Specifically,  Alice maps the state of the quantum message system $M$ and the CSI systems $E_0^n$ to the state of the channel input systems $A'^n$ in  such a manner that limits the leakage-rate of Bob's information on $C^n$ from $B^n$,
where the systems  $E_0^n$ and $C^n$ are entangled with the channel state systems $E^n$ (see Figure~\ref{fig:MSKsiCode}a). 
Another significant distinction from the classical case is that the leakage requirement involves Bob's share of the entanglement resources, since the decoder has access to both the output systems and his part of the entangled pairs (see Figure~\ref{fig:MSKsiCode}b). In the classical setting,   shared randomness does not need to be included in the leakage constraint as it cannot help the decoder. On the other hand, we know that Bob can extract quantum information by performing measurements on his entanglement resources,  using the teleportation protocol for example.
We note that in the quantum information literature, the term `masking' is sometimes used in a different context of an invertible process that distributes a quantum state to two receivers such that each receiver cannot gain information on the original quantum state \cite{ModiPatiSenSen:18p,LieJeong:19a,LieKwonKimJeong:19a}. In particular, it was shown in \cite{ModiPatiSenSen:18p} that  a universal unitary masker that satisfies this property \emph{for every} input state does not exist. Our setting is fundamentally different as we consider a system with a fixed quantum state $|\phi_{E E_0 C}\rangle^{\otimes n}$ that is known to all parties and controls a communication channel with a single output. 

Analogously to the classical model, we consider channel state systems $C^n$ that  store 
undesired quantum information which leaks
to the receiver \cite{MerhavShamai:07p}. This could model a leakage in the system of secret information, or could stand for another transmission to another receiver (Charlie), with a product state, out of our control, and which is not intended to our receiver (Bob),
and is therefore to be concealed from him. Thus, the goal of the transmitter (Alice) now is to try and mask this undesired information as
much as possible on the one hand, and to transmit reliable independent information rate on the other. 
The systems $E_0^n$ can be thought of as part of the environment of both our transmitter and the transmitter of $C^n$, possibly entangled if those transmitters had previous interaction, while $E^n$ belong to the channel's environment. Dupuis' interprtation  \cite{Dupuis:09c} for the entanglement between $E_0^n$ and $E^n$ is that Alice shares entanglement with the channel itself.   

A full characterization is established for the entanglement-assisted masking equivocation region with maximally correlated channel state systems, and a regularized formula is given for the quantum masking region without assistance. 
We also derive a single-letter outer bound on the unassisted masking region for Hadamard channels, and verify that the inner and outer bounds coincide in the standard case of a channel that does not depend on its state. 
To prove the direct part, we first determine an achievable masking region with rate-limited entanglement. 
%
%
%
Here, we are most interested in the asymptotic characterization of achievable communication rates. On the other hand, in previous work,
the decoupling approach typically produces such characterizations as a consequence of results for the one-shot setting, where the blocklength is $n=1$ \cite{HaydenHorodeckiWinterYard:08p,Dupuis:08a,Dupuis:10z}.
Therefore, we derive an asymptotic version of the decoupling theorem that can be applied directly, without considering the one-shot counterpart. 
While the derivation follows from the one-shot decoupling theorem using familiar arguments, it provides an analytic tool that
is easier to combine with classical techniques,  without a one-shot proxy.
Here, the decoupling approach is used such that both Bob's environment and the channel state systems $E^n$ and $C^n$ are decoupled from Alice's purifying reference system. 
In order to establish the masking requirement, we approximate the leakage rate using  the decoupled state that results from the decoupling theorem.  The approximation relies on the Alicki-Fannes-Winter inequality  \cite{AlickiFannes:04p,Winter:16p},   as the decoupled state is close to the actual output state and its leakage rate has a simpler bound. This demonstrates how the decoupling approach is suitable to our needs. Further explanation on the decoupling nature of our problem is given in Section~\ref{sec:summ}.

%

Our result with entanglement assistance requires the assumption that the channel state
systems $E$, $E_0$, and $C$ are maximally correlated.
Analytically, the presence of three channel state systems poses a difficulty that does not exist in the classical setting of Merhav and Shamai \cite{MerhavShamai:07p}, 
and this is where the maximal correlation assumption comes into play.
We note that the maximal correlation assumption holds in the special case of a classical channel state,
 yet our setting is more general.
The converse proof without assistance is based on different considerations from those in the classical converse proof by Merhav and Shamai \cite{MerhavShamai:07p}.
In the classical proof, the derivation of the bounds on both the communication and leakage rates begins with Fano's inequality, followed by arguments that do not hold in our model since conditional quantum entropies can be negative.
Hence, we bound the leakage rate  in a different manner using the coherent information bound on the communication rate.

\section{Definitions and Related Work}
\subsection{Notation, States, and Information Measures}
 We use the following notation conventions. 
Calligraphic letters $\Xset,\Yset,\Zset,...$ are used for finite sets.
Lowercase letters $x,y,z,\ldots$  represent constants and values of classical random variables, and uppercase letters $X,Y,Z,\ldots$ represent classical random variables.  
 The distribution of a  random variable $X$ is specified by a probability mass function (pmf) 
	$p_X(x)$ over a finite set $\Xset$. 
 We use $x^j=(x_1,x_{2},\ldots,x_j)$ to denote  a sequence of letters from $\Xset$. 
 A random sequence $X^n$ and its distribution $p_{X^n}(x^n)$ are defined accordingly. 
%

The quantum state of a system $A$ is a density operator $\rho$ on the Hilbert space $\Hset_A$.
A density operator is an Hermitian, positive semidefinite operator, with unit trace, \ie 
 $\rho^\dagger=\rho$, $\rho\succeq 0$, and $\trace(\rho)=1$.
The state is said to be pure if $\rho=\kb{\psi}$, for some vector $|\psi\rangle\in\Hset_A$, where
$\langle \psi |$ 
is the Hermitian conjugate of $|\psi\rangle$. 
In general, a density operator has a spectral decomposition of the following form,
\begin{align}
\rho=\sum_{z\in\Zset} p_Z(z) \kb{ \psi_z } 
\end{align}
where $\Zset=\{1,2,\ldots,|\Hset_A|\}$, $p_Z(z)$ is a probability distribution over $\Zset$, and $\{ |\psi_z\rangle \}_{z\in\Zset}$ forms an orthonormal basis of the Hilbert space $\Hset_A$.
 The density operator can thus be thought of as an average of pure states.
A measurement of a quantum system is any set of operators $\{\Lambda_j \}$ that forms a positive operator-valued measure (POVM), \ie
the operators are positive semi-definite and 
$\sum_j \Lambda_j=\identity$, where $\identity$ is the identity operator (see 
\cite[Definition 4.2.1]{Wilde:17b}). According to the Born rule, if the system is in state $\rho$, then the probability of the measurement outcome $j$ is given by $p_A(j)=\trace(\Lambda_j \rho)$.
The trace distance between two density operators $\rho$ and $\sigma$ is $\norm{\rho-\sigma}_1$ where $\norm{F}_1=\trace(\sqrt{F^\dagger F})$.

Define the quantum entropy of the density operator $\rho$ as
\begin{align}
H(\rho) \triangleq& -\trace[ \rho\log(\rho) ]
\end{align}
which is the same as the Shannon entropy 
associated with the eigenvalues of $\rho$.
We may also consider the state of a pair of systems $A$ and $B$ on the tensor product $\Hset_A\otimes \Hset_B$ of the corresponding Hilbert spaces.
Given a bipartite state $\sigma_{AB}$, 
define the quantum mutual information as
\begin{align}
I(A;B)_\sigma=H(\sigma_A)+H(\sigma_B)-H(\sigma_{AB}) \,. 
\end{align} 
Furthermore, conditional quantum entropy and mutual information are defined by
$H(A|B)_{\sigma}=H(\sigma_{AB})-H(\sigma_B)$ and
$I(A;B|C)_{\sigma}=H(A|C)_\sigma+H(B|C)_\sigma-H(A,B|C)_\sigma$, respectively.
The coherent information is then defined as
\begin{align}
I(A\rangle B)_\sigma=-H(A|B)_\sigma  
\end{align}
and $I(A\rangle B|C)_\sigma=I(A\rangle BC)_\sigma=-H(A|BC)_\sigma$ accordingly.

A pure bipartite state 
is called \emph{entangled} if it cannot be expressed as the tensor product 
of two states 
in $\Hset_A$ and $\Hset_B$. 
The maximally entangled state 
between two systems 
of dimension $D$ 
is defined by
$
| \Phi_{AB} \rangle = \frac{1}{\sqrt{D}} \sum_{j=0}^{D-1} |j\rangle_A\otimes |j\rangle_B 
$, where $\{ |j\rangle_A \}_{j=0}^{D-1}$ and $\{ |j\rangle_B \}_{j=0}^{D-1}$  
are respective orthonormal bases. 
Note that $I(A;B)_{\kb{\Phi}}=2\cdot \log(D)$ and $I(A\rangle B)_{\kb{\Phi}}= \log(D)$.

\subsection{Quantum Channel}
\label{subsec:Qchannel}
A quantum channel maps a quantum state at the sender system to a quantum state at the receiver system. 
Here, we consider a channel with two inputs, where one of the inputs, which is referred to as the channel state, is not controlled by the encoder.
Formally, a quantum state-dependent channel $(\channel_{E A'\rightarrow B} ,|\phi_{E E_0 C}\rangle)$ is defined by a   linear, completely positive, trace preserving map 
$
\channel_{E A'\rightarrow B}  
$ 
and a quantum state $|\phi_{E E_0 C}\rangle$.
This model can be interpreted as if the channel is entangled with the systems $E$, $E_0$, and $C$.
A quantum channel has a Kraus representation
\begin{align}
\channel_{EA'\rightarrow B}(\rho_{EA'})=\sum_j N_j \rho_{EA'} N_j^\dagger 
\end{align}
for all $\rho_{EA'}$, and for some set of operators $N_j$ such that $\sum_j N_j^\dagger N_j=\identity$. 
Every quantum channel $\channel_{E A'\rightarrow B}$ has an isometric extension $\Uset^{\channel}_{E A'\rightarrow BK}$, also called  a Stinespring dilation, such that 
\begin{align}
&\Uset^{\channel}_{E A'\rightarrow BK}(\rho_{EA'})=U\rho_{EA'} U^\dagger \\
& {\channel}_{E A'\rightarrow B}(\rho_{EA'})=\trace_K( U\rho_{EA'} U^\dagger )
\end{align}
where the operator $U$ is an isometry, \ie $ U^\dagger U=\identity$  \cite[Section VII]{BocheCaiCaiDeppe:14p}. 
The system $K$ is often associated with the decoder's environment, or with a malicious eavesdropper in  the wiretap channel model  \cite{Devetak:05p},
and the channel $\widehat{\channel}_{EA'\rightarrow K}(\rho_{EA'})=\trace_B( U\rho_{EA'} U^\dagger )$ is called the complementary channel for
${\channel}_{E A'\rightarrow B}$. 

We assume that both the channel state systems and the quantum channel have a product form. That is, the joint state of 
the systems $E^n=(E_1,\ldots,E_n)$, $E_0^n=(E_{0,1},\ldots,E_{0,n})$ and $C^n=(C_1,\ldots,C_n)$ is $
|\phi_{E E_0 C}\rangle^{\otimes n}$, and if the systems $A'^n=(A_1',\ldots,A_n')$ are sent through $n$ channel uses, then the input state $\rho_{ E^n A'^n}$ undergoes the tensor product mapping
$
\channel_{E^n A'^n\rightarrow B^n}\equiv  \channel_{E A'\rightarrow B}^{\otimes n} 
$. 
Given CSI, the transmitter has access to the systems $E_0^n$, which are entangled with the channel state systems $E^n$.
We will further consider a secrecy requirement that limits the information that the receiver can obtain on $C^n$.
The sender and the receiver are often referred to as Alice and Bob. 

\begin{remark}
\label{rem:mixed}
Our results apply to the case where $E$, $E_0$, and $C$ are in a mixed state as well. Specifically, given a mixed state $\varphi_{E E_0 C}$, there exists a purification $| \phi_{TE E_0 C} \rangle$, such that the reduced density operator for this purification is $\varphi_{E E_0 C}$. Hence, we can redefine the channel as follows. First, replace the channel state system $E$   by $\tilde{E}=(T,E)$, and then consider the quantum state-dependent channel
$\widetilde{\channel}_{\tilde{E} A'\rightarrow B}$, where
\begin{align}
\widetilde{\channel}_{TE  A'\rightarrow B}(\rho_{TEA'})=\channel_{EA'\rightarrow B}(\trace_T(\rho_{TEA'})) \,.
\label{eq:channelTm}
\end{align}
\end{remark}

\subsection{Less Noisy, Degradable, and Hadamard Channels}
\label{subsec:LessNHadamard}
In the unassisted setting, we will also be interested in the following special cases. 
\subsubsection{Less Noisy Output}
\label{subsec:LessN}
First, we define the class of state-dependent channels with a less noisy output.
\begin{definition}
\label{def:LessN}
A quantum state-dependent channel $(\channel_{EA'\rightarrow B},|\phi_{EE_0C}\rangle)$ is said to have a less noisy output if there exists
an isometric extension $\Uset^{\channel}_{E A'\rightarrow BK}$ such that for every $\rho_{AA'EC}$ with $\rho_{EC}=\phi_{EC}$,
\begin{align}
H(A|B)_\rho\leq H(A| KC)_\rho 
\end{align}
where $\rho_{ABKC}=\Uset^{\channel}_{E A'\rightarrow BK}(\rho_{AEA'C})$.
\end{definition}
The definition for a channel with a less noisy output can be equivalently stated as
\begin{align}
I(A;B)_\rho\geq& I(A; KC)_\rho
\intertext{or}
 I(A\rangle B)_\rho\geq& I(A\rangle K|C)_\rho 
\label{eq:IcLess}
\end{align}
for all $\rho_{AA'EC}$ with $\rho_{EC}=\phi_{EC}$. Intuitively, the channel output is less noisy than its environment. Specifically, if we could use $\Uset^{\channel}_{E A'\rightarrow BK}$ as a broadcast channel, with Receiver $B$ and Receiver $K$, then  (\ref{eq:IcLess}) would imply that quantum information can be reliably sent to Receiver $B$ at a higher rate than it can be sent to Receiver $K$, even if Receiver $K$ has complete knowledge of $C^n$, i.e. Receiver $K$ has $C^n$ as CSI. 
%
For a quantum channel $\Pset_{A'\rightarrow B}$ that does have a state, the definition above coincides with the standard definition of a less noisy broadcast channel \cite[Section II.C]{YardHaydenDevetak:11p}. 
 
A stronger requirement is that of a degradable channel \cite{YardHaydenDevetak:11p,DevetakShor:05p}.
\begin{definition}
A quantum state-dependent channel $(\channel_{EA'\rightarrow B},|\phi_{EE_0C}\rangle)$ is said to be degradable if there exists  an isometric extension $\Uset^{\channel}_{E A'\rightarrow BC_1 K}$ such that the complementary channel $\widehat{\channel}_{EA'\rightarrow C_1 K}$ is a concatenation of the main channel $\channel_{EA'\rightarrow B}$ and a degrading channel $\Dset_{B\rightarrow C_1K}$, i.e.
\begin{align}
\widehat{\channel}_{EA'\rightarrow C_1 K}=&\Dset_{B\rightarrow C_1K} \circ \channel_{EA'\rightarrow B}
\intertext{and for every $\rho_{AA'EC}$ with $\rho_{EC}=\phi_{EC}$,}
\rho_{ABKC_1}=&\rho_{ABKC}
\end{align}
where $\rho_{ABC_1 K C}\equiv \Uset^{\channel}_{E A'\rightarrow BC_1 K}(\rho_{AEA'C})$.
\end{definition}
Based on the data processing theorem, 
the conditions of the definition above imply that $I(A;B)_\rho\geq I(A;KC_1)_\rho=I(A;KC)_\rho$. Thereby, if a channel is degradable, then it has a less noisy output. Similarly, the intuition is that the state of the decoder's environment is a noisy version of the channel output state.

\subsubsection{Hadamard Channels}
\label{subsec:HadamardC}
Next, we consider the special case of Hadamard channels, which are defined as channels with an
entanglement-breaking complementary \cite{KingRuskaiNathanson:07p}. Here, we will use the following definition.
Consider an isometric channel
\begin{align}
\Vset_{EA'\rightarrow C_1 K B}(\rho_{EA'})=V\rho_{EA'} V^\dagger
\end{align}
with
\begin{align}
V
\equiv \sum_{x\in\Xset} |\eta_{C_1 K}^x \rangle\langle \zeta_{EA'}^x| \otimes |\psi_B^x\rangle 
\label{eq:iV}
\end{align}
for some pure states $|\eta_{C_1 K}^x \rangle$, $| \zeta_{EA'}^x\rangle$, and $|\psi_B^x\rangle$, such that
$\sum_x\kb{ \zeta_{EA'}^x}=\identity_{EA'}$, where
 $\{|\psi_B^x\rangle\}_{x\in\Xset}$ is an orthonormal basis for the output Hilbert space $\Hset_B$. 
Given a state $\rho_{AA'EC}$ at the input, the output state is then 
\begin{align}
\rho_{AC_1 K B C }= \Vset_{EA'\rightarrow C_1 K B} (\rho_{AEA'C}) \,.
\end{align}
The definition of a Hadamard channel is given below. 
\begin{definition}
\label{def:HadamardI}
A Hadamard state-dependent channel $(\channel^H_{EA'\rightarrow B},|\phi_{EE_0 C}\rangle)$ is a channel of the form
\begin{align}
\channel^H_{EA'\rightarrow B}(\rho_{EA'})=&\trace_{C_1 K}\left( V (\rho_{EA'}) V^\dagger  \right)
\end{align} 
with the isometry $V$ as in (\ref{eq:iV}), and
such that for every input state $\rho_{AA'EC}$ with $\rho_{EC}=\phi_{EC}$, 
 the output state  satisfies 
\begin{align}
\rho_{AKBC_1}=\rho_{AKBC}
\label{eq:hC1C}
\end{align}
where $\rho_{AC_1 K B C }= V \rho_{AEA'C} V^\dagger$.
\end{definition}
It can be shown that 
the definition of a Hadamard channel above coincides with the definition of a channel whose complementary is entanglement breaking (see detailed proof in \cite[Section II.C.2]{BradlerHaydenTouchetteWilde:10p}).
 
Observe that the complementary channel $\widehat{\channel}^H_{EA'\rightarrow C_1 K}$ 
can be simulated as follows. 
First, Bob performs a projective measurement on the channel output $B$ in the
basis $\{|\psi_B^x\rangle\}_{x\in\Xset}$. Then, given the measurement outcome $x^*$, the state $|\eta_{C_1 K}^{x^*} \rangle$ is prepared. 
It follows that a Hadamard channel is degradable, and thus has a less noisy output. 

\subsection{Coding}
\label{subsec:Mcoding}
We define 
a secrecy code to transmit quantum information given entanglement resources. We denote Alice and Bob's entangled systems by
$G_A$ and $G_B$, respectively.
With non-causal CSI, Alice has acess to the systems $E_0^n$, which are entangled with  the channel state sequence $E^n$.

\begin{definition} 
\label{def:EAcapacity}
A $(2^{nQ},2^{nR_e},n)$ quantum masking  code with rate-limited entanglement assistance and CSI at the encoder consists of the following:    
A quantum message state $\rho_{M}$, where $M$ is a system of dimension $|\Hset_M|=2^{nQ}$,  
a pure entangled state $\Psi_{G_A,G_B}$, where $|\Hset_{G_A}|=|\Hset_{G_B}|=2^{nR_e}$,
an encoding channel $\Fset_{ M G_A E_0^n \rightarrow A'^n}$, and a decoding channel 
$ \Dset_{B^n G_B\rightarrow \hM}  $.
We denote the code by $(\Fset,\Psi,\Dset)$.

The communication scheme is depicted in Figure~\ref{fig:MSKsiCode}b.  
The sender Alice has the systems  $M$, $G_A$, $E_0^n$, and $A'^n$, and the receiver Bob has the systems $B^n$,  $G_B$, and $\hM$. Alice encodes the quantum state of the message system $M$ using her share of the entangled resources $G_A$ and her access to the systems $E_0^n$ which are entangled with the channel state systems. To this end, she applies the encoding map $\Fset_{  M G_A E_0^n \rightarrow A'^n}$, which results in the input state
\begin{align}
\rho_{C^n E^n  A'^n G_B}= \Fset_{ E_0^n M G_A \rightarrow A'^n}(\phi_{C E E_0}^{\otimes n}\otimes \rho_M \otimes \Psi_{G_A G_B} ) 
\end{align}
 and transmits the systems $A'^n$ over 
$n$ channel uses of $\channel_{EA'\rightarrow B}$. Hence, the output state is
\begin{align}
\rho_{C^n B^n G_B}=\channel_{E^n A'^n\rightarrow B^n} (\rho_{C^n E^n A'^n  G_B}) \,.
\end{align}

 Bob receives the channel output and applies the decoding map $\Dset_{B^n G_B\rightarrow \hM}$ to the output systems $B^n$ and to his share of the entangled resources $G_B$,
 such that the state of $\hM$ is an estimate of the original state of the message system $M$. The estimation error is given by 
\begin{align}
e^{(n)}(\Fset,\Psi,\Dset,\rho_M)= 
\frac{1}{2}\norm{\rho_{M}-\Dset_{B^n G_B\rightarrow \hM}(\rho_{B^n G_B}) }_1
\end{align}
where $\rho_{B^n G_B}=\trace_{C^n}(\rho_{C^n B^n G_B})$.
The masking leakage rate of the code $(\Fset,\Psi,\Dset)$ is defined as
\begin{align}
\ell^{(n)}(\Fset,\Psi,\Dset,\rho_M)\triangleq  \frac{1}{n} I(C^n;B^n G_B)_\rho \,.
\label{eq:elln}
\end{align}
A $(2^{nQ},2^{nR_e},n,\eps,L)$ quantum masking code satisfies 
$
e^{(n)}(\Fset,\Psi,\Dset,\rho_M)\leq\eps $ 
and $\ell^{(n)}(\Fset,\Psi,\Dset,\rho_M)\leq L$ for 
all $\rho_M$.  
%
A triplet $(Q,L,R_e)$, where $Q,L,R_e\geq 0$, is called achievable  if for every $\eps,\delta>0$ and sufficiently large $n$, there exists a 
$(2^{nQ},2^{nR_e},n,\eps,L+\delta)$ quantum masking code. 

Next, we define the masking equivocation region with and without entanglement assistance.
A rate-leakage pair $(Q,L)$ is called achievable with entanglement assistance  if $(Q,L,R_e)$ is achievable for some $R_e\geq 0$.
 The entanglement-assisted masking region $\opR_{\text{Q}}^{\text{ea}}(\channel)$ is defined as the set of achievable pairs $(Q,L)$ with entanglement assistance and CSI at the encoder. 
Alternatively, one may fix the leakage rate and consider the optimal transmission rate. The quantum capacity-leakage function 
$\opC_{\text{Q}}^{\text{ea}}(\channel,L)$ is defined as the supremum of achievable rates $Q$ for a given leakage $L$. 
Note that $\opC_{\text{Q}}^\text{ea}(\channel,\infty)$ reduces to the standard definition of the entanglement-assisted capacity, without a masking requirement.

Furthermore, a rate-leakage pair $(Q,L)$ is called achievable without assistance  if $(Q,L,R_e=0)$ is achievable.
 The masking region $\opR_\text{Q}(\channel)$ and quantum capacity-leakage function $\opC_\text{Q}(\channel,L)$ without assistance are defined in a similar manner. 

One may also consider the transmission of classical information, where the message system is limited to states
$|m\rangle$ for $m=1,2,\ldots,2^{nR}$. 
In this case, we denote the classical masking regions
and capacity-leakage functions by $\opR_{\text{Cl}}^\text{ea}(\channel)$, $\opR_{\text{Cl}}(\channel)$ and
$\opC_{\text{Cl}}^\text{ea}(\channel,L)$, $\opC_{\text{Cl}}(\channel,L)$, respectively. 
\end{definition}

Note that $\opC_{\text{Q}}(\channel,L)$ and $\opC_{\text{Q}}^\text{ea}(\channel,L)$ have the units of \emph{qubits} per channel use, whereas the units of $\opC_{\text{Cl}}(\channel,L)$ and $\opC_{\text{Cl}}^\text{ea}(\channel,L)$ are \emph{classical bits} per channel use.

\begin{remark}
Notice that with entanglement assistance, the leakage rate (\ref{eq:elln}) includes Bob's share $G_B$ of the entanglement resources, since the decoder has access to both $B^n$ and $G_B$. This is another significant distinction from the classical case. In the classical setting, the leakage constraint does not need to include  shared randomness, as it cannot help the decoder. On the other hand, in our quantum model, we know that Bob can extract quantum information by performing measurements on $G_B$,  using the teleportation protocol for example.
\end{remark}

\begin{remark}
\label{rem:nonTmaskL}
Observe that if $L\geq 2\cdot H(C)_\phi$, then the masking requirement trivially holds because $I(C^n;B^n G_B)_\rho\leq
2H(C^n)_\rho=2nH(C)_\phi$. That is, if $L\geq 2H(C)_\phi$, then
$\opC_{\text{Q}}(\channel,L)=\opC_{\text{Q}}(\channel,\infty)$, and similarly for $\opC_{\text{Cl}}(\channel,L)$, $\opC_{\text{Q}}^\text{ea}(\channel,L)$, and $\opC_{\text{Cl}}^\text{ea}(\channel,L)$.
\end{remark}

\begin{remark}
Note that quantum state-dependent channels have in general a complicated behavior with respect to quantum information transmission and we cannot necessarily  expect that the region of achievable rate-leakage pairs $(Q,L)$ without entanglement assistance is equal to the limit of achievable rate-leakage pairs for $R_e\rightarrow 0$ \cite{BocheNotzel:14p1}.
\end{remark}

\subsection{Related Work}
\label{subsec:Previous}
We briefly review known results for the case where there is no masking requirement.
First, consider a quantum channel which is not affected by a channel state, \ie $\channel_{EA'\rightarrow B}(\rho_{EA'})=
\Pset_{A'\rightarrow B}(\trace_E(\rho_{EA'}))$. 
\begin{theorem} [see {\cite{BennettShorSmolin:99p,BennettShorSmolin:02p}}]
\label{theo:CeaNoSI}
The entanglement-assisted quantum capacity of a quantum channel $\Pset_{A'\rightarrow B}$ that does not depend on a channel state, without a masking requirement, is given by 
\begin{align}
\opC_\text{Q}^{\text{ea}}(\Pset,\infty)= \max_{|\phi_{A A'}\rangle } \frac{1}{2} I(A;B)_\rho 
\label{eq:QeaNoMsk}
\end{align}
with $\rho_{AB}\equiv   \Pset_{A'\rightarrow B}( \kb{ \phi_{A A' }})$, where $A$ is an auxiliary system of dimension $|\Hset_{A}|\leq |\Hset_{A'}|$.
\end{theorem}
Without assistance, a single letter characterization is an open problem for a general quantum channel.
Yet, a regularized formula for the quantum capacity was given in \cite{BarnumNielsenSchumacher:98p,Loyd:97p,Shor:02l,Devetak:05p}, in terms of the coherent information. Although calculation of such a formula is intractable in general, it provides a computable lower bound, and there are special cases where the capacity can be computed exactly
\cite{DevetakShor:05p}.
Define 
\begin{align}
\inC_\text{Q}(\Pset,\infty)=&   \max_{|\phi_{A A'}\rangle }  I(A\rangle B)_\rho
\label{eq:QNoMsk}
\end{align}
 with $\rho_{AB}\equiv   \Pset_{A'\rightarrow B}( \kb{ \phi_{A A' }})$ and $|\Hset_{A}|\leq |\Hset_{A'}|$.
\begin{theorem} [see {\cite{BarnumNielsenSchumacher:98p,Loyd:97p,Shor:02l,Devetak:05p,DevetakShor:05p}}]
\label{theo:CNoSI}
The quantum capacity of a quantum channel $\Pset_{A'\rightarrow B}$ that does not depend on a channel state, without assistance and without a masking requirement, is given by 
\begin{align}
\opC_\text{Q}(\Pset,\infty)= \lim_{k\rightarrow\infty}\frac{1}{k} \inC_\text{Q}(\Pset^{\otimes k},\infty) \,.
\label{eq:CqNosi}
\end{align}
Furthermore, if $\Pset_{A'\rightarrow B}$ has a less noisy output, then
\begin{align}
\opC_\text{Q}(\Pset,\infty)= \inC_\text{Q}(\Pset,\infty) \,.
\end{align} 
\end{theorem}
A multi-letter characterization as in (\ref{eq:CqNosi}) is often referred to as a regularized formula.
We note that in some cases, the entanglement-assisted capacity can be significantly higher than the capacity without assistance. For example,  the entanglement-assisted quantum capacity of a qubit erasure channel $\Pset_{A'\rightarrow B}(\rho)=(1-\eps)\rho+\eps \kb{e}$, where the erasure state $|e\rangle$ is orthogonal to the qubit space, is  $\opC_\text{Q}^{\text{ea}}(\Pset,\infty)=1-\eps$. 
On the other hand, without assistance, the quantum capacity 
is $\opC_\text{Q}(\Pset,\infty)=1-2\eps$ for $0\leq \eps< \frac{1}{2}$, and
$\opC_\text{Q}(\Pset,\infty)=0$ for $\eps\geq\frac{1}{2}$ \cite{BennettDiVincenzoSmolin:97p}.
\begin{remark}
\label{rem:bennettEA}
Theorem~\ref{theo:CeaNoSI} is an interesting example for a general phenomenon in quantum information theory. As was pointed out in
\cite{Bennett:19t}, using entanglement resources has two benefits:
\begin{enumerate}[1)]
\item
Entanglement-assisted protocols can accomplish a performance increase compared to unassisted protocols. 
\item
Introducing entanglement resources transforms the capacity evaluation from an uncomputable task to  an optimization that can be easily performed (numerically).
\end{enumerate}
%
\end{remark}

\begin{remark}
Among other important aspects for the design and development of communication systems, it is crucial to evaluate the current performance, how close it is to the optimum, and whether it is worth to invest in further development of a particular technology \cite{AroaraSinghRandahawa:19p,CostelloForney:07c}. For those purposes, given an estimate of the channel parameters, it can be useful to calculate the capacity as a number, and the general formula may be less interesting for such purposes.
At the time of writing, a realization of a full-scale quantum communication system that approaches the Shannon-theoretic limits does not exist, and we can only hope that future systems of quantum communication will reach the level of maturity of classical commercial systems today, which already employ sophisticated error correction codes with near-Shannon limit performance \cite{NisiotiThomos:20a,RichardsonKudekar:18p}.
\end{remark}

\begin{remark}
\label{rem:AhlswedeReg}
As Ahlswede remarked in \cite{Ahlswede:06t}, for the purpose of computing the capacity,
 a regularized characterization as in Theorem~\ref{theo:CNoSI}  is not necessarily a problem. 
Given a specific quantum channel, e.g. an optical fiber channel with specific parameters, a practitioner is usually interested in computing the channel capacity as a number (see previous remark). 
Following Ahlswede's argument in \cite{Ahlswede:06t}, given a fixed channel $\Pset_{A'\rightarrow B}$, if
the sequence $\{\frac{1}{n} \inC_Q(\Pset^{\otimes n},\infty)\}_{n\geq 1}$ has a sufficiently high convergence rate, say exponentially fast, then the quantum capacity can be approximated numerically up to any desired precision.
%
Whereas, from a theoretical perspective, a single-letter formula usually offers a lot more insight.
We will come back to this in Section~\ref{sec:summ}.
%
%
\end{remark}

Next, we move to Dupuis' result on a quantum state-dependent channel $\channel_{EA'\rightarrow B}$ with entanglement assistance and CSI at the encoder.
Denote the reduced density matrix of the channel state system by $\phi_E\triangleq \trace_{E_0 C}(\phi_{E E_0 C})$.
%
\begin{theorem} [see {\cite{Dupuis:08a,Dupuis:09c}}]
\label{theo:CeaNC}
The entanglement-assisted quantum capacity of a quantum channel $(\channel_{EA'\rightarrow B},\phi_{EE_0})$, with CSI at the encoder and without a masking requirement, is given by
\begin{align}
\opC_Q^{\text{ea}}(\channel,\infty)= \sup_{ \rho_{A E A' } \,:\; \rho_E=\phi_E } \frac{1}{2} [ I(A;B)_\rho - I(A;E)_\rho ] 
\label{eq:QeaNoMskE}
\end{align}
 with $\rho_{AB}\equiv   \channel_{EA'\rightarrow B}( \rho_{A E A'})$.
\end{theorem}

 %

\section{Information Theoretic Tools}
In this section, we present tools that will be useful in the analysis. We begin with the decoupling theorem. 
We establish an \emph{i.i.d.} version of the decoupling theorem, so that we will not have to worry about the one-shot setting in the achievability proof for our capacity theorems. 

We use the following definitions. An operator $V_{A\rightarrow B}$ that has $0$-$1$ singular values is called a partial isometry.
 For every pair of Hilbert spaces $\Hset_{A}$ and $\Hset_B$ with orthonormal bases $\{ |i_A\rangle \}$ and $\{ |j_B\rangle \}$, respectively, define the operator $\text{op}_{A \rightarrow B}(|\psi_{AB}\rangle)$ by
\begin{align}
\text{op}_{A \rightarrow B}(|i_A\rangle \otimes |j_B\rangle )\equiv |j_B \rangle\langle i_A| \,.
\label{eq:op}
\end{align}
While the operation above depends on the choice of bases, we will not specify those since it is not important for our purposes.
To generalize this definition to any state $|\psi_{AB}\rangle$, consider its decomposition 
$|\psi_{AB}\rangle=\sum_{i,j} a_{i,j} |i_A\rangle \otimes |j_B\rangle $, and define 
 $\text{op}_{A \rightarrow B}(|\psi_{AB}\rangle)=\sum_{i,j} a_{i,j} \text{op}_{A \rightarrow B}(|i_A\rangle \otimes |j_B\rangle )$.
Before presenting the decoupling theorem, we give the following useful properties of $\text{op}_{A \rightarrow B}(|\psi_{AB}\rangle)$, as stated in \cite{Dupuis:10z}.
\begin{lemma}[{\cite[Lemma 2.7]{Dupuis:10z}}]
\label{lemm:opCommut}
For every pure states $|\psi_{AB}\rangle$ and $|\theta_{AC}\rangle$,
\begin{align}
 \text{op}_{A\rightarrow B}(|\psi_{AB}\rangle)\cdot |\theta_{A C}\rangle= \text{op}_{A\rightarrow C}(|\theta_{AC}\rangle)\cdot |\psi_{A B}\rangle \,.
\end{align}
\end{lemma}

\begin{lemma}[{\cite[Lemma 2.8]{Dupuis:10z}}]
\label{lemm:opPhi}
For every pure state $|\psi_{AB}\rangle$,
\begin{align}
\sqrt{|\Hset_A|} \text{op}_{A\rightarrow B}(|\psi_{AB}\rangle)\cdot |\Phi_{A A'}\rangle= |\psi_{A'B}\rangle \,.
\end{align}
\end{lemma}

We give our i.i.d. version of the decoupling theorem below.
\begin{theorem}[The i.i.d. decoupling theorem]
\label{theo:decoupling}
Let $|\omega_{ABK}\rangle$ be a pure state, and $S$, $R$, $G_1$, $G_2$ be quantum systems at state
\begin{align}
|\sigma_{S R G_1 G_2}\rangle = |\Psi_{S R}\rangle \otimes |\Phi_{G_1 G_2} \rangle 
\end{align}
in the product Hilbert space $\Hset_S^{\otimes 2}\otimes \Hset_G^{\otimes 2}$.
Let $W_{S G_1 \rightarrow A^n}$ be a full-rank partial isometry, and denote
\begin{align}
|\sigma_{A^n R G_2}\rangle=W_{S G_1\rightarrow A^n} |\sigma_{S R G_1 G_2} \rangle \,.
\end{align}  
Define the quantum channel $\Tset_{A\rightarrow K}$ by
\begin{align}
\Tset_{A\rightarrow K}(\rho_A)=|\Hset_A|\trace_B\left[ \text{op}_{A\rightarrow BK}(|\omega_{ABK}\rangle)(\rho_A)  \right] \,.
\end{align}
Then,
\begin{align}
\int_{\mathbb{U}_{A^n}} dU_{A^n} \, \norm{
\Tset_{A\rightarrow K}^{\otimes n}(U_{A^n}  \sigma_{A^n R})-\omega_K\otimes\sigma_R
}_1 \leq \sqrt{ \frac{|\Hset_S|}{|\Hset_{G}|} 2^{-n H(A|K)_{\omega}+n\eps(n)} }
\label{eq:Decoup1}
\end{align}
and
\begin{align}
\int_{\mathbb{U}_{A^n}} dU_{A^n} \, \norm{
\Tset_{A\rightarrow K}^{\otimes n}(U_{A^n}  \sigma_{A^n R G_2})-\omega_K\otimes\sigma_{RG_2}
}_1 \leq \sqrt{ |\Hset_S||\Hset_{G}| 2^{-n H(A|K)_{\omega}+n\eps(n)} }
\label{eq:Decoup2}
\end{align}
where the integral is over the Haar measure on all unitaries $U_{A^n}$, and $\eps(n)$ tends to zero as $n\rightarrow\infty$.
\end{theorem}
The proof of Theorem~\ref{theo:decoupling} is given in Appendix~\ref{app:decoupling}, based on the one-shot decoupling theorem along  with arguments from \cite{Dupuis:10z}.
Intuitively, the theorem above shows that by choosing a unitary $U_{A^n}$ uniformly at random, we can  decouple between  $K$ and $R$ provided that the dimensions satisfy 
\begin{align}
\frac{1}{n}\log \frac{|\Hset_S|}{|\Hset_{G}|} < H(A|K)_{\omega}-\eps(n) \,.
\intertext{Similarly, $K$ and $(R,G_2)$ can be decoupled if}
\frac{1}{n}\log (|\Hset_S||\Hset_{G}|) < H(A|K)_{\omega}-\eps(n) \,.
\end{align}


Uhlmann's theorem \cite{Uhlmann:76p} is often used along the decoupling approach to establish the existence of proper encoding and decoding operations.
\begin{theorem}[Uhlmann's theorem {\cite{Uhlmann:76p}\cite[Corollary 3.2]{Dupuis:10z}}]
For every pair of pure states $|\psi_{AB}\rangle$ and $|\theta_{AC}\rangle$ that satisfy $\norm{\psi_A-\theta_A}_1 \leq \eps$, there exists an isometry $F_{B\rightarrow C}$ such that
$ \norm{ (\identity\otimes F_{B\rightarrow C})\psi_{AB}-\theta_{AC} }_1 \leq 2\sqrt{\eps} $.
\end{theorem}

\begin{remark}
\label{rem:decoupIt}
We give a rough explanation, in the spirit of \cite[Section 24.10]{Wilde:17b}, to demonstrate how decoupling can be useful in an achievability proof for quantum communication.
Consider a quantum channel $\Pset_{A'\rightarrow B}$ that does not depend on a channel state, without entanglement assistance.
Let $|\Psi_{MR}\rangle$ be a purification of the message state $\rho_M$, where $R$ is Alice's reference system.
Suppose that $|\sigma_{RB^n K^n J_1}\rangle$ is a purification of the joint state of Alice's reference system $R$, the channel output  $B^n$, and Bob's environment $K^n$, with a purifying system $J_1$. Observe that if the reduced state $\sigma_{R K^n J_1}$ 
is  a product state, i.e. $\sigma_{RK^n J_1}=\psi_R \otimes \xi_{K^n J_1}$, then it has a purification of the form
$|\Psi_{MR}\rangle\otimes |\xi_{K^n J_1 J_2}\rangle$. Since all purifications are related by isometries \cite[Theorem 5.1.1]{Wilde:17b}, there exists an isometry $D_{B^n\rightarrow M  J_2}$ such that  $|\Psi_{MR}\rangle\otimes |\xi_{K^n J_1 J_2}\rangle=   D_{B^n\rightarrow M  J_2} |\sigma_{RB^n K^n J_1}\rangle$.
Tracing out $R$, $K^n$, $J_1$, and $J_2$, it follows that there exists a decoding map $\Dset_{B^n\rightarrow M}$ that recovers the message state, i.e.
$\rho_M=\Dset_{B^n\rightarrow M}(\rho_{B^n})$. Therefore, in order to show that there exists a reliable coding scheme, it is sufficient to encode in such a manner that approximately decouples between Alice's reference system and Bob's environment, i.e., such that
$\sigma_{RK^n J_1}\approx\psi_R \otimes \xi_{K^n J_1}$.
\end{remark}

\section{Main Results}
We state our results on the quantum state-dependent channel $\channel_{E A'\rightarrow B}$ with masking.

\subsection{Rate-Limited Entanglement Assistance}
First, we consider communication with rate-limited entanglement assistance. We give an achievability result which will be used in the sequel to prove the direct part for the quantum masking region, both with and without entanglement assistance.

\begin{theorem}
\label{theo:MskAchiev}
Let $\left(\channel_{EA'\rightarrow B}, |\phi_{E E_0 C}\rangle \right)$ be a quantum state-dependent channel.
Let $\rho_{ E A' A C}$ be any mixed state with $\rho_{EC}=\phi_{EC}$. Then, any rate point $(Q,L,R_e)$ such that
\begin{align}
  Q+R_e \leq& H(A|E C)_\rho  			\label{eq:MskAchiev1} \\
  Q-R_e \leq& I(A\rangle B)_\rho 	\label{eq:MskAchiev2} \\
 	 L   \geq& I(C;AB)_\rho 					\label{eq:MskAchiev3} 
\end{align}
is achievable for transmission with rate-limited entanglement assistance and CSI at the encoder, where the auxiliary system $A$ is arbitrary, with $\rho_{A  B C}=\channel_{E A'\rightarrow B}(\rho_{ A E A'  C})$.
That is, for every $\eps,\delta>0$ and sufficiently large $n$, there exists a $(2^{nQ},2^{nR_e},n,\eps,L+\delta)$ quantum masking code with CSI $E_0^n$ at the encoder, and such that $C^n$ is masked from the decoder.
\end{theorem}
The proof of Theorem~\ref{theo:MskAchiev} is given in Appendix~\ref{app:MskAchiev}.
The theorem above provides an achievability result that takes into account the  tradeoff between communication and resource rates.
 As a byproduct, the coding scheme executes state merging \cite{HorodeckiOppenheim:07p}, as Alice effectively sends her share $G_A$ to Bob.
 Namely, as can be seen in Appendix~\ref{app:MskAchiev}, we begin the protocol with an entangled state $\Psi_{G_A G_B}$, where Alice has the system $G_A$ and Bob has $G_B$; and when the protocol has been completed, Bob ends up with the systems $G_A'$ and $G_B'$ at state $\approx \Psi_{G_A' G_B'} $.

\subsection{Entanglement-Assisted Masking Region}
Next, we consider entanglement-assisted masking, where Alice and Bob have unlimited entanglement resources. In this section, we assume that the channel state systems are in maximally correlated
state 
\begin{align}
\varphi_{EE_0 C}=\sum_{s\in\Sset} q(s) \kb{s}_E\otimes \kb{s}_{E_0} \otimes \kb{s}_C 
\label{eq:CondSchmidt}
\end{align}
where $q(s)$ is a probability distribution, and $\{|s\rangle_E\}$, $\{|s\rangle_{E_0}\}$, $\{|s\rangle_{C}\}$ each form an orthonormal basis of the respective Hilbert space. Notice that the state above is separable and not entangled.
We note that in general, one can always apply the spectral theorem to an individual system and obtain a decomposition of the form
$
\varphi_{E}=\sum
q(s) \kb{s}_E
$, 
and similarly for $\varphi_{E_0}$ and $\varphi_C$. Yet, the assumption in (\ref{eq:CondSchmidt}) implies that $E$, $E_0$, and $C$ have the same spectrum.
In addition, if Alice performs a projective measurement on the CSI systems in the basis $\{|s\rangle_{E_0}\}_{s\in\Sset}$, then the problem reduces to that of a quantum channel that depends on a classical random variable $S\sim q(s)$. Hence, this assumption holds in the special case of a classical channel state.
However, in our setting, Alice may perform any quantum operation on the CSI systems $E_0^n$. Thus, given the restriction (\ref{eq:CondSchmidt}), the setting is less general than our original model, and yet it is more general than that of a classical channel state. The masking problem given entanglement assistance for a general quantum state $\varphi_{EE_0 C}$ remains open.
 
We determine the entanglement-assisted masking region and 
capacity-leakage function, for the transmission of either quantum information or classical information.
Define
\begin{align}
\mathcal{R}_{\text{Q}}^\text{ea}(\channel)=&
\bigcup_{ \rho_{ E A' A C} \,:\: \rho_{EC}=\varphi_{EC}}
\left\{ \begin{array}{rl}
  (Q,L) \,:\;
	0\leq Q \leq& \frac{1}{2}[ I(A;B)_\rho- I(A; EC)_\rho]  \\
  L   \geq& I(C;AB)_\rho
	\end{array}
\right\}
\intertext{and}
\mathcal{R}_{\text{Cl}}^\text{ea}(\channel)=&
\bigcup_{ \rho_{ E A' A C} \,:\: \rho_{EC}=\varphi_{EC}}
\left\{ \begin{array}{rl}
  (R,L) \,:\;
	0\leq R \leq&  I(A;B)_\rho- I(A; EC)_\rho  \\
  L   \geq& I(C;AB)_\rho
	\end{array}
\right\}
\label{eq:calRClea}
\end{align}
with $\rho_{A  B C}=\channel_{E A'\rightarrow B}(\rho_{ A E A'  C})$.
\begin{theorem}
\label{theo:MskEA}
Let $\left(\channel_{EA'\rightarrow B}, \varphi_{E E_0 C}\right)$ be a quantum state-dependent channel with CSI at the encoder, with maximally correlated channel state systems, as in (\ref{eq:CondSchmidt}).
Then, the entanglement-assisted quantum masking region and classical masking region  
are given by
\begin{align}
\mathbb{R}_{\text{Q}}^\text{ea}(\channel)=&\mathcal{R}_{\text{Q}}^\text{ea}(\channel) 
\intertext{and}
\mathbb{R}_{\text{Cl}}^\text{ea}(\channel)=&\mathcal{R}_{\text{Cl}}^\text{ea}(\channel)
\end{align}
respectively. 
\end{theorem}
The proof of Theorem~\ref{theo:MskEA} is given in Appendix~\ref{app:MskEA}.
The direct part is based on Theorem~\ref{theo:MskAchiev}. As can be seen in Appendix~\ref{app:MskEA}, the entanglement-assisted capacity can be achieved if the entanglement rate is higher than
$ \frac{1}{2} I(A\rangle B)_\rho-\frac{1}{2} H(A|EC)_\rho $. 
The converse proof requires more attention.
As we have three channel state systems, namely, $E_0^n$, $E^n$, and $C^n$, we need to choose the auxiliary system $A$ 
such that both the communication and leakage rate constraints are met. Thereby, the assumption in (\ref{eq:CondSchmidt}) is only required for the converse proof.

Equivalently, we can characterize the  capacity-leakage function with entanglement assistance. 
%
 The following corollary is an immediate consequence of Theorem~\ref{theo:MskEA}.
\begin{corollary}
\label{coro:MskEA}
Given $\left(\channel_{EA'\rightarrow B}, \varphi_{E E_0 C}\right)$ as in Theorem~\ref{theo:MskEA},
%
the entanglement-assisted quantum capacity-leakage function and classical capacity-leakage function  
are given by
\begin{align}
\opC_{\text{Q}}^\text{ea}(\channel,L)=&
\sup_{ \substack{ \rho_{A E A' C} \,:\;  I(C;AB)_\rho \leq L \\ \rho_{EC}=\phi_{EC}  }} \,\;\frac{1}{2}[ I(A;B)_\rho- I(A;E C)_\rho]
\label{eq:CLfEAq}
\intertext{and}
\opC_{\text{Cl}}^\text{ea}(\channel,L)=&
\sup_{ \substack{ \rho_{A E A' C} \,:\;  I(C;AB)_\rho \leq L \\ \rho_{EC}=\phi_{EC}  }} \,\;[ I(A;B)_\rho- I(A;E C)_\rho]
\label{eq:CLfEAcl}
\end{align}
respectively, with $\rho_{A  B C}=\channel_{E A'\rightarrow B}(\rho_{ A E A'  C})$. 
\end{corollary}

\begin{remark}
\label{rem:Uncomput}
It was mentioned in Remark~\ref{rem:bennettEA}, point 2), that in various settings entanglement assistance leads to a characterization that is easy to compute.
Unfortunately,  this goal was not accomplished in the present work nor in the previous results by Dupuis \cite{Dupuis:09c}. Clearly, the characterization of the masking region and the capacity-leakage function has a single-letter form with respect to the channel dependency. However, there is no upper bound on the necessary dimension of the auxiliary system $A$ in Theorems~\ref{theo:CeaNC}, \ref{theo:MskAchiev}, and \ref{theo:MskEA}, and in Corollary~\ref{coro:MskEA}.
If we could restrict the optimization to pure states $|\psi_{EA'AC}\rangle$, then we would argue that the dimension 
of $A$ need not be larger than the Schmidt rank of $|\psi_{EA'AC}\rangle$, hence optimizing over a Hilbert space of dimension
$|\Hset_A|=|\Hset_{A'}||\Hset_{E}||\Hset_{C}|$ is sufficient.
Note that one can always compute achievable rates by choosing an arbitrary dimension, but the optimal rates cannot be  computed with absolute precision in general. 
Yet, in analogy to Remark~\ref{rem:AhlswedeReg}, for a fixed channel $\channel_{EA'\rightarrow B}$, state $\varphi_{EE_0 C}$, and leakage rate $L$, the values of (\ref{eq:CLfEAq}) and
(\ref{eq:CLfEAcl}) can be approximated if there exists a computable function to upper bound the dimension of the auxiliary system in the optimization problem as a function of the required precision.
 %
\end{remark}

\subsection{Unassisted Masking Region}
In this section, we consider masking without assistance. 
 We establish a regularized formula for the quantum masking region and 
capacity-leakage function for the transmission of quantum information. For the class of Hadamard channels, we obtain single-letter inner and outer bounds, which coincide in the standard case of a channel that does not depend on the state.
Define
\begin{align}
\mathcal{R}_{\text{Q,in}}(\channel)=&
\bigcup_{ \rho_{ E A' A C} \,:\: \rho_{EC}=\phi_{EC}}
\left\{ \begin{array}{rl}
  (Q,L) \,:\;
	0\leq Q \leq& \min\{ I(A\rangle B)_\rho \,,\; H(A|E C)_\rho \}  \\
  L   \geq& I(C;AB)_\rho
	\end{array}
\right\}
\intertext{
with $\rho_{A  B C}=\channel_{E A'\rightarrow B}(\rho_{ A E A'  C})$. Furthermore, given an isometric extension 
$\Uset^{\channel}_{EA'\rightarrow B K}$, define
}
\mathcal{R}_{\text{Q,out}}(\Uset^{{\channel}})=&
\bigcup_{ \rho_{ E A' A C} \,:\: \rho_{EC}=\phi_{EC}}
\left\{ \begin{array}{rl}
  (Q,L) \,:\;
	0\leq Q \leq&  H(A| C K)_\rho   \\
  L   \geq& I(C;AB)_\rho
	\end{array}
\right\}
\end{align}
with $\rho_{A  B K C}=\Uset^\channel_{E A'\rightarrow B K}(\rho_{ A E A'  C})$.
Recall that we have defined the class of Hadamard channels in Subsection~\ref{subsec:HadamardC},  in terms of an isometric extension 
$\Vset^{\,\text{H}}_{EA'\rightarrow BC_1 K}$ of a particular form (see Definition~\ref{def:HadamardI}).
Our main result on channel state masking without assistance is given below.
\begin{theorem}
\label{theo:MskQ}
Let   $\left(\channel_{EA'\rightarrow B}, |\phi_{E E_0 C}\rangle \right)$ be a quantum state-dependent channel with CSI at the encoder.
Then,
\begin{enumerate}[1)]
\item
the quantum masking region  
is given by
\begin{align}
\mathbb{R}_{\text{Q}}(\channel)=  \bigcup_{k=1}^{\infty} \frac{1}{k} \mathcal{R}_{\text{Q,in}}(\channel^{\otimes k}) \,.
\end{align}
\item
For a Hadamard channel $\channel^{\,\text{H}}_{EA'\rightarrow B}$, 
the quantum masking region is bounded by
\begin{align}
\mathcal{R}_{\text{Q,in}}(\channel^{\,\text{H}}) \subseteq
\mathbb{R}_{\text{Q}}(\channel^{\,\text{H}})\subseteq  \mathcal{R}_{\text{Q,out}}(\Vset^{\,\text{H}}) \,.
\end{align}
\end{enumerate}
\end{theorem}
The proof of Theorem~\ref{theo:MskQ} is given in Appendix~\ref{app:MskQ}. Our converse proof is based on different arguments from those in the classical converse proof by Merhav and Shamai \cite{MerhavShamai:07p}.
In the classical proof, the derivation of the bounds on both communication rate $Q$ and leakage rate $L$ begins with Fano's inequality.
Here, on the other hand, entangled states may have a negative conditional entropy; hence the leakage bound is derived in a different manner, using the coherent information bound on the rate. The direct part is a consequence of our previous result on masking with rate-limited entanglement assistance (see Theorem~\ref{theo:MskAchiev}).
We derive a single-letter outer bound for Hadamard channels using the special properties of those channels. To bound the communication rate $Q$, we only need to use the fact that Hadamard channels are degradable. As for the bound on the leakage rate $L$, here we observe that for Hadamard channels, there exists a channel from the output $B$ to $BC_1K$, i.e. the channel output combined with the decoder's environment.

\begin{remark}
Observe that for a pure input state $\rho_{ E A' A C}=\kb{\psi_{E A' A C}}$, the extended output systems $A,B,C,K$ are in a pure state as well,
which in turn implies that
\begin{align}
H(A|CK)_\rho=& H(ACK)_\rho-H(CK)_\rho \nonumber\\
=& H(B)_\rho-H(AB)_\rho 
=I(A\rangle B)_\rho 
\end{align}
where $K$ is part of the output of the isometric extension $\Vset^{\,\text{H}}_{EA'\rightarrow BC_1 K}$ (see Definition~\ref{def:HadamardI}).
It follows that the quantum masking region is bounded by
\begin{align}
\mathbb{R}_{\text{Q}}(\channel)\supseteq \mathcal{R}_{\text{Q,in}}(\channel) \supseteq 
\bigcup_{ |\psi_{ E A' A C}\rangle \,:\: \psi_{EC}=\phi_{EC}}
\left\{ \begin{array}{rl}
  (Q,L) \,:\;
	0\leq Q \leq& I(A\rangle B)_\rho   \\
  L   \geq& I(C;AB)_\rho
	\end{array}
\right\} 
\end{align}
with $\rho_{A  B C}=\channel_{E A'\rightarrow B}(\kb{\psi_{ A E A'  C}})$.
In the trivial case of a quantum channel $\Pset_{A'\rightarrow B}$ that does not depend on a state, 
the masking region can be achieved with pure product states $|\psi_{EC A A' }\rangle=|\phi_{EC}\rangle\otimes |\theta_{AA'}\rangle$, hence the inner bound and the outer bound coincide, i.e.
\begin{align}
\mathcal{R}_{\text{Q,in}}(\Pset) =\mathcal{R}_{\text{Q,out}}(\Uset^\Pset) =
\bigcup_{ |\theta_{  A A' } \rangle}
\left\{ \begin{array}{rl}
  (Q,L) \,:\;
	0\leq Q \leq&  I(A\rangle B)_\rho   \\
  L   \geq& 0
	\end{array}
\right\}\,.
\end{align}
Then, if $\Pset_{A'\rightarrow B}$ is a Hadamard channel, 
 the quantum masking region is 
$\mathbb{R}_{\text{Q}}(\Pset)=\mathcal{R}_{\text{Q,in}}(\Pset) =\mathcal{R}_{\text{Q,out}}(\Uset^\Pset)$.
\end{remark}
As an immediate consequence of Theorem~\ref{theo:MskQ}, we obtain the following characterization of the capacity-leakage function. 
\begin{corollary}
\label{coro:MskQ}
Let   $\left(\channel_{EA'\rightarrow B}, |\phi_{E E_0 C}\rangle \right)$ be a quantum state-dependent channel with CSI at the encoder.
\begin{enumerate}[1)]
\item
The quantum capacity-leakage function  
is given by
\begin{align}
\opC_{\text{Q}}(\channel,L)=  \lim_{k\rightarrow\infty} \frac{1}{k} 
\sup_{
\substack{
\rho_{ E^k A'^k A^k C^k} \,:\: \rho_{E^k C^k}=\phi_{EC}^{\otimes k} \\
L   \geq \frac{1}{k} I(C^k;A^k B^k)_\rho
}
}  \min\{ I(A^k\rangle B^k)_\rho \,,\; H(A^k|E^k C^k)_\rho \}
\end{align}
with $\rho_{A^k  B^k C^k}=\channel_{E A'\rightarrow B}^{\otimes k}(\rho_{ A^k E^k A'^k  C^k})$.
\item
For a Hadamard channel $\channel^{\,\text{H}}_{EA'\rightarrow B}$, 
the quantum masking region is bounded by
\begin{align}
\opC_{\text{Q},L}(\channel^{\,\text{H}})\geq&
\sup_{
\substack{
\rho_{ E A' A C} \,:\: \rho_{E C}=\phi_{EC} \\
L   \geq  I(C;A B)_\rho
}
}  \min\{ I(A\rangle B)_\rho \,,\; H(A|E C)_\rho \} 
\intertext{and}
\opC_{\text{Q},L}(\channel^{\,\text{H}})\leq&  
\sup_{
\substack{
\rho_{ E A' A C} \,:\: \rho_{E C}=\phi_{EC} \\
L   \geq  I(C;A B)_\rho
}
}   H(A| C K)_\rho 
\end{align}
with $\rho_{A  BC_1 K C}=\Vset^{\text{H}}_{E A'\rightarrow BC_1 K}(\rho_{ A E A'  C})$.
\end{enumerate}
\end{corollary}
The computational issues that were raised in Remarks \ref{rem:AhlswedeReg} and \ref{rem:Uncomput}  apply to the results in Theorem~\ref{theo:MskQ} and
Corollary~\ref{coro:MskQ} as well.

\subsection{Example: State-Dependent Dephasing Channel}
To illustrate our results, we consider a quantum dephasing channel that depends on a classical state and compute achievable rate-leakage regions. Consider a pair of qubit dephasing channels
\begin{align}
\Pset_{A'\rightarrow B}^{(s)}(\rho)=(1-\eps_s)\rho +\eps_s Z\rho Z \,,\; s=0,1
\end{align}
where $Z$ is the phase-flip Pauli matrix, and
 $\eps_0,\eps_1$ are given parameters, with
$0\leq \eps_s\leq 1$ for $s\in\{0,1\}$.
Suppose the channel state systems $E$, $C$, and $E_0$ contain a copy of a classical random bit $S\sim \text{Bernoulli}(q)$, with
$0\leq q\leq \frac{1}{2}$. 
Then, the qubit state-dependent channel $\channel_{EA'\rightarrow B}$ is defined such that
given an input state 
\begin{align}
\rho_{EA'}=(1-q)\kb{0}_E\otimes \sigma_0+q\kb{1}_E\otimes \sigma_1
\end{align}
 the output state is
\begin{align}
\channel_{EA'\rightarrow B}(\rho_{EA'})=(1-q)\Pset_{A'\rightarrow B}^{(0)}(\sigma_0)+q\Pset_{A'\rightarrow B}^{(1)}(\sigma_1) \,.
\end{align}
Observe that the dephasing channel can also be viewed as a controlled phase-flip gate that is controlled by a classical random bit.
In particular,  the state-dependent channel above is "controlled" by a random variable $W_S$ such that given $S=s$,
\begin{align}
W_s\sim\text{Bernoulli}(\eps_s) \,.
\end{align}

Consider the transmission of classical information while masking the channel state sequence from the receiver.
In the special case of $\eps_0=0$ and $\eps_1=1$, we have $W_S=S$. That is, the channel acts as a controlled-$Z$ gate where the channel state system $E$ (or $S$) is the controlling qubit.  The entanglement-assisted masking region in this case is 
\begin{align}
\opR_{\text{Cl}}^{\text{ea}}(\channel) =
\bigcup_{ 0\leq \lambda\leq 1}
\left\{ \begin{array}{rl}
  (R,L) \,:\;
	0\leq R \leq&  2  \\
  L   \geq& 0
	\end{array}
\right\} \,.
\end{align}
To understand why, observe that given CSI at the encoder, Alice can first perform the controlled phase-flip operation on her entangled qubit, and then use the super-dense coding protocol. Doing so, she effectively eliminates the phase flip operation of the channel.  Subsequently, Bob receives the information perfectly, at rate of $2$ classical bits per channel use, regardless of the values of $S^n$. Hence, there is no leakage.

Now, let $ \eps_0\leq \frac{1}{2} \leq \eps_1 $, and define
\begin{align}
\bar{\eps}=&(1-q)\eps_0 +q\eps_1  \label{eq:barEps} \\
\hat{\eps}=&(1-q)\eps_0 +q(1-\eps_1) \,.
\label{eq:hatEps}
\end{align}
Without CSI, the channel can be reduced to a standard dephasing channel that does not depend on a state, with the average phase-flip parameter $\bar{\eps}$.
%
%
First, we use Theorem~\ref{theo:MskEA} to show that the entanglement-assisted masking region is bounded by
\begin{align}
\opR_{\text{Cl}}^{\text{ea}}(\channel) \supseteq \inR_{0} =
\left\{ \begin{array}{rl}
  (R,L) \,:\;
	0\leq R \leq&  2-h_2(\lambda*\bar{\eps} )  \\
  L   \geq& h_2(\lambda*\bar{\eps} )-(1-q)h_2(\lambda*\eps_0)-q h_2(\lambda*\eps_1)
\end{array}
\right\}
\end{align}
where $h_2(x)=-x\log_2 x-(1-x)\log_2(1-x)$ is the binary entropy function, and $a*b=(1-a)b+a(1-b)$.
To show achievability of the region above, suppose that Alice performs phase-flip operation controlled by a random variable
 $Y\sim\text{Bernoulli}(\lambda)$ which is statistically independent of $S$. 
That is, $\rho_{EA'A}=\phi_E\otimes \rho_{A'A}$, with
\begin{align}
\rho_{A'A}= [(1-\lambda)\Phi_{AA'}+\lambda (1\otimes Z)\Phi_{AA'}(1\otimes Z)] \,.
\end{align}
Then, Bob receives the output of a phase-flip gate that is controlled by $(W_S+Y) \mod 2$, which is distributed according to 
$\text{Bernoulli}(\lambda*\bar{\eps})$ (see (\ref{eq:barEps})). Thus, for the output state $\rho_{SBA}=\channel_{EA'\rightarrow B}(\rho_{SEA'A})$, we have 
\begin{align}
&I(A;B)_\rho-I(A;S)_\rho=I(A;B)_\rho=H(A)_\rho+H(B)_\rho-H(AB)_\rho=1+1-h_2(\lambda*\bar{\eps})
\\
&I(S;AB)_\rho=H(AB)_\rho-H(AB|S)_\rho=h_2(\lambda*\bar{\eps})-[(1-q)h_2(\lambda*\eps_0)+q h_2(\lambda*\eps_1)] \,.
\end{align}
We note that as Alice's input is in a product state with the channel state system $E$,  this rate-leakage region can also be achieved without CSI. 

Next, we derive achievability of the following region,  
\begin{align}
\opR_{\text{Cl}}^{\text{ea}}(\channel) \supseteq \inR_{1} \equiv
\bigcup_{ 0\leq \lambda\leq \frac{1}{2}}
\left\{ \begin{array}{rl}
  (R,L) \,:\;
	0\leq R \leq&  2-h_2(\lambda*\hat{\eps} )  \\
  L   \geq& h_2(\lambda*\hat{\eps} )-(1-q)h_2(\lambda*\eps_0)-q h_2(\lambda*\eps_1)
	\end{array}
\right\} \,.
\end{align}
 Therefore, higher communication rates can be achieved with CSI at the encoder 
at the expense of leaking information on the channel state sequence to the receiver. 
To obtain the region above from Theorem~\ref{theo:MskEA}, suppose that Alice performs phase-flip operation controlled by the random variable
$S+Y$, with addition modulo $2$, where $Y\sim\text{Bernoulli}(\lambda)$ is statistically independent of $S$. Precisely,
\begin{align}
\rho_{EA'A}=(1-q)\kb{0}\otimes [(1-\lambda)\Phi_{AA'}+\lambda (1\otimes Z)\Phi_{AA'}(1\otimes Z)]+q\kb{1}\otimes [(1-\lambda)(1\otimes Z)\Phi_{AA'}(1\otimes Z)+\lambda \Phi_{AA'}] \,.
\end{align}
Then, Bob receives the output of a phase-flip gate that is controlled by $(W_S+S+Y)$, which is distributed according to 
$\text{Bernoulli}(\lambda*\hat{\eps})$ (see (\ref{eq:hatEps})). Hence, achievability  for the region $\inR_{1}$ follows in a similar manner as for $\inR_{0}$.

Similarly, without entanglement assistance, the quantum masking region is bounded by 
\begin{align}
\opR_{\text{Q}}(\channel) \supseteq 
\bigcup_{ 0\leq \lambda\leq \frac{1}{2}}
\left\{ \begin{array}{rl}
  (Q,L) \,:\;
	0\leq Q \leq&  1-h_2(\lambda*\hat{\eps} )  \\
  L   \geq& h_2(\lambda*\hat{\eps} )-(1-q)h_2(\lambda*\eps_0)-q h_2(\lambda*\eps_1)
	\end{array}
\right\} \,.
\end{align}

\section{Summary and Concluding Remarks}
\label{sec:summ}
In this section, we summarize our results and compare between the techniques in our work and in previous work.
We consider a quantum channel $\channel_{EA'\rightarrow B}$ that depends on quantum state $|\phi_{E E_0 C}\rangle$, when the encoder has the CSI systems $E_0^n$ and is required to mask the channel state systems $C^n$ from the decoder.
First, we established an achievability result for a setting where Alice and Bob share entanglement resources at a limited rate $R_e$.
That is, before communication begins, Alice and Bob are provided with $2^{nR_e}$-dimension systems $G_A$ and $G_B$, respectively, in an entangled state 
$\Psi_{G_A G_B}$ of their choosing.

A significant distinction from the classical case is that the leakage requirement 
\begin{align}
\frac{1}{n} I(B^n G_B;C^n)_\rho \leq L
\end{align}
includes Bob's entangled share $G_B$, since the decoder has access to both the output systems and his part of the entangled pairs. In the classical setting,   shared randomness does not need to be included in the leakage constraint as it cannot help the decoder. On the other hand, we know that Bob can extract quantum information by performing measurements on $G_B$,  using the teleportation protocol for example.

Given a small leakage constraint $L\rightarrow 0$, we must ensure that Bob's systems $B^n G_B$ are decoupled from the channel state systems $C^n$. In this sense, masking can be viewed as a \emph{decoupling problem}, and thus it seems natural to solve the problem using the decoupling approach.
Here, we are most interested in the asymptotic characterization of achievable communication rates. 
Therefore, we have derived an asymptotic version of the decoupling theorem that can be applied directly, without considering the one-shot counterpart. 
While the derivation of our i.i.d. decoupling theorem, Theorem~\ref{theo:decoupling},  follows from the one-shot decoupling theorem using familiar arguments, it provides an analytic tool that is easier to combine with classical techniques,  without a one-shot proxy.

We presented an achievability result for channel state masking with rate-limited entanglement assistance in Theorem~\ref{theo:MskAchiev}, taking into account the tradeoff between the entanglement and communication resources.
The proof of our achievability theorem is based on the i.i.d. decoupling theorem along with Uhlmann's theorem \cite{Uhlmann:76p}. To establish the masking requirement, we approximate the leakage rate using  the decoupled output state that results from the decoupling theorem, and which approximates the actual output state. This approximation relies on the Alicki-Fannes-Winter inequality  \cite{AlickiFannes:04p,Winter:16p}, as the decoupled state is close to the actual output state and its leakage rate has a simpler bound. 

We determined the entanglement-assisted masking equivocation region and the capacity-leakage function
in Theorem~\ref{theo:MskEA} and Corollary~\ref{coro:MskEA}, respectively, 
 under the assumption that the channel state systems $E$, $E_0$, and $C$ are maximally correlated, i.e.
\begin{align}
\varphi_{EE_0 C}=\sum_{s\in\Sset} q(s) \kb{s}_E\otimes \kb{s}_{E_0} \otimes \kb{s}_C 
\label{eq:CondSchmidt1}
\end{align}
where $q(s)$ is a probability distribution and the vectors form an orthonormal basis for each of the corresponding Hilbert spaces. 
Analytically, the presence of three channel state systems poses a difficulty in choosing the auxiliary system $A$ that would satisfy both communication and leakage rate bounds. This difficulty does not exist in the classical setting of Merhav and Shamai \cite{MerhavShamai:07p}, since in the classical setting, $C$, $E$, and $E_0$ are simply copies of the same random variable. 
The direct part follows from our achievability result with rate-limited entanglement, and does not require the assumption above.

Next, we established a regularized formula for the quantum masking region and capacity-leakage function \emph{without} assistance
in Theorem~\ref{theo:MskQ} and Corollary~\ref{coro:MskQ}, respectively. The direct part here also follows from our achievability result with rate-limited entanglement. 
Our converse proof is based on different arguments compared to those of the classical proof by Merhav and Shamai \cite{MerhavShamai:07p}.
In both classical and quantum converse proofs, the leakage rate is bounded by an expression of the form
\begin{align}
L+\delta\geq \frac{1}{n}( I(C^n;M B^n)_\rho- H(M|B^n)_\rho+ H(M|B^n C^n)_\rho) 
\end{align}
(see (\ref{eq:Clineq}) and Eq. (21) in \cite{MerhavShamai:07p}). The next step in the classical proof in \cite{MerhavShamai:07p} is to use Fano's inequality in order to bound the second term by
\begin{align}
H(M|B^n)_\rho\leq n\eps_n
\label{eq:HfanoS}
\end{align}
and to eliminate the last term, as $H(M|B^n C^n)_\rho\geq 0$. In the quantum setting, we can still write (\ref{eq:HfanoS}), but it would not lead to the desired result because the last term $H(M|B^n C^n)_\rho$ is negative and could not be eliminated (see Remark~\ref{rem:Hnegative}).
Hence, we bound the leakage rate  in a different manner using the coherent information bound on the communication rate.

We also derived single-letter inner and outer bounds for Hadamard channels, using the special properties of those channels, and showed that the bounds coincide in the standard case of a channel that does not depend on a state. To bound the communication rate $Q$, we only needed to use the fact that Hadamard channels are degradable. To bound the leakage rate $L$, we observed that for Hadamard channels, there exists a channel from the channel output to the combined system of the  output and its environment.

A shortcoming of our results, as well as the previous results by Dupuis \cite{Dupuis:09c},
is that we do not have a bound on the dimension of the auxiliary system $A$, as  mentioned in Remark~\ref{rem:Uncomput}. 
Although one can always compute an achievable region by simply choosing the dimension of $A$, the optimal rates cannot be 
 computed exactly in general.
%
%
If we could restrict the optimization to pure states $|\psi_{EA'AC}\rangle$, then we would argue that the dimension 
of $A$ need not be larger than the Schmidt rank of $|\psi_{EA'AC}\rangle$, hence optimizing over a Hilbert space of dimension
$|\Hset_A|=|\Hset_{A'}||\Hset_{E}||\Hset_{C}|$ is sufficient.
A similar difficulty appears in other quantum models such as the  broadcast channel (see Discussion section in \cite{DupuisHaydenLi:10p}), wiretap channel \cite[Remark 5]{QiSharmaWilde:18p},
and squashed entanglement \cite[Section 1]{LiWinter:14p}.
Considering the setting where entanglement assistance is not available, we mentioned in Remark~\ref{rem:AhlswedeReg} that regularization does not necessarily pose a problem for practical purposes. 
Whereas, from a theoretical perspective, 
a single-letter formula usually offers a lot more insight than a multi-letter characterization since the latter is not unique (see e.g. \cite[Section 13.1.3]{Wilde:17b}). Nonetheless, remarkable properties such as super-activation \cite{SmithYard:08p} were derived from the multi-letter characterization as well.

\section*{Acknowledgment}
%
%
U. Pereg, C. Deppe, and H. Boche were supported by the Bundesministerium f\"ur Bildung und Forschung (BMBF) through Grants
16KIS0856 (Pereg, Deppe), 16KIS0858 (Boche), and the Viterbi scholarship of the Technion (Pereg).
This work of H. Boche was supported in part by the German Federal
Ministry of Education and Research (BMBF) within the national initiative for
``Post Shannon Communication (NewCom)" under Grant 16KIS1003K and in
part by the German Research Foundation (DFG) within the Gottfried Wilhelm
Leibniz Prize under Grant BO 1734/20-1 and within Germany’s Excellence
Strategy EXC-2092 – 390781972 and EXC-2111 – 390814868.

%

\begin{appendices}
\section{Proof of Theorem~\ref{theo:decoupling}}
\label{app:decoupling}

We prove the i.i.d. decoupling theorem using the one-shot counterpart in \cite{Dupuis:10z} along with arguments therein.
To this end, we need the following definitions from \cite{Renner:08a}. 
 Define the conditional min-entropy by 
\begin{align}
H_{\min}(\rho_{AB}|\sigma_B)=&-\log \inf \left\{ \lambda\in\mathbb{R} \,:\;  \rho_{AB} \preceq \lambda \cdot (\identity_A\otimes \sigma_B)  \right\}
\nonumber\\
H_{\min}(A|B)_\rho=& \sup_{\sigma_B} H_{\min}(\rho_{AB}|\sigma_B)
 \,.
\end{align}
where the supremum is over quantum states of the system $B$.
In general, the conditional min-entropy is bounded by
\begin{align}
-\log |\Hset_B| \leq H_{\min}(A|B)_\rho \leq \log |\Hset_A| \,.
\label{eq:H2bound}
\end{align}
To see this, observe that if we choose $\sigma_B=\frac{\identity_B}{|\Hset_B|}$, then the matrix inequality
$\rho_{AB} \preceq \lambda (\identity_A\otimes \sigma_B)$ holds for $\lambda=|\Hset_B|$,
hence $H_{\min}(\rho_{AB}|\sigma_B)\geq -\log|\Hset_B|$.
As for the upper bound, 
the matrix inequality implies that
$1=\trace(\rho_{AB})\leq \lambda |\Hset_A|\trace(\sigma_B)=\lambda |\Hset_A|$, hence $H_{\min}(\rho_{AB}|\sigma_B)\leq \log|\Hset_A|$.
Furthermore, 
the lower  bound is saturated when the joint state of $A$ and $B$ is 
$|\Phi_{AB}\rangle$,
whereas the upper bound for a product state $\frac{\identity_A}{|\Hset_A|}\otimes \rho_B$.

Then, define the smoothed min-entropy by 
\begin{align}
H_{\min}^{\eps}(A|B)_\rho=\max_{ \sigma_{AB} \,:\; d_F(\rho_{AB},\sigma_{AB})\leq \eps } H_{\min}^{\eps}(A|B)_\sigma \,.
\end{align}
for arbitrarily small $\eps>0$, where $d_F(\rho,\sigma)=\sqrt{ 1-\norm{\sqrt{\rho}\sqrt{\sigma}}_1^2 }$ is the fidelity distance between the states. 
The theorem below follows from Lemma 2.3 and Theorem 3.8 in \cite{Dupuis:10z}.
\begin{theorem}[The one-shot decoupling theorem {\cite{Dupuis:10z}}] 
\label{theo:OSdecoupling}
Let $\rho_{A R}$ be a quantum state, $\Tset_{A\rightarrow K}$ be a quantum channel, and 
$\zeta_{A' K}=\Tset_{A\rightarrow K} (\Phi_{A' A})$. Then, for arbitrarily small $\eps>0$,
\begin{align}
\int_{\mathbb{U}_A} dU_{A} \, \norm{
\Tset_{A\rightarrow K}(U_{A} \rho_{A R})-\zeta_K\otimes\rho_R
}_1 \leq  2^{-\frac{1}{2}H_{\min}^{\eps}(A'|K)_\zeta-\frac{1}{2}H_{\min}^{\eps}(A|R)_\rho}  +8\eps
\end{align}
where the integral is over the Haar measure on all unitaries $U_{A}$.
\end{theorem}


Now, in order to prove the i.i.d. decoupling theorem, we use Theorem~\ref{theo:OSdecoupling} as follows. To show (\ref{eq:Decoup1}),    plug 
\begin{align}
&A\leftarrow A^n \,,\; \rho_{AR}\leftarrow W_{S G_1\rightarrow A^n}(\sigma_{S G_1 R}) \,,\; \Tset\leftarrow \Tset^{\otimes n} \,.
\end{align}
Then, by Theorem~\ref{theo:OSdecoupling},
\begin{align}
\int_{\mathbb{U}_{A^n}} dU_{A^n} \, \norm{
\Tset_{A\rightarrow K}^{\otimes n}(U_{A^n} \sigma_{A^n R})-\zeta_K^{\otimes n}\otimes\sigma_R
}_1 \leq 2^{-\frac{1}{2}H_{\min}^{\eps}(A'^n|K^n)_{\zeta^{\otimes n}}-\frac{1}{2}H_{\min}^{\eps}(S, G_1|R)_\sigma} +8\eps
\label{eq:ineqDecoupP1}
\end{align}
with arbitrarily small $\eps>0$, $\sigma_{A^n R}=W_{S G_1\rightarrow A^n}(\sigma_{S G_1 R})$, and
\begin{align}
\zeta_{A' K}=&\Tset_{A\rightarrow K} (\Phi_{A' A})=|\Hset_A|\trace_B\left[ \text{op}_{A\rightarrow BK}(|\omega_{ABK}\rangle)(\Phi_{A' A})  \right]
\nonumber\\
=& \trace_B\left( \omega_{A'BK} \right) =\omega_{A'K} 
\label{eq:ineqDecoupP2}
\end{align}
where the second line follows from Lemma~\ref{lemm:opPhi}.
Hence, it follows that
\begin{align}
H_{\min}^{\eps}(A'^n|K^n)_{\zeta^{\otimes n}}=H_{\min}^{\eps}(A^n|K^n)_{\omega^{\otimes n}} \geq n(H(A|K)_\omega - \delta_1(n) )
\label{eq:ineqDecoupP3}
\end{align}
where the last inequality is due to the quantum asymptotic equipartition property (see \cite[Theorem 9]{TomamichelColbeckRenner:09p} and \cite[Lemma 2.3]{Dupuis:10z}), and where $\delta_1(n)\rightarrow 0$ as $n\rightarrow\infty$. 
Thus, by (\ref{eq:ineqDecoupP1})-(\ref{eq:ineqDecoupP3}),
\begin{align}
\int_{\mathbb{U}_{A^n}} dU_{A^n} \, \norm{
\Tset_{A\rightarrow K}^{\otimes n}(U_{A^n} \sigma_{A^n R})-\omega_K^{\otimes n}\otimes\sigma_R
}_1 \leq 2^{-n\frac{1}{2}H(A|K)_{\omega}-\frac{1}{2}H_{\min}^{\eps}(S,G_1|R)_\sigma+n\delta_2(n)}
\label{eq:ineqDecoupP4}
\end{align}
where $\delta_2(n)=\delta_1(n)+\frac{\log(8\eps)}{n}$. Since $S R$ and $G_1 G_2$ are 
in a product state $|\Psi_{SR}\rangle\otimes |\Phi_{G_1 G_2}\rangle$ over $\Hset_S^{\otimes 2}\otimes \Hset_G^{\otimes 2}$, we have that
\begin{align}
H_{\min}^{\eps}(S,G_1|R)_\sigma\geq H_{\min}(S|R)_\sigma + H_{\min}(G_1)_\Phi \geq -\log|\Hset_S|+\log|\Hset_G| 
\end{align}
where the last inequality holds by (\ref{eq:H2bound}).
Hence, (\ref{eq:Decoup1}) follows.

To show (\ref{eq:Decoup2}),  apply Theorem~\ref{theo:OSdecoupling} in the same manner with $(R,G_2)$ instead of $R$, which yields
\begin{align}
\int_{\mathbb{U}_{A^n}} dU_{A^n} \, \norm{
\Tset_{A\rightarrow K}^{\otimes n}(U_{A^n} \sigma_{A^n R G_2})-\omega_K^{\otimes n}\otimes\sigma_{R,G_2}
}_1 \leq 2^{-n\frac{1}{2}H(A|K)_{\omega}-\frac{1}{2}H_{\min}^{\eps}(S,G_1|R,G_2)_\sigma+n\delta_2(n)}
\label{eq:ineqDecoupP2b}
\end{align}
with
\begin{align}
H_{\min}^{\eps}(S,G_1|R,G_2)_\sigma\geq H_{\min}(S|R)_\Psi + H_{\min}(G_1|G_2)_\Phi \geq -\log|\Hset_S|-\log|\Hset_G| \,.
\end{align}
Thus, (\ref{eq:Decoup2}) follows as well. This completes the proof of Theorem~\ref{theo:decoupling}.
\qed

\begin{center}
\begin{figure}[ht!]
\includegraphics[scale=0.75,trim={3.5cm 2cm 0 3.5cm},clip]{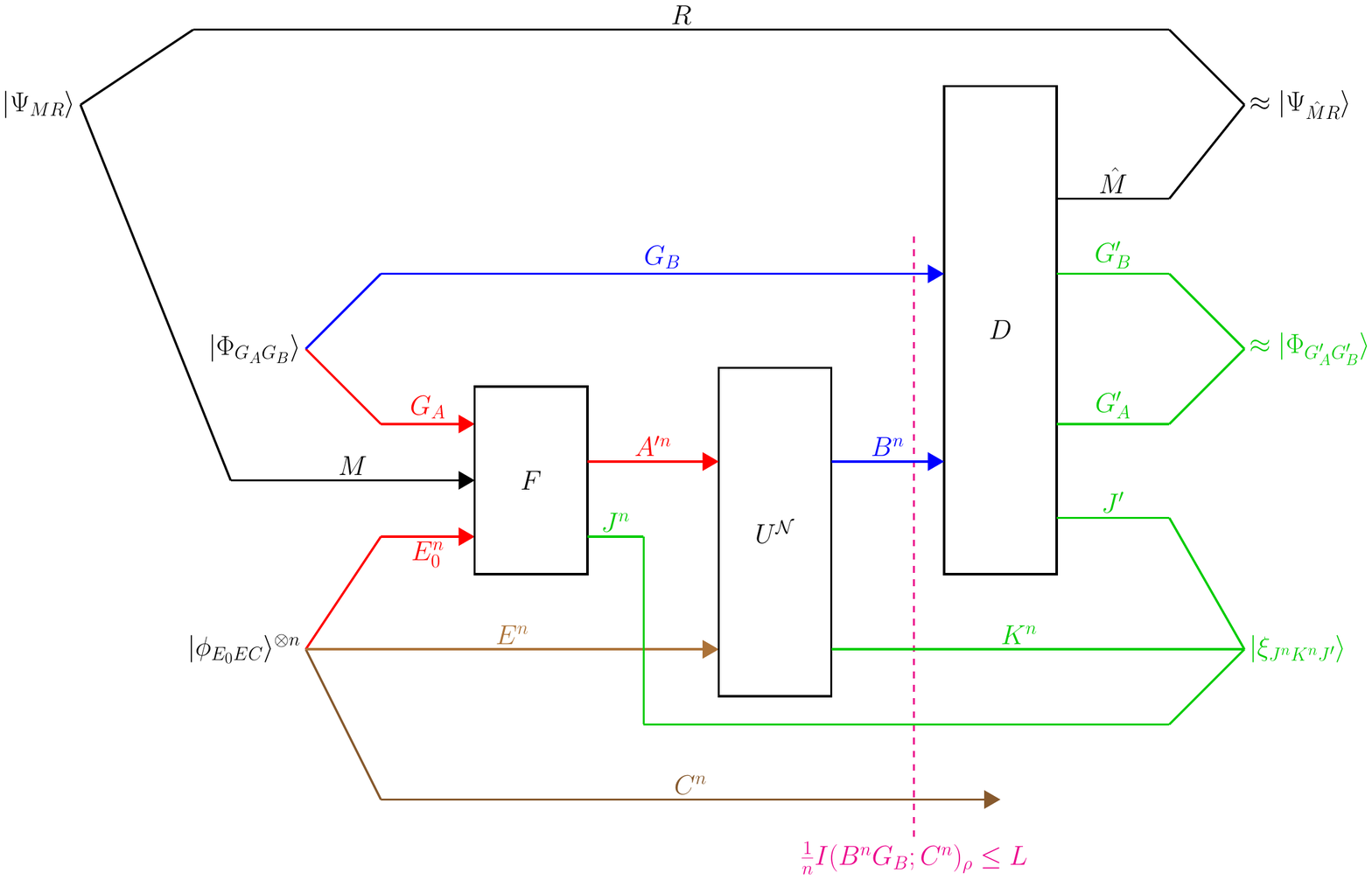} 
\caption{Coding scheme for a state-dependent quantum channel $\channel_{EA'\rightarrow B}$ with state information at the encoder and masking from the decoder, given rate-limited  entanglement assistance. The quantum systems of Alice and Bob are marked in red and blue, respectively; the channel state systems $E^n$ and $C^n$ are marked in brown; and the purifying systems are marked in green.
The quantum message is stored in a $2^{nQ}$-dimension system $M$, which is purified by the reference system $R$ of the same dimension, while Alice and Bob's entanglement resources are in the quantum systems $G_A$ and $G_B$, respectively, each of dimension $2^{nR_e}$.
The input state is thus $|\Psi_{RM}\rangle \otimes |\Phi_{G_A,G_B}\rangle\otimes |\phi_{E_0 EC}\rangle^{\otimes n}$.  
 Alice encodes the quantum message using her share of the entanglement resources, $G_A$, along with her access to the side information systems $E_0^n$, which are entangled with the channel state systems $E^n$ and $C^n$. To this end, she applies the encoding isometry $F_{  M G_A E_0^n \rightarrow A'^n J^n}$, where 
$J^n$ are purifying reference systems. 
Then, she transmits the systems $A'^n$ over $n$ channel uses of the  isometric extension $U^\channel_{EA'\rightarrow BK}$ of the channel $\channel_{EA'\rightarrow B}$, where $K$ is the receiver's environment. 
 Bob receives the channel output systems $B^n$, combines them with his share $G_B$ of the entanglement resources, and applies the decoding isometry $D_{B^n G_B\rightarrow \hM G_A' G_B' J^n K^n J'}$. Using the i.i.d. decoupling theorem and Uhlmann's theorem, it is shown that the resulting state is close in trace distance to $|\Psi_{\hM R}\rangle\otimes |\Phi_{G_A G_B}\rangle \otimes |\xi_{J^n K^n J'}\rangle$. 
Given $L\geq I(C;AB)_\omega+\delta$, it is shown that the leakage requirement
 $
 \frac{1}{n} I(C^n;B^n G_B)_\rho \leq L 
$ is satisfied as well.
}
\label{fig:MSKsiCodeProofA}
\end{figure}
\end{center}

\section{Proof of Theorem~\ref{theo:MskAchiev}}
\label{app:MskAchiev}
The achievability proof is based on the i.i.d. decoupling theorem along with Uhlmann's theorem. To establish the masking requirement, we approximate the leakage rate using  the decoupled output state that results from the decoupling theorem, and which approximates the actual output state. This approximation relies on the Alicki-Fannes-Winter inequality  \cite{AlickiFannes:04p,Winter:16p}, as the decoupled state is close to the actual output state and its leakage rate is easier to evaluate.

%

Consider  a quantum state-dependent  channel $\channel_{EA'\rightarrow B}$ with state information at the encoder and masking from the decoder, given rate-limited  entanglement assistance. The elements of the coding scheme are displayed in Figure~\ref{fig:MSKsiCodeProofA}, where the quantum systems of Alice and Bob are marked in red and blue, respectively; the channel state systems $E^n$ and $C^n$ are marked in brown; and the purifying systems are marked in green.
Before we step into the formal proof, we describe the coding scheme in a nutshell.
The quantum message is stored in a system $M$, which is purified by a reference system $R$. Alice and Bob's entanglement resources are in the quantum systems $G_A$ and $G_B$, respectively.
Now, Alice encodes the quantum message using her share of the entanglement resources, $G_A$, along with her access to the side information systems, $E_0^n$, which in turn are entangled with the channel state systems $E^n$ and $C^n$.  
To this end, she applies an encoding isometry 
and transmits the systems $A'^n$ over $n$ channel uses of the  isometric extension  of the channel, $U^\channel_{EA'\rightarrow BK}$, where $K$ is the receiver's environment.  Bob receives the channel output systems $B^n$ and decodes by applying an isometry  to $B^n$ and $G_B$.
We will show that there exist encoding and decoding  isometries, $F_{  M G_A E_0^n \rightarrow A'^n J^n}$ and $D_{B^n G_B\rightarrow \hM G_A' G_B' J^n K^n J'}$, respectively, that recover the quantum message state
and satisfy the leakage requirement, where $J^n$ and $J'$ are purifying reference systems. 
In our proof, the decoupling approach is used such that both Bob's environment and the channel state systems $E^n$ and $C^n$ are decoupled from Alice's purifying reference system (see Remark~\ref{rem:decoupIt}). 
To show the leakage requirement, we approximate the leakage rate using  the decoupled output state that results from the decoupling theorem, as the decoupled state is close to the actual output state and its leakage rate is easier to evaluate.
 The details are given below.

Let $|\theta_{A C E A'  J}\rangle$ be any pure state with $\theta_{CE}=\phi_{CE}$, where $A$ is an arbitrary system. In the proof below, we will use auxiliary quantum systems $A^n$ such that the channel input systems $A'^n$ are entangled with $A^n$.
Given a quantum message state $\rho_M$, let
$R$ be a reference system that purifies the message system $M$, \ie such that the systems $M$ and $R$ have a pure joint state
$|\Psi_{MR}\rangle$, with $|\Hset_R|=|\Hset_M|=2^{nQ}$. 
%
Suppose that Alice and Bob share an entangled state $|\Phi_{G_A G_B}\rangle$ of dimension $|\Hset_{G_A}|=|\Hset_{G_B}|=2^{nR_e}$.
Then, the joint state is
\begin{align}
|\psi_{RM G_A G_B }\rangle \equiv |\Psi_{RM}\rangle \otimes |\Phi_{G_A,G_B}\rangle \,.
\end{align}
Let $\Uset^{\channel}_{E A'\rightarrow B K}$ be an isometric extension of the channel $\channel_{EA'\rightarrow B}$, with
\begin{align}
\Uset^{\channel}_{E A'\rightarrow B K}(\rho_{EA'})=U^{\channel}_{E A'\rightarrow B K} \rho_{E A'} (U^{\channel }_{E A'\rightarrow B K})^\dagger
\end{align} 
 and let
\begin{align}
|\omega_{A C B K J}\rangle = U^{\channel}_{E A'\rightarrow B K} |\theta_{A C E A'  J}\rangle \,.
\label{eq:Oacbkj}
\end{align}
Denote
\begin{align}
\Delta_1(n)\equiv&  2^{-n[H(A|EC)_\omega -Q-R_e-\eps]/2}  
\label{eq:Delta1}
\\
\Delta_2(n)\equiv&  2^{-n[H(A|KJ)_\omega -Q+R_e-\eps]/2} 
\label{eq:Delta2}
\end{align}
where $\eps>0$ is arbitrarily small.  
Observe that $\Delta_1(n) $ tends to zero exponentially as $n\rightarrow \infty$ provided that $Q+R_e<H(A|E C)_\theta-\eps$.
As for $\Delta_2(n)$, given a pure quantum state $|\omega_{A C B K J}\rangle$, 
 we have $H(AKJ)_\omega=H(BC)_\omega$ and $H(KJ)_\omega=H(BCA)_\omega$, hence
\begin{align}
H(A|KJ)_\omega=H(BC)_\omega-H(BCA)_\omega=-H(A|BC)_\omega \geq -H(A|B)_\omega=I(A\rangle B)_\omega
\end{align}
where the last inequality holds since conditioning does not increase entropy \cite[Theorem 11.4.1]{Wilde:17b}.
Thus, $\Delta_2(n)\leq 2^{-n(I(A\rangle B)_\omega-Q+R_e-\eps)}$, which tends to zero exponentially as $n\rightarrow \infty$ provided that $Q-R_e<I(A\rangle B)_\omega-\eps$.

First, we show that there exist encoding and decoding operations such that the decoding error
vanishes. 
Consider a full-rank partial isometry $W_{M  G_A  \rightarrow A^n}$, \ie an operator with $0$-$1$ singular values and rank
$2^{n(Q+R_e)}$, and let
\begin{align}
\Pi_{A\rightarrow C E A' J} \equiv 
 |\Hset_{A}| \text{op}_{A\rightarrow C E A' J} (|\theta_{A C E A'  J}\rangle)  \,.
\label{eq:Pidef}
\end{align}
Then, define a quantum channel $\Tset_{A\rightarrow K J}$ by
\begin{align}
\Tset_{A\rightarrow K J}(\rho_{A})=  \trace_{C,B} \left( U_{EA'\rightarrow BK}^{\channel }\left( \Pi_{A\rightarrow  C E A' J} ( \rho_{A}) \right)  \right) \,.
\end{align}
According to the first part of Theorem~\ref{theo:decoupling}, the i.i.d. decoupling theorem, applying a random unitary $U_{A^n}$ decouples between the systems $(K^n,J^n)$ and $R$ in the sense that
\begin{align}
\int_{\mathbb{U}_{A^n}} dU_{A^n}\, \norm{
\Tset_{A^n\rightarrow K^n J^n}(U_{A^n} W_{M G_A \rightarrow A^n} \psi_{RM G_A })-\omega_{K J}^{\otimes n}\otimes \psi_R
}_1
\leq  2^{-n[ H(A|KJ)_\omega -Q+R_e-\eps_1(n) ]/2} 
\label{eq:decoupIn1}
\end{align}
with $\Tset_{A^n\rightarrow K^n J^n}\equiv \Tset_{A\rightarrow K J}^{\otimes n}$,
where $\eps_1(n)$ tends to zero as $n\rightarrow \infty$.

Similarly, the second part of Theorem~\ref{theo:decoupling} with
$
\Tset_{A\rightarrow C E}'(\rho_{A})=  \trace_{A' J} \left[ \Pi_{A\rightarrow C E A' J} ( \rho_{A})   \right] 
$ 
yields 
\begin{multline}
\int_{\mathbb{U}_{A^n}} dU_{A^n}\, \norm{
\trace_{A'^n J^n} \left[ \Pi_{A^n\rightarrow 
C^n E^n A'^n J^n}  U_{A^n} W_{M G_A \rightarrow A^n}\psi_{M G_A G_B R}   \right]
-\psi_{G_B R} \otimes \phi_{CE}^{\otimes n}
}_1
\\
\leq 2^{-n[H(A|CE)_\omega -Q-R_e-\eps_2(n)]/2} 
\label{eq:decoupIn2}
\end{multline}
with $\Pi_{A^n\rightarrow C^n E^n A'^n J^n}\equiv \Pi_{A\rightarrow C E A' J}^{\otimes n}$,
where $\eps_2(n)$ tends to zero as $n\rightarrow \infty$.
%
Thus, it can be inferred from (\ref{eq:decoupIn1})-(\ref{eq:decoupIn2}) that there exists a unitary $U_{A^n}$ such that both of the following inequalities hold,
\begin{align}
 \norm{
\Tset_{A^n\rightarrow K^n J^n}(U_{A^n} W_{M G_A \rightarrow A^n} \psi_{RM G_A })-\omega_{K J}^{\otimes n}\otimes \psi_R
}_1
\leq \Delta_2(n) 
\label{eq:ineqD2}
\end{align}
and 
\begin{align}
 \norm{
\trace_{A'^n J^n} \left[ \Pi_{A^n\rightarrow C^n E^n A'^n J^n} \cdot U_{A^n} W_{M G_A \rightarrow A^n} \psi_{MG_A G_B R}   \right]
-\psi_{G_B R}\otimes \phi_{EC}^{\otimes n}
}_1
\leq \Delta_1(n)  
\label{eq:ineqD1}
\end{align}
where $\Delta_1(n)$ and $\Delta_2(n)$ are as defined in (\ref{eq:Delta1})-(\ref{eq:Delta2}).
In words, there exists a unitary $U_{A^n}$ that decouples both $(K^n,J^n)$ from $R$, and also
$(C^n,E^n)$ from $(G_B,R)$.
The existence of a unitary that satisfies both inequalities simultaneously follows from the union of events bound and Markov's inequality, as 
$\prob{f_1(U)>\Delta_1 \vee f_2(U)>\Delta_2}\leq \frac{\mathbb{E} f_1(U) }{\Delta_1}+\frac{\mathbb{E} f_2(U) }{\Delta_2} $.
 We note that such a unitary $U_{A^n}$ need not be unique.

According to Uhlmann's theorem, (\ref{eq:ineqD1}) implies that there exists an isometry $F_{M G_A  E_0^n \rightarrow A'^n J^n}$ such that
\begin{align}
 \norm{
  \Pi_{A^n\rightarrow C^n E^n A'^n J^n} \cdot U_{A^n} W_{M G_A \rightarrow A^n} \psi_{MG_A G_B R}
- F_{M G_A E_0^n\rightarrow A'^n J^n}( \psi_{G_B R M G_A }\otimes \phi_{E_0  E C}^{\otimes n})
}_1
\leq 2\sqrt{\Delta_1(n)}
\label{eq:Fcodeineq1}
\end{align}
(see Figure~\ref{fig:MSKsiCodeProofA}).
Hence, by applying the isometric extension of the channel and using the triangle inequality and the monotonicity of the trace distance under quantum channels, we obtain
\begin{multline}
 \norm{
\Tset_{A^n\rightarrow K^n J^n}(U_{A^n} W_{M G_A \rightarrow A^n} \psi_{RM G_A })
-\trace_{C^n B^n G_B}\left( (U_{EA'\rightarrow BK}^{\channel})^{\otimes n} F_{M G_A E_0^n\rightarrow A'^n J^n}( \psi_{ G_B  R M G_A }\otimes \phi_{E_0 E C}^{\otimes n}) \right)
}_1
\\
\leq 2\sqrt{\Delta_1(n)} \,.
\end{multline}
Together with (\ref{eq:ineqD2}), this implies that
\begin{align}
\norm{
\trace_{C^n B^n G_B} \left((U_{EA'\rightarrow B K}^{\channel})^{\otimes n} F_{M G_A E_0^n\rightarrow A'^n J^n}( \psi_{G_B R M G_A }\otimes \phi_{E_0 E C}^{\otimes n})\right) - \omega_{K J}^{\otimes n}\otimes \psi_R
}_1 \leq
2\sqrt{\Delta_1(n)}+\Delta_2(n) \,.
\end{align}
Next, by Uhlmann's theorem, there exists a decoding operator $D_{B^n  G_B\rightarrow \hM G_A'   G_B' J' }$
such that
\begin{multline}
\norm{
D_{B^n  G_B\rightarrow \hM G_A'   G_B' J' }
(U^{\channel}_{EA'\rightarrow BK})^{\otimes n} F_{M G_A E_0^n\rightarrow A'^n J^n}( \psi_{ G_B R M G_A }^{\otimes n}\otimes \phi_{E_0 E C}^{\otimes n}) - \xi_{K^n J^n J'} \otimes \psi_{M G_A G_B R}
}_1 \\ \leq
2\sqrt{
2\sqrt{\Delta_1(n)}+\Delta_2(n) }
\end{multline}
for some $\xi_{ K^n J^n J'}$. By tracing out $K^n$, $J^n$, $C^n$, $G_A'$, $G_B'$, and $J'$, we have that there exist an encoding map 
$\Fset_{M G_A E_0^n\rightarrow A'^n }$ and a decoding map $\Dset_{B^n  G_B\rightarrow \hM}$ such that the estimation error is bounded by
\begin{align}
e^{(n)}(\Fset,\Phi,\Dset,\rho_M)=
\norm{
\Dset_{B^n  G_B\rightarrow \hM }
\channel_{EA'\rightarrow B}^{\otimes n} \Fset_{M G_A E_0^n\rightarrow A'^n }( \psi_{ G_B R M G_A }^{\otimes n}\otimes \phi_{E_0 E}^{\otimes n}) - \Psi_{RM}
}_1  \leq
2\sqrt{
2\sqrt{\Delta_1(n)}+\Delta_2(n) } \,.
\end{align}


As for the leakage requirement, let $\delta>0$ be arbitrarily small. Observe that the joint state of the output systems 
is given by
\begin{align}
|\sigma_{G_B R  K^n J^n B^n C^n}\rangle=
(U^{\channel}_{EA'\rightarrow BK})^{\otimes n} F_{M G_A E_0^n\rightarrow A'^n J^n}( \psi_{ G_B R M G_A }^{\otimes n}\otimes \phi_{E_0 E C}^{\otimes n}) \,.
\end{align}
By (\ref{eq:Fcodeineq1}),
$
\norm{ \sigma-\eta }_1 \leq 2\sqrt{\Delta_1(n)}
$, with
\begin{align}
|\eta_{G_B R  K^n J^n B^n C^n}\rangle=&
 (U^{\channel}_{EA'\rightarrow BK})^{\otimes n} \Pi_{A^n\rightarrow C^n E^n A'^n J^n}  (U_{A^n} W_{M G_A \rightarrow A^n} \psi_{MG_A G_B R})
\nonumber\\
\stackrel{(a)}{=}& \left( U^{\channel}_{EA'\rightarrow BK} |\Hset_{A}|\text{op}_{A\rightarrow C E A' J} (|\theta_{A C E A'  J}\rangle) \right)^{\otimes n}   (U_{A^n} W_{M G_A \rightarrow A^n} \psi_{MG_A G_B R})
\nonumber\\
\stackrel{(b)}{=}& \left( |\Hset_{A}|\text{op}_{A\rightarrow C BK J} (|\omega_{A C BK  J}\rangle) \right)^{\otimes n}   (U_{A^n} W_{M G_A \rightarrow A^n} \psi_{MG_A G_B R})
\label{eq:etaAchiev}
\end{align}
where $(a)$ follows from the definition of $\Pi_{A^n\rightarrow C^n E^n A'^n J^n}$ in (\ref{eq:Pidef}), and $(b)$ follows from the definitions of $\text{op}_{A \rightarrow B}(\cdot)$ and $|\omega_{A C BK  J}\rangle$ in (\ref{eq:op}) and (\ref{eq:Oacbkj}), respectively.
Next, by Lemma~\ref{lemm:opCommut}, we have 
\begin{align}
|\eta_{G_B R  K^n J^n B^n C^n}\rangle=&
\left( |\Hset_{A}|^n\text{op}_{A^n\rightarrow G_B R} (U_{A^n} W_{M G_A \rightarrow A^n} \psi_{MG_A G_B R}) \right) |\omega_{A C BK  J}\rangle^{\otimes n} \,.
\label{eq:etaAchiev1}
\end{align}
Hence,
\begin{align}
\eta_{G_B R B^n C^n}= \Pi_{A^n\rightarrow G_B R}' (\omega_{A B C}^{\otimes n})
\label{eq:etaAchiev2}
\end{align}
with $\Pi_{A^n\rightarrow G_B R}'\equiv |\Hset_{A}|^n\text{op}_{A^n\rightarrow G_B R} (U_{A^n} W_{M G_A \rightarrow A^n} \psi_{MG_A G_B R})$.

By the Alicki-Fannes-Winter inequality, the mutual information is continuous in the joint state \cite{AlickiFannes:04p,Winter:16p}. In particular,
$
\norm{ \sigma-\eta }_1 \leq 2\sqrt{\Delta_1(n)}
$ implies that
\begin{align}
|I(C^n;B^n G_B)_\sigma - I(C^n;B^n G_B)_\eta| \leq 4n\log |\Hset_B| \sqrt{\Delta_1(n)}  +2(1+\sqrt{\Delta_1(n)})
\end{align}
(see \cite[Theorem 11.10.3]{Wilde:17b}).
Since $\Delta_1(n)$ tends to zero as $n\rightarrow\infty$, it follows that for sufficiently large $n$, the leakage rate is bounded by
\begin{align}
\ell^{(n)}(\Fset,\Psi,\Dset,\rho_M)=\frac{1}{n} I(C^n;B^n G_B)_\sigma\leq& \frac{1}{n} I(C^n;B^n G_B)_\eta+\delta
\nonumber\\
\leq& \frac{1}{n} I(C^n;B^n G_B R)_\eta+\delta
\nonumber\\
\leq& \frac{1}{n} I(C^n;A^n B^n)_{\omega^{\otimes n}}+\delta
\nonumber\\
=& 
I(C;AB)_{\omega}+\delta
\end{align}
where the third inequality follows from (\ref{eq:etaAchiev2}) and the data processing theorem for the quantum mutual information \cite[Theorem 11.9.4]{Wilde:17b}. Thus, the secrecy requirement  holds with leakage rate $L$ provided that
$I(C;AB)_\omega \leq L-\delta$. 
\qed

\section{Proof of Theorem~\ref{theo:MskEA}}
\label{app:MskEA}

Given unlimited supply of entanglement resources, a qubit is exchangeable with two classical bits. This follows by applying 
the teleportation protocol and the super-dense coding protocol (see \cite[Sections 1.3.7, 2.3]{NielsenChuang:02b}). 
Therefore, the characterization of the classical masking region follows from that of the quantum masking region, and vice versa. In particular, we prove the theorem by showing achievability for the quantum masking region, and the converse part for the classical masking region. As can be seen below, the maximal correlation assumption in (\ref{eq:CondSchmidt}) is only required for the converse proof.

\subsection{Achievability Proof}

First, consider the direct part for the quantum masking region. Let $(Q,L)\in \mathcal{R}_{\text{Q}}^\text{ea}(\channel)$.
Then, for some $\rho_{ E A' A C}$ with $\rho_{EC}=\varphi_{EC}$, we have  $Q \leq \frac{1}{2}[ I(A;B)_\rho- I(A;EC)_\rho]$ and  $L   \geq I(C;AB)_\rho$. We need to show that there exists $R_e\geq 0$ such that $(Q,R_e,L)$ is achievable.

As mentioned in Remark~\ref{rem:mixed}, given a mixed state $\varphi_{E E_0 C}$, we can simply consider the  channel
$\widetilde{\channel}_{\tilde{E}  A'\rightarrow B}$, with the augmented channel state system $\tilde{E}=(T,E)$, as defined in (\ref{eq:channelTm}), where  $| \phi_{TE E_0 C} \rangle$ is a purification of the mixed state $\varphi_{E E_0 C}$. Given the maximal correlation assumption (\ref{eq:CondSchmidt}), the standard purification is
\begin{align}
| \phi_{TE E_0 C} \rangle=\sum_{s\in\Sset} \sqrt{q(s)} |s\rangle_T\otimes |s\rangle_E\otimes |s\rangle_{E_0} \otimes |s\rangle_C \,.
\end{align}
Let $\rho_{TEA'AC}$ be an extension of $\rho_{ E A' A C}$ with $\rho_{TEC}=\phi_{TEC}$.
Then, we can write
\begin{align}
\rho_{CTEAA'}=\sum_{s\in\Sset} q(s) \kb{s}_C \otimes \kb{s}_T \otimes \kb{s}_E \otimes \rho_{AA'}^s 
\end{align}
for some $\rho_{AA'}^s $. Since the eigenvalues of $\rho_{CEAA'}$ are the same as those of $\rho_{CTEAA'}$, it follows that 
$  I(A;TEC)_\rho=I(A;EC)_\rho$.

We now claim that the inequalities (\ref{eq:MskAchiev1})-(\ref{eq:MskAchiev3}) hold for 
\begin{align}
R_e &\equiv   \frac{1}{2} H(A|\tilde{E}C)_\rho- \frac{1}{2}I(A\rangle B)_\rho 
   \,.
\label{eq:ReDirEA}
\end{align}
Indeed, 
\begin{align}
Q+R_e\leq& \frac{1}{2}[ I(A;B)_\rho- I(A;\tilde{E}C)_\rho] +\frac{1}{2} H(A|\tilde{E}C)_\rho-\frac{1}{2} I(A\rangle B)_\rho 
\nonumber\\
=& H(A|\tilde{E}C)_\rho  \,.
\intertext{and} 
Q-R_e\leq& \frac{1}{2}[ I(A;B)_\rho- I(A;\tilde{E}C)_\rho] -\frac{1}{2} H(A|\tilde{E}C)_\rho+\frac{1}{2} I(A\rangle B)_\rho
\nonumber\\
=& I(A\rangle B)_\rho
\end{align}
since $I(A;D)_\rho=H(A)_\rho-H(A|D)_\rho$, and due to the definition of the coherent information as 
$I(A\rangle B)_\rho\equiv -H(A|B)_\rho $. 

We also need to verify that $R_e\geq 0$.  Let $|\theta_{AC\tilde{E}A'J}\rangle$ be a purification of $\rho_{AC\tilde{E}A'}$ and define
$|\omega_{ACBKJ}\rangle$ as in (\ref{eq:Oacbkj}). 
Since the state of $ACBKJ$ is pure, we have 
$H(A|KJC)_\omega=-H(A|B)_\omega=I(A\rangle B)_\rho$, hence
\begin{align}
0\leq I(A;KJ|C)_\omega=& H(A|C)_\rho-I(A\rangle B)_\rho 
\,.
\label{eq:nonNegRe}
\end{align}
As 
$H(A|C)_\rho=H(A|\tilde{E}C)_\rho$, 
(\ref{eq:nonNegRe}) implies that the assignment of $R_e$ in (\ref{eq:ReDirEA}) is non-negative as required.
It follows that the conditions of Theorem~\ref{theo:MskAchiev} are satisfied, hence $(Q,R_e,L)$ is achievable.
We deduce that $\mathbb{R}_{\text{Q}}^\text{ea}(\channel)\supseteq\mathcal{R}_{\text{Q}}^\text{ea}(\channel) $. 

Given unlimited amount of entanglement resources, if Alice can send $nQ$ qubits to Bob with estimation error $\eps$ and leakage rate 
$L$, then she can send $2nQ$ classical bits with the same error and leakage rate 
using the superdense coding protocol  \cite[Section 2.3]{NielsenChuang:02b}. Thus, for the transmission of classical bits, rate-leakage pairs $(R,L)$ such that $R \leq I(A;B)_\rho- I(A;E,C)_\rho$ and  $L   \geq I(C;AB)_\rho$ are achievable.
We deduce that $\mathbb{R}_{\text{Cl}}^\text{ea}(\channel) \supseteq \mathcal{R}_{\text{Cl}}^\text{ea}(\channel) $ as well.

\subsection{Converse Proof}
Next, we move to the converse part. While extending the classical arguments, we need to be careful since conditional entropies can be negative in the quantum setting, and since we have three channel state systems, $C$, $E$, and $E_0$. This poses a challenge in defining the auxiliary system $A$ that would satisfy both communication and leakage rate bounds. Here, we will use the assumption that the
 channel state systems are maximally correlated, as in (\ref{eq:CondSchmidt}). 

 Again, due to the superdense coding protocol, if Alice \emph{cannot} send $nR$ classical bits to Bob with estimation error $\eps$ and leakage rate $L$, then she \emph{cannot} send $\frac{1}{2} nR$ qubits with the same error and leakage rate. Thus, it suffices to consider the classical masking region.
%
%
%
Suppose that Alice and Bob are trying to distribute randomness. An upper bound on the rate at which Alice can distribute randomness to Bob also serves as an upper bound on the rate at which they can communicate classical bits. In this task, Alice and Bob share an entangled state $\Psi_{G_A G_B}$. Alice first prepares a maximally corrleated state
\begin{align}
\pi_{MM'} \equiv \frac{1}{2^{nR}}\sum_{m=1}^{2^{nR}} \kb{ m }_M \otimes \kb{ m }_{M'} \,.
\end{align}
locally, where $M$ and $M'$ are classical registers that store the message. Denote the joint state at the beginning by
\begin{align}
\psi_{MM' G_A G_B   E_0^n E^n C^n}= \pi_{MM'}\otimes \Psi_{G_A G_B} \otimes \phi_{ E_0 E C}^{\otimes n}
\end{align}
where 
$E^n$ are the channel state systems, $E_0^n$ are the CSI systems that are available to Alice, and $C^n$ are the systems that are masked from Bob (see Figure~\ref{fig:MSKsiCode}).
 Then, Alice applies an encoding channel $\Fset_{M' G_A E_0^n \rightarrow A'^n}$ to the classical system $M'$, her share $G_A$ of the entangled state $\Psi_{G_A G_B}$, and the CSI systems $E_0^n$. 
The resulting state is 
\begin{align}
\rho_{M  A'^n G_B  E^n C^n}\equiv \Fset_{M' G_A E_0^n \rightarrow A'^n}( \psi_{MM' G_A E_0^n G_B   E^n  C^n} 
) \,.
\end{align}
After Alice sends the systems $A'^n$ through the channel, Bob receives the systems $B^n$ at state
\begin{align}
\rho_{M B^n G_B  C^n }\equiv \channel^{\otimes n}_{E A'\rightarrow B} (\rho_{  M E^n A'^n G_B  C^n}) \,.
\end{align}
Then, Bob performs a decoding channel $\Dset_{B^n G_B\rightarrow \hM}$, producing 
\begin{align}
\rho_{  M \hM  C^n}\equiv \Dset_{B^n G_B\rightarrow \hM}(\rho_{M B^n G_B  C^n}) \,.
\label{eq:DecConv1}
\end{align}

Consider a sequence of codes $(\Fset_n,\Psi_n,\Dset_n)$ for randomness distribution, such that
\begin{align}
\frac{1}{2} \norm{ \rho_{M\hM} -\pi_{MM'} }_1 \leq& \alpha_n \label{eq:randDconv} \\
\frac{1}{n} I(C^n;B^n G_B)_\rho \leq& L+\beta_n \label{eq:randDconvLeak}
\end{align}
where 
$\alpha_n,\beta_n$ tend to zero as $n\rightarrow\infty$.
By the Alicki-Fannes-Winter inequality \cite{AlickiFannes:04p,Winter:16p} \cite[Theorem 11.10.3]{Wilde:17b}, (\ref{eq:randDconv}) implies that
\begin{align}
|H(M|\hM)_\rho - H(M|M')_{\pi} |\leq n\eps_n
\label{eq:AFW}
\end{align}
where $\eps_n$ tends to zero as $n\rightarrow\infty$.
Now, observe that $H(\pi_{M M'})=H(\pi_{M})=H(\pi_{ M'})=nR$, hence $I(M;\hM)_{\pi}=nR$.
Also,   $H(\rho_{M})=H(\pi_{ M})=nR$ implies that 
$I(M;M')_{\pi} - I(M;\hM)_{\rho}= H(M|\hM)_\rho - H(M|M')_{\pi}$. Therefore, by (\ref{eq:AFW}),
\begin{align}
nR=&I(M;\hM)_{\pi} \nonumber\\
\leq& I(M;\hM)_{\rho}+n\eps_n \nonumber\\
\leq& I(M;B^n G_B)_{\rho}+n\eps_n 
\label{eq:ConvIneq1}
\end{align}
where the last line follows from (\ref{eq:DecConv1}) and the quantum data processing inequality \cite[Theorem 11.5]{NielsenChuang:02b}.

As in the classical setting, the chain rule for the quantum mutual information states that 
$I(A;B,C)_\sigma=I(A;B)_\sigma+I(A;C|B)_\sigma$ for all $\sigma_{ABC}$ (see \eg \cite[Property 11.7.1]{Wilde:17b}). 
As a straightforward consequence, this leads to the Cisz\'ar sum identity,
\begin{align}
\sum_{i=1}^n I(A_{i+1}^n;B_i|B^{i-1})_\sigma=\sum_{i=1}^n I(B^{i-1};A_i|A_{i+1}^n)_\sigma
\label{eq:CsiszarIdentity}
\end{align}
for every sequence of systems $A^n$ and $B^n$.
Returning to (\ref{eq:ConvIneq1}), we apply the chain rule and rewrite the inequality as
\begin{align}
nR \leq& I(G_B M;B^n)_{\rho}+I(M;G_B)_{\rho}-I(G_B;B^n)_{\rho}+n\eps_n \nonumber\\
\leq& I(G_B M;B^n)_{\rho}+I(M;G_B)_{\rho}+n\eps_n \nonumber\\
=& I(G_B M;B^n)_{\rho}+n\eps_n 
\label{eq:ConvIneq2}
\end{align}
where the equality holds since the systems $M$ and $G_B$ are in a product state. The chain rule further 
implies that 
\begin{align}
I(G_B M;B^n)_{\rho}=&\sum_{i=1}^n I(G_B M;B_i| B^{i-1})_\rho 
\nonumber\\
\leq& \sum_{i=1}^n I(G_B M B^{i-1};B_i)_\rho \nonumber\\
=& \sum_{i=1}^n I(G_B M B^{i-1} C_{i+1}^n;B_i)_\rho -\sum_{i=1}^n I(B_i;C_{i+1}^n|G_B M B^{i-1})_\rho \nonumber\\
=& \sum_{i=1}^n I(G_B M B^{i-1} C_{i+1}^n;B_i)_\rho -\sum_{i=1}^n I(B^{i-1};C_i|G_B M C_{i+1}^n)_\rho
\label{eq:ConvIneq3}
\end{align}
where the last line follows from the quantum version of the Csisz\'ar sum identity in (\ref{eq:CsiszarIdentity}).
Since the systems $C_i$ and $(G_B,M,C_{i+1}^n)$ are in a product state, $I(B^{i-1};C_i|G_B M C_{i+1}^n)_\rho=
I(G_B M C_{i+1}^n B^{i-1};C_i)_\rho$. 
Therefore, defining  
\begin{align}
A_i=(G_B,M,B^{i-1},C_{i+1}^n)
\label{eq:EAconvAi}
\end{align}
we obtain 
\begin{align}
I(G_B M;B^n)_{\rho}\leq& \sum_{i=1}^n I(A_i;B_i)_\rho -\sum_{i=1}^n I(A_i;C_i)_\rho \,.
\label{eq:ConvIneq4}
\end{align}

Next, we claim that based on our assumption that $\varphi_{E_0 E C}$ is as in (\ref{eq:CondSchmidt}), we have $I(A_i;C_i)_\rho=I(A_i;E_i C_i )_\rho$.
To see this, consider the joint state of the systems $A_i$, $C_i$, and $E_i$,
\begin{multline}
\rho_{M B^{i-1} G_B C_{i+1}^n C_i E_i }=\sum_{s^n\in\Sset^n} q^n(s^n) \frac{1}{2^{nR}}\sum_{m=1}^{2^{nR}} \kb{m}_M 
\otimes \channel^{\otimes i-1}_{E A'\rightarrow B} \left( \kb{ s^{i-1} }_{E^{i-1}}\otimes
\rho_{A'^{i-1} G_B}^{m,s^n} \right) 
%
\otimes \kb{ s_{i+1}^n }_{C_{i+1}^n}\\ \otimes \kb{ s_i }_{C_i} \otimes \kb{ s_i }_{E_i}
\end{multline}
with $\rho_{A'^n G_B}^{m,s^n}\equiv \Fset_{M' G_A E_0^n \rightarrow A'^n}\left( \kb{m}_{M'}\otimes \Psi_{G_A G_B} \otimes \kb{ s^n }_{E_0^n}\right)$. 
Observing that the eigenvalues of the state $\rho_{A_i C_i E_i }$ are the same as those of 
$\rho_{A_i C_i }$, it follows that
$
H(A_i C_i E_i )_\rho=H(A_i C_i )_\rho
$ and 
$
H( C_i  E_i )_\rho=H( C_i )_\rho 
$, thus, 
\begin{align}
I(A_i;C_i)_\rho=I(A_i; E_i C_i)_\rho \,.
\label{eq:ConvIneq5}
\end{align}

Now, 
 let  $Y$ be a classical random variable with a uniform distribution over
$\{ 1,\ldots, n \}$, in a product state with the previous quantum systems, \ie $C^n$, $E^n$, $E_0^n$,  $M$, $M'$, $G_A$, $G_B$, $A'^n$, and
$B^n$. 
Then, by (\ref{eq:ConvIneq2}), (\ref{eq:ConvIneq4}), and (\ref{eq:ConvIneq5}), 
\begin{align}
R-\eps_n\leq& 
\frac{1}{n}\sum_{i=1}^n [ I(A_i;B_i)_\rho-I(A_i;E_i C_i)_\rho ] 
\nonumber\\
=& I(A_Y;B_Y|Y)_\rho-I(A_Y;E_Y C_Y|Y)_\rho  \nonumber\\
=& I(A_Y,Y;B_Y)_\rho-I(Y;B_Y)_\rho-I(A_Y,Y;E_Y C_Y)_\rho+I(Y;E_Y C_Y)_\rho \nonumber\\
\leq& I(A_Y,Y;B_Y)_\rho-I(A_Y,Y;E_Y C_Y)_\rho+I(Y;E_Y C_Y)_\rho \nonumber\\
=& I(A_Y,Y;B_Y)_\rho-I(A_Y,Y;E_Y C_Y)_\rho
\end{align}
with 
$
\rho_{YA_Y E_Y C_Y  A'_Y}=\frac{1}{n} \sum_{i=1}^n \kb{i} \otimes \rho_{A_i E_i C_i  A'_i} 
$ and $
\rho_{Y A_Y C_Y B_Y }=\channel_{EA'\rightarrow B}(\rho_{Y A_Y C_Y E_Y A_Y'}) 
$, 
where the last equality holds since $E^n$ and $C^n$  are in a product state $\phi_{EC}^{\otimes n}$, hence
$I(Y; E_Y C_Y)_\rho=H(E_Y C_Y)_\rho-H(E_Y C_Y|Y)_\rho=H(EC)_\phi-H(EC)_\phi=0$. Thus, defining
\begin{align}
A\equiv (A_Y,Y) \,,\; E\equiv E_Y \,,\;  C\equiv C_Y \,,\; A'\equiv A'_Y 
\label{eq:ConvDsin}
\end{align}
 and $B$ such that 
$
\rho_{ABC}=\channel_{EA'\rightarrow B}(\rho_{AEA' C}) 
$, we have that 
\begin{align}
R-\eps_n\leq I(A;B)_\rho-I(A;E C)_\rho \,.
\end{align}
We have thus shown the desired bound on the coding rate.

As for the leakage rate, by (\ref{eq:randDconvLeak}),
\begin{align}
n(L+\beta_n) \geq&   I(C^n;B^n G_B)_\rho 
\nonumber\\
=&  I(C^n;B^n G_B M)_\rho -I(C^n; M |B^n G_B)_\rho 
\nonumber\\
=& I(C^n;B^n  G_B  M)_\rho -H( M |B^n G_B)_\rho + H( M |C^n B^n G_B)_\rho \,.
\label{eq:ineqL1}
\end{align}
Note that the conditional entropy of a classical-quantum state $\rho_{XA}=\sum_{x\in\Xset}p_X(x)\kb{x}\otimes \rho_A^x$ is always  nonnegative, since $H(A|X)_\rho=\sum_x p_X(x) H(\rho_A^x)\geq 0 $ and $H(X|A)_\rho\geq H(X|A,X)=0$, as conditioning cannot increase quantum entropy \cite[Theorem 11.15]{NielsenChuang:02b}. Since $M$ is  classical, the last term in the RHS of (\ref{eq:ineqL1}) is nonnegative, i.e. 
\begin{align}
H( M |C^n, B^n, G_B)_\rho\geq 0 \,.
\label{eq:ineqL1a}
\end{align}
Furthermore, we have by (\ref{eq:ConvIneq1}) that the second term is bounded by 
\begin{align}
H( M |B^n G_B)_\rho = H(M)_{\pi} - I( M ;B^n G_B)_\rho\leq n\eps_n
\label{eq:ineqL1b}
\end{align}
Thus, by (\ref{eq:ineqL1})-(\ref{eq:ineqL1b}),
 \begin{align}
n(L+\beta_n+\eps_n) \geq&    I(C^n;B^n G_B M)_\rho 
\nonumber\\
=& \sum_{i=1}^n I(C_i;B^n G_B M|C_{i+1}^n)_\rho 
\nonumber\\
\geq& \sum_{i=1}^n I(C_i;B_i B^{i-1} G_B M|C_{i+1}^n)_\rho \,.
\intertext{
Then, since  $C_i$ and $C_{i+1}^n$ are in a product state, $I(C_i;C_{i+1}^n)_\rho=0$, hence
} 
L+\beta_n+\eps_n \geq&     \frac{1}{n}\sum_{i=1}^n I(C_i;B_i B^{i-1}  G_B  M C_{i+1}^n)_\rho \nonumber\\
=&     \frac{1}{n}\sum_{i=1}^n I(C_i;A_i B_i)_\rho=      I(C_Y;A_Y B_Y|Y)_\rho \nonumber\\
=&      I(C_Y;A_Y Y B_Y)_\rho=I(C;A B)_\rho
\end{align} 
where the first equality is due to our definition of $A_i$ in (\ref{eq:EAconvAi}), the second holds as the classical variable $Y$ is uniformly distributed over $\{1,\ldots,n\}$, the third since $I(C_Y;Y)_\rho=H(C_Y)_\rho-H(C_Y|Y)_\rho=H(C)_\phi-H(C)_\phi=0$, and the last equality follows from (\ref{eq:ConvDsin}).
This concludes the proof of Theorem~\ref{theo:MskEA}.
\qed 

\section{Proof of Theorem~\ref{theo:MskQ}}
\label{app:MskQ}
Let $\channel_{EA'\rightarrow B}$ be a quantum state-dependent  channel  with state information at the encoder and masking from the decoder, as in Theorem~\ref{theo:MskEA}. We now consider quantum communication without assistance.  
The converse proof without assistance is based on different considerations from those in the classical converse proof by Merhav and Shamai \cite{MerhavShamai:07p}.
In the classical proof, the derivation of the bounds on both the communication and leakage rates begins with Fano's inequality, followed by arguments that do not hold in our model since conditional quantum entropies can be negative.
Hence, we bound the leakage rate  in a different manner using the coherent information bound on the communication rate.
The direct part is a consequence of our previous result on masking with rate-limited entanglement assistance (see Theorem~\ref{theo:MskAchiev}).
In the second part, we derive a single-letter outer bound for Hadamard channels using the special properties of those channels. To bound the communication rate $Q$, we only need to use the fact that Hadamard channels are degradable. As for the bound on the leakage rate $L$, here we observe that for Hadamard channels, there also exists a channel from the output $B$ to $BC_1K$, i.e. the channel output combined with the decoder's environment.

\subsection*{Part 1}
Achievability of rate-leakage pairs in $\mathcal{R}_{\text{Q}}(\channel)$ immediately follows from Theorem~\ref{theo:MskAchiev}, taking $R_e=0$.
To show that rate-leakage pairs in $\frac{1}{k}  \mathcal{R}_{\text{Q}}(\channel^{\otimes k})$ are achievable as well,  employ the coding scheme in the proof of Theorem~\ref{theo:MskAchiev} in Appendix~\ref{app:MskAchiev} for the product channel $\channel^{\otimes k}$, where $k$ is arbitrarily large.

Next, we move the converse part. 
Suppose that Alice and Bob are trying to generate entanglement between them. 
An upper bound on the rate at which Alice and Bob can generate entanglement also serves as an upper bound on the rate at which they can communicate qubits, since a noiseless quantum channel can be used to generate entanglement by sending one part of an entangled pair. In this task, Alice locally prepares a maximally entangled state,
\begin{align}
|\Phi_{M M'}\rangle = \frac{1}{\sqrt{2^{nQ}}}\sum_{m=1}^{2^{nQ}} | m \rangle_{M} \otimes | m \rangle_{M'} \,.
\end{align}
 Denote the joint state at the beginning by
\begin{align}
|\theta_{M M'   E_0^n E^n C^n}\rangle= |\Phi_{M M'}\rangle\otimes  |\phi_{ E_0 E C}\rangle^{\otimes n}
\end{align}
where 
$E^n$ are the channel state systems, $E_0^n$ are the CSI systems that are available to Alice, and $C^n$ are the systems that are masked from Bob. 
 Then, Alice applies an encoding channel $\Fset_{M' E_0^n \rightarrow A'^n}$ to the quantum system $M'$ and the CSI systems $E_0^n$. 
The resulting state is 
\begin{align}
\rho_{M  A'^n   E^n C^n}\equiv \Fset_{M' E_0^n \rightarrow A'^n }( \theta_{M M'   E_0^n E^n C^n} 
) \,.
\label{eq:QconvI1}
\end{align}
After Alice sends the systems $A'^n$ through the channel, Bob receives the systems $B^n$ in the state
\begin{align}
\rho_{M B^n   C^n }\equiv \channel^{\otimes n}_{E A'\rightarrow B} (\rho_{  M E^n A'^n   C^n}) \,.
\end{align}
Then, Bob performs a decoding channel $\Dset_{B^n \rightarrow \hM}$, producing 
\begin{align}
\rho_{  M \hM  C^n}\equiv \Dset_{B^n \rightarrow \hM}(\rho_{A B^n   C^n}) \,.
\label{eq:DecConv1Q}
\end{align}

Consider a sequence of codes $(\Fset_n,\Dset_n)$ for entanglement generation, such that
\begin{align}
\frac{1}{2} \norm{ \rho_{M \hM} -\Phi_{M M'} }_1 \leq& \alpha_n \label{eq:randDconvQ} \\
\frac{1}{n} I(C^n;B^n)_\rho \leq& L+\beta_n \label{eq:randDconvLeakQ}
\end{align}
where 
$\alpha_n,\beta_n$ tend to zero as $n\rightarrow\infty$.

By the Alicki-Fannes-Winter inequality \cite{AlickiFannes:04p,Winter:16p} \cite[Theorem 11.10.3]{Wilde:17b}, (\ref{eq:randDconvQ}) implies that $|H(M|\hM)_\rho - H(M|M')_{\Phi} |\leq n\eps_n$, or equivalently,
\begin{align}
|I(M \rangle \hM)_\rho - I(M \rangle M')_{\Phi} |\leq n\eps_n 
\label{eq:AFWq}
\end{align}
where $\eps_n$ tends to zero as $n\rightarrow\infty$. 
Observe that $I(M \rangle M')_\Phi=H(M)_\Phi-H(MM')_\Phi=nQ-0=nQ$. Thus,
\begin{align}
nQ=& I(M\rangle M')_\Phi \nonumber\\
  \leq& I(M\rangle \hM)_\rho+n\eps_n \nonumber\\
	\leq& I(M\rangle B^n)_\rho+n\eps_n 
	\label{eq:ineqQuna1}
\end{align}
where the last line follows from (\ref{eq:DecConv1Q}) and the data processing inequality for the coherent information 
\cite[Theorem 11.9.3]{Wilde:17b}.
In addition,
\begin{align}
nQ=& H(M)_\Phi=H(M)_\theta \nonumber\\
  =& H(M|E^n C^n)_\theta \nonumber\\
	=& H(M|E^n C^n)_\rho
\end{align}
where the second line follows since $M$ and $(E^n,C^n)$ are in a product state.
Hence, $Q\leq \frac{1}{n} \min\{ I(M\rangle B^n)_\rho ,   H(M|E^n C^n)_\rho \}+\eps_n$. 
Let $A^n$ be quantum systems such that for some isometry $W_{M\rightarrow A^n}$, 
\begin{align}
\rho_{A^n  A'^n   E^n C^n}=W_{M\rightarrow A^n}\rho_{M  A'^n   E^n C^n} W_{M\rightarrow A^n}^\dagger \,.
\end{align}
 Since the von Neumann entropy is isometrically invariant \cite[Property 11.1.5]{Wilde:17b}, it follows that
\begin{align}
Q\leq \frac{1}{n} \min\{ I(A^n\rangle B^n)_\rho ,   H(A^n|E^n C^n)_\rho \}+\eps_n \,.
\end{align}

As for the leakage rate, by (\ref{eq:randDconvLeakQ}),
\begin{align}
n(L+\beta_n)\geq& I(C^n;B^n)_\rho 
\nonumber\\
=& I(C^n;M B^n)_\rho- I(C^n;M|B^n)_\rho
\nonumber\\
=& I(C^n;M B^n)_\rho- H(M|B^n)_\rho+ H(M|B^n C^n)_\rho
\label{eq:Clineq}\\ 
=& I(C^n;M B^n)_\rho+I(M\rangle B^n)_\rho+ H(M|B^n C^n)_\rho
\nonumber\\
\geq& I(C^n;M B^n)_\rho+n(Q-\eps_n)+ H(M|B^n C^n)_\rho
\end{align}
where the last line follows from (\ref{eq:ineqQuna1}). Since
\begin{align}
H(M|B^n C^n)_\rho\geq -\log|\Hset_M|=-nQ
\end{align}
(see \cite[Theorem 11.5.1]{Wilde:17b}), we have
\begin{align}
L+\beta_n+\eps_n\geq& \frac{1}{n} I(C^n;M B^n)_\rho \label{eq:randDconvLeakQf}
\\
=&\frac{1}{n} I(C^n;A^n B^n)_\rho  \,.
\end{align}
This completes the proof for the regularized capacity-leakage characterization.

\begin{remark}
\label{rem:Hnegative}
We note that in the classical converse proof  in \cite{MerhavShamai:07p}, the authors obtain an inequality that is similar to (\ref{eq:Clineq}) (see Eq. (21) in \cite{MerhavShamai:07p}). The next step in their proof is to use Fano's inequality in order to bound the second term by
\begin{align}
H(M|B^n)_\rho\leq n\eps_n
\label{eq:Hfano}
\end{align}
and to eliminate the third term, as $H(M|B^n C^n)_\rho\geq 0$. In the quantum setting, we can still write (\ref{eq:Hfano}), however, the last term is negative and could not be eliminated, as 
$H(M|B^n C^n)_\rho\leq H(M|B^n)_\rho\leq -n(Q-\eps_n)<0$. 
\end{remark}

\subsection*{Part 2}
Suppose that $\channel^{\,\text{H}}_{EA'\rightarrow B}$ is a Hadamard channel with an isometric extension 
$\Vset^{\,\text{H}}_{EA'\rightarrow BC_1 K}$ (see Definition~\ref{def:HadamardI}). 
The direct part follows from Theorem~\ref{theo:MskAchiev} as in part 1.
It remains to prove the single-letter converse part.

Returning to the entanglement generation protocol which we started with in part 1,  we now define
\begin{align}
A_i=(M,B^{i-1},K^{i-1},C^{i-1},C_{i+1}^n) \,.
\label{eq:NoAstAi}
\end{align}
%
%
 For every $i\in \{ 1,\ldots, n  \}$, consider the spectral representation   
\begin{align}
\rho_{ M E^i A'^i C_{i+1}^n    }=\sum_{x_i\in\Xset_i} p_{X_i}(x_i) \psi^{x_i}_{ME^i A'^i C_{i+1}^n}
\end{align}
where  $p_{X_i}(x_i)$ is a probability distribution and $\{ |\psi^{x_i}_{ME^i A'^i C_i^n} \rangle \}_{x_i\in\Xset_i}$ form an orthonormal basis, hence
\begin{align}
\rho_{ M B^i C_1^i K^i C_{i+1}^n    }=\sum_{x_i\in\Xset_i} p_{X_i}(x_i) \psi^{x_i}_{MB^i C_1^i K^i C_{i+1}^n}
\label{eq:jointMBnKnCn1}
\end{align}
where $ |\psi^{x_i}_{MB^i C_1^i K^i C_{i+1}^n} \rangle=(\Vset^{\,\text{H}}_{EA'\rightarrow BC_1 K})^{\otimes i}|\psi^{x_i}_{ME^i A'^i 
C_{i+1}^n} \rangle$. By (\ref{eq:hC1C}),
we also have $\rho_{ M B^i  K^i C_1^i C_{i+1}^n    }=\rho_{ M B^i  K^i C^i C_{i+1}^n    }$, hence
\begin{align}
\rho_{ M B^i K^i C^n    }=\sum_{x_i\in\Xset_i} p_{X_i}(x_i) \psi^{x_i}_{MB^i K^i C^n}
\label{eq:jointMBnKnCn}
\end{align}
with $ |\psi^{x_i}_{MB^i C^i K^i C_{i+1}^n} \rangle=(\Vset^{\,\text{H}}_{EA'\rightarrow BC K})^{\otimes i}|\psi^{x_i}_{ME^i A'^i 
C_{i+1}^n} \rangle$.
Given  a sequence of codes $(\Fset_n,\Dset_n)$ that satisfy (\ref{eq:randDconvQ})-(\ref{eq:randDconvLeakQ}), 
\begin{align}
n(Q-\eps_n) \leq& -H(M| B^n)_\rho \nonumber\\
  \leq& -H(M| B^n X^n )_\rho  \nonumber\\
     =& H(B^n| X^n  )_\rho-H(M B^n| X^n )_\rho  
\end{align}		
where the first inequality is due to (\ref{eq:ineqQuna1}), and the second inequality holds since  conditioning does not increase entropy \cite[Theorem 11.4.1]{Wilde:17b}. By (\ref{eq:jointMBnKnCn}),  the state of $M,B^n,K^n,C^n$ is pure when conditioned on 
 $X^n=x^n$, hence $H(M B^n| X^n )_\rho  =H( K^n  C^n| X^n )_\rho  $. Thus, we can write the last bound as
\begin{align}		
n(Q-\eps_n)		 \leq& H(B^n| X^n)_\rho-H(K^n  C^n | X^n)_\rho =H(B^n X^n)_\rho-H(K^n  C^n  X^n)_\rho \nonumber\\
		 =& \sum_{i=1}^n[ H(B_i X_i|B^{i-1}  X^{i-1})_\rho-H(K_i  C_i  X_i|K^{i-1}  C^{i-1}  X^{i-1} )_\rho]  \nonumber\\
		 =& \sum_{i=1}^n\left[ H(B_i X_i)_\rho-H(K_i  C_i X_i)_\rho -\left( I(B_i X_i;B^{i-1} X^{i-1})_\rho-I(K_i  C_i X_i;K^{i-1} C^{i-1} X^{i-1})_\rho \right) \right]   \nonumber\\
		 \leq& \sum_{i=1}^n\left[ H(B_i X_i)_\rho-H(K_i  C_i X_i)_\rho \right]
	\label{eq:ineqQuna1deg}
\end{align}
where the last inequality holds since Hadamard channels are degradable (see Subsection~\ref{subsec:LessNHadamard}), and thus 
\begin{align}
I(B_i X_i;B^{i-1} X^{i-1})_\rho\geq I(K_i  C_{1,i} X_i;K^{i-1} C_1^{i-1} X^{i-1})_\rho=I(K_i  C_{i} X_i;K^{i-1} C^{i-1} X^{i-1})_\rho
\end{align}
by the data processing theorem for the quantum mutual information \cite[Theorem 11.9.4]{Wilde:17b} and due to (\ref{eq:hC1C}).

 Now, according to (\ref{eq:jointMBnKnCn}),  the state of $M,B^i,K^i,C^n$ is pure for a  given $X_i=x_i$, hence 
\begin{align}
H(B_i | X_i )_\rho=H(M B^{i-1} K^{i} C^n| X_i )_\rho  =  H(A_i K_i C_i|X_i)
\end{align} 
(see (\ref{eq:NoAstAi})).
Then, (\ref{eq:ineqQuna1deg}) implies
\begin{align}
	  n(Q-\eps_n) \leq&  \sum_{i=1}^n H(A_i|C_i K_i X_i)_\rho \nonumber\\
								\leq&  \sum_{i=1}^n H(A_i|C_i K_i)_\rho
\label{eq:ineqQuna2deg}
\end{align}
since  conditioning does not increase entropy. 
%
%
Defining $Y$ to be a classical random variable of uniform distribution over
$\{ 1,\ldots, n \}$, in a product state with the previous systems, 
we have 
\begin{align}
Q-\eps_n\leq& 
\frac{1}{n} \sum_{i=1}^n H(A_i|C_i K_i )_\rho \nonumber\\
=& H(A_Y|C_Y K_Y Y)_\rho\leq H(A|C K )_\rho  
\end{align}
with 
$
\rho_{YA_Y E_Y C_Y  A'_Y}=\frac{1}{n} \sum_{i=1}^n \kb{i} \otimes \rho_{A_i E_i C_i  A'_i} 
$, $
\rho_{Y A_Y C_Y B_Y K_Y}=\Uset^\channel_{EA'\rightarrow B K}(\rho_{Y A_Y C_Y E_Y A_Y'}) 
$, 
and then 
\begin{align}
A\equiv (A_Y,Y) \,,\; E\equiv E_Y \,,\;  C\equiv C_Y \,,\; A'\equiv A'_Y 
\label{eq:ConvDsinQ}
\end{align}
 and $B,C_1,K$ such that 
$
\rho_{ABC_1 KC}=\Vset^{\,\text{H}}_{EA'\rightarrow B C_1 K}(\rho_{AEA' C}) 
$. 

As for the leakage rate, we begin with an observation that follows from our definition of Hadamard state-dependent channels in Subsection~\ref{subsec:LessNHadamard}.
Observe that given a Hadamard channel which is extended by $\Vset_{EA'\rightarrow C_1 K B}$, there exists a channel from $B$ to $BC_1 K$.
Specifically, if we define a channel $\Lset_{B\rightarrow BC_1 K}$ as the mapping $\psi_B^x \mapsto \psi_B^x\otimes \eta_{C_1 K}^x $, then we have 
\begin{align}
\rho_{ABC_1 KC}=\Lset_{B\rightarrow BC_1 K} (\rho_{ABC}) 
\label{eq:Lchannel}
\end{align}
or explicitly, 
\begin{align}
\Vset_{EA'\rightarrow C_1 K B}(\rho_{AEA'C}) =(\Lset_{B\rightarrow BC_1 K} \circ \channel_{EA'\rightarrow B})(\rho_{AEA'C})
\end{align}
 for all $\rho_{AA'EC}$ with $\rho_{EC}=\phi_{EC}$.

By (\ref{eq:ineqQuna1}),
\begin{align}
n(L+\beta_n)\geq& 
 I(C^n;M B^n )_\rho+n(Q-\eps_n)+ H(M|B^n K^n C^n)_\rho
\nonumber\\
\geq& I(C^n;M B^n )_\rho-n\eps_n
\label{eq:ineqQuna1H}
\end{align}
since
$
H(M|B^n K^n C^n)_\rho\geq -\log|\Hset_M|=-nQ
$ 
(see \cite[Theorem 11.5.1]{Wilde:17b}).
Next, we apply the chain rule and write
 \begin{align}
n(L+\beta_n+\eps_n) \geq&    I(C^n;B^n  M)_\rho 
\nonumber\\
=& \sum_{i=1}^n I(C_i;B^n   M|C_{i+1}^n)_\rho 
\nonumber\\
\geq& \sum_{i=1}^n I(C_i;B_i B^{i-1}  M|C_{i+1}^n)_\rho  
\nonumber\\
=& \sum_{i=1}^n I(C_i;B_i B^{i-1}  M C_{i+1}^n)_\rho 
\label{eq:ineqQuna2H}
\end{align} 
where the last equality holds since $C_i$ and $C_{i+1}^n$ are in a product state, hence $I(C_i;C_{i+1}^n)_\rho=0$. 
Using the fact that there exists a channel from $B^{i-1}$ to $B^{i-1}C_1^{i-1}K^{i-1}$ (see (\ref{eq:Lchannel})), along with the data processing theorem for the quantum mutual information, we deduce that 
\begin{align}
I(C_i;B_i M B^{i-1}   C_{i+1}^n)_\rho \geq& I(C_i;B_i M B^{i-1} K^{i-1} C_1^{i-1} C_{i+1}^n)_\rho \nonumber\\
																			=& I(C_i;B_i M B^{i-1} K^{i-1} C^{i-1} C_{i+1}^n)_\rho \nonumber\\
																			=& I(C_i;A_i B_i )_\rho
\label{eq:ineqQuna3H}
\end{align}
where the first equality follows from our definition of a Hadamard state-dependent channel (see (\ref{eq:hC1C})), and the last line is due to (\ref{eq:NoAstAi}).
Thus, by (\ref{eq:ineqQuna2H}) and (\ref{eq:ineqQuna3H}),
\begin{align}
L+\beta_n+\eps_n \geq&        \frac{1}{n}\sum_{i=1}^n I(C_i;A_i B_i)_\rho=      I(C_Y;A_Y B_Y|Y)_\rho \nonumber\\
=&     I(C_Y;Y A_Y B_Y)_\rho= I(C;A  B)_\rho
\end{align} 
where the first equality holds as the classical variable $Y$ is uniformly distributed over $\{1,\ldots,n\}$, the second since $I(C_Y;Y)_\rho=H(C_Y)_\rho-H(C_Y|Y)_\rho=H(C)_\phi-H(C)_\phi=0$, and the last equality follows from (\ref{eq:ConvDsinQ}).
This concludes the proof of Theorem~\ref{theo:MskQ}.
\qed
\end{appendices}

 
\bibliography{references2}{}

\ifdefined\bibstar\else\newcommand{\bibstar}[1]{}\fi
\begin{thebibliography}{100}
\providecommand{\url}[1]{#1}
\csname url@samestyle\endcsname
\providecommand{\newblock}{\relax}
\providecommand{\bibinfo}[2]{#2}
\providecommand{\BIBentrySTDinterwordspacing}{\spaceskip=0pt\relax}
\providecommand{\BIBentryALTinterwordstretchfactor}{4}
\providecommand{\BIBentryALTinterwordspacing}{\spaceskip=\fontdimen2\font plus
\BIBentryALTinterwordstretchfactor\fontdimen3\font minus
  \fontdimen4\font\relax}
\providecommand{\BIBforeignlanguage}[2]{{%
\expandafter\ifx\csname l@#1\endcsname\relax
\typeout{** WARNING: IEEEtran.bst: No hyphenation pattern has been}%
\typeout{** loaded for the language `#1'. Using the pattern for}%
\typeout{** the default language instead.}%
\else
\language=\csname l@#1\endcsname
\fi
#2}}
\providecommand{\BIBdecl}{\relax}
\BIBdecl

\bibitem{BHCPDA:13p}
E.~{Bou-Harb}, C.~{Fachkha}, M.~{Pourzandi}, M.~{Debbabi}, and C.~{Assi},
  ``Communication security for smart grid distribution networks,'' \emph{IEEE
  Commun. Mag.}, vol.~51, no.~1, pp. 42--49, January 2013.

\bibitem{LRBW:17p}
J.~Lopez, R.~Rios, F.~Bao, and G.~Wang, ``Evolving privacy: From sensors to the
  internet of things,'' \emph{Future Generation Computer Systems}, vol.~75, pp.
  46--57, 2017.

\bibitem{PiquerasJoverMarojevic:19p}
R.~{Piqueras Jover} and V.~{Marojevic}, ``Security and protocol exploit
  analysis of the 5g specifications,'' \emph{IEEE Access}, vol.~7, pp.
  24\,956--24\,963, 2019.

\bibitem{WYDP:19p}
H.~{Wang}, Q.~{Yang}, Z.~{Ding}, and H.~V. {Poor}, ``Secure short-packet
  communications for mission-critical iot applications,'' \emph{IEEE Trans.
  Wireless Commun.}, vol.~18, no.~5, pp. 2565--2578, May 2019.

\bibitem{Wyner:75p}
A.~D. Wyner, ``The wire-tap channel,'' \emph{Bell Syst. Tech. J}, vol. 54(8),
  pp. 1355--1387, 1975.

\bibitem{LiangPoorShamai:09n}
Y.~Liang, H.~V. Poor, and S.~Shamai, ``Information theoretic security,''
  \emph{Foundations and Trends{\textregistered} in Communications and
  Information Theory}, vol.~5, no. 4--5, pp. 355--580, 2009.

\bibitem{LiangKramerPoorShamai:05p}
Y.~Liang, G.~Kramer, H.~V. Poor, and S.~Shamai, ``Compound wiretap channels,''
  \emph{EURASIP J. on Wireless Commun. Netw.}, vol.~1, no. 2009, pp. 1--12,
  2005.

\bibitem{BarrosRodrigues:06c}
J.~Barros and M.~R.~D. Rodrigues, ``Secrecy capacity of wireless channels,'' in
  \emph{Proc. IEEE Int. Symp. Inf. Theory (ISIT'2006)}, Seattle, WA, USA, July
  2006, pp. 356--360.

\bibitem{BellareTessaroVardy:12a}
M.~Bellare, S.~Tessaro, and A.~Vardy, ``A cryptographic treatment of the
  wiretap channel,'' \emph{\textup{\texttt{arXiv:1201.2205}}}, 2012.

\bibitem{BocheSchaeferPoor:15p}
H.~Boche, R.~F. Schaefer, and H.~V. Poor, ``On the continuity of the secrecy
  capacity of compound and arbitrarily varying wiretap channels,'' \emph{IEEE
  Trans. Inf. Foren. Secur.}, vol.~10, no.~12, pp. 2531--2546, Dec 2015.

\bibitem{GoldfeldCuffPermuter:16p}
Z.~{Goldfeld}, P.~{Cuff}, and H.~H. {Permuter}, ``Semantic-security capacity
  for wiretap channels of type ii,'' \emph{IEEE Trans. Inf. Theory}, vol.~62,
  no.~7, pp. 3863--3879, July 2016.

\bibitem{XingLiuZhang:16p}
H.~{Xing}, L.~{Liu}, and R.~{Zhang}, ``Secrecy wireless information and power
  transfer in fading wiretap channel,'' \emph{IEEE Trans. Vehic. Tech.},
  vol.~65, no.~1, pp. 180--190, Jan 2016.

\bibitem{BGPSCP:18p}
A.~{Bunin}, Z.~{Goldfeld}, H.~H. {Permuter}, S.~{Shamai}, P.~{Cuff}, and
  P.~{Piantanida}, ``Key and message semantic-security over state-dependent
  channels,'' \emph{IEEE Trans. Inf. Theory}, pp. 1--1, 2018.

\bibitem{BocheCaiNotzelDeppe:19p}
H.~Boche, M.~Cai, J.~N{\"o}tzel, and C.~Deppe, ``Secret message transmission
  over quantum channels under adversarial quantum noise: Secrecy capacity and
  super-activation,'' \emph{J. Math. Phys.}, vol.~60, no.~6, p. 062202, 2019.

\bibitem{LiLiangPoorShamai:19p}
C.~{Li}, Y.~{Liang}, H.~V. {Poor}, and S.~{Shamai}, ``Secrecy capacity of
  colored gaussian noise channels with feedback,'' \emph{IEEE Trans. Inf.
  Theory}, vol.~65, no.~9, pp. 5771--5782, Sep. 2019.

\bibitem{MerhavShamai:07p}
N.~{Merhav} and S.~{Shamai}, ``Information rates subject to state masking,''
  \emph{IEEE Trans. Inf. Theory}, vol.~53, no.~6, pp. 2254--2261, June 2007.

\bibitem{LeTreustBloch:20p}
M.~{Le Treust} and M.~R. {Bloch}, ``State leakage and coordination with causal
  state knowledge at the encoder,'' \emph{IEEE Trans. Inf. Theory}, 2020.

\bibitem{KoyluogluSoundararajanVishwanath:16p}
O.~O. {Koyluoglu}, R.~{Soundararajan}, and S.~{Vishwanath}, ``State
  amplification subject to masking constraints,'' \emph{IEEE Trans. Inf.
  Theory}, vol.~62, no.~11, pp. 6233--6250, Nov 2016.

\bibitem{KoyluogluSoundararajanVishwanath:11c}
------, ``State amplification under masking constraints,'' in \emph{Proc.
  Allerton Conf. Commun., Control, Computing}, Monticello, IL, USA, 2011, pp.
  936--943.

\bibitem{DikshteinSomekhBaruchShamai:19c}
M.~{Dikshtein}, A.~{Somekh-Baruch}, and S.~{Shamai}, ``Broadcasting information
  subject to state masking over a mimo state dependent gaussian channel,'' in
  \emph{Proc. IEEE Int. Symp. Inf. Theory (ISIT'2019)}, Paris, France, July
  2019, pp. 275--279.

\bibitem{AsoodehDizaAlajajiLinder:16p}
S.~Asoodeh, M.~Diaz, F.~Alajaji, and T.~Linder, ``Information extraction under
  privacy constraints,'' \emph{Information}, vol.~7, no.~1, p.~15, 2016.

\bibitem{LeTreustBloch:16c}
M.~{Le Treust} and M.~{Bloch}, ``Empirical coordination, state masking and
  state amplification: Core of the decoder's knowledge,'' in \emph{Proc. IEEE
  Int. Symp. Inf. Theory (ISIT'2016)}, Barcelona, Spain, July 2016, pp.
  895--899.

\bibitem{TutunchuogluOzelYenerUlukus:14c}
K.~{Tutuncuoglu}, O.~{Ozel}, A.~{Yener}, and S.~{Ulukus}, ``State amplification
  and state masking for the binary energy harvesting channel,'' in \emph{2014
  IEEE Information Theory Workshop (ITW 2014)}, Hobart, TAS, Australia, July
  2014, pp. 336--340.

\bibitem{Courtade:12c}
T.~A. {Courtade}, ``Information masking and amplification: The source coding
  setting,'' in \emph{Proc. IEEE Int. Symp. Inf. Theory (ISIT'2012)},
  Cambridge, MA,USA, 2012, pp. 189--193.

\bibitem{NCLMSYH:20p}
S.~X. {Ng}, A.~{Conti}, G.~L. {Long}, P.~{Muller}, A.~{Sayeed}, J.~{Yuan}, and
  L.~{Hanzo}, ``Guest editorial advances in quantum communications, computing,
  cryptography, and sensing,'' \emph{IEEE J. Selected Areas in Commun.},
  vol.~38, no.~3, pp. 405--412, 2020.

\bibitem{DowlingMilburn:03p}
J.~P. Dowling and G.~J. Milburn, ``Quantum technology: the second quantum
  revolution,'' \emph{Philos. Trans. Royal Soc. London. Series A: Math., Phys.
  and Eng. Sciences}, vol. 361, no. 1809, pp. 1655--1674, 2003.

\bibitem{JKLGD:13p}
P.~Jouguet, S.~Kunz-Jacques, A.~Leverrier, P.~Grangier, and E.~Diamanti,
  ``Experimental demonstration of long-distance continuous-variable quantum key
  distribution,'' \emph{Nature Photonics}, vol.~7, no.~5, p. 378, 2013.

\bibitem{BennettBrassard:14p}
C.~H. Bennett and G.~Brassard, ``Quantum cryptography: public key distribution
  and coin tossing.'' \emph{Theor. Comput. Sci.}, vol. 560, no.~12, pp. 7--11,
  2014.

\bibitem{BecerraFanMigdall:15p}
F.~E. Becerra, J.~Fan, and A.~Migdall, ``Photon number resolution enables
  quantum receiver for realistic coherent optical communications,''
  \emph{Nature Photonics}, vol.~9, no.~1, p.~48, 2015.

\bibitem{YCLZRC:17p}
J.~Yin, Y.~Cao, Y.~H. Li, S.~K. Liao, L.~Zhang, J.~G. Ren, W.~Q. Cai, W.~Y.
  Liu, B.~Li, H.~Dai, G.~B. Li, Q.~M. Lu, Y.~H. Gong, Y.~Xu, S.~L. Li, F.~Z.
  Li, Y.~Y. Yin, Z.~Q. Jiang, M.~Li, J.~J. Jia, G.~Ren, D.~He, Y.~L. Zhou,
  X.~X. Zhang, N.~Wang, X.~Chang, Z.~C. Zhu, N.~L. Liu, Y.~A. Chen, C.~Y. Lu,
  R.~Shu, C.~Z. Peng, J.~Y. Wang, and J.~W. Pan, ``Satellite-based entanglement
  distribution over 1200 kilometers,'' \emph{Science}, vol. 356, no. 6343, pp.
  1140--1144, 2017.

\bibitem{ZDSZSG:17p}
W.~Zhang, D.~S. Ding, Y.~B. Sheng, L.~Zhou, B.~S. Shi, and G.~C. Guo, ``Quantum
  secure direct communication with quantum memory,'' \emph{Phys. Rev. Lett.},
  vol. 118, no.~22, p. 220501, 2017.

\bibitem{PERLHPCVV:20p}
L.~Petit, H.~G.~J. Eenink, M.~Russ, W.~I.~L. Lawrie, N.~W. Hendrickx, S.~G.~J.
  Philips, J.~S. Clarke, L.~M.~K. Vandersypen, and M.~Veldhorst, ``Universal
  quantum logic in hot silicon qubits,'' \emph{Nature}, vol. 580, pp. 355--359,
  april 2020.

\bibitem{BouwmeesterZeilinger:00b}
D.~Bouwmeester and A.~Zeilinger, ``The physics of quantum information: basic
  concepts,'' in \emph{The physics of quantum information}.\hskip 1em plus
  0.5em minus 0.4em\relax Springer, 2000, pp. 1--14.

\bibitem{ImreGyongyosi:12b}
S.~Imre and L.~Gyongyosi, \emph{Advanced quantum communications: an engineering
  approach}.\hskip 1em plus 0.5em minus 0.4em\relax John Wiley \& Sons, 2012.

\bibitem{Kitaev:97b}
A.~Y. Kitaev, ``Quantum error correction with imperfect gates,'' in
  \emph{Quantum Communication, Computing, and Measurement}.\hskip 1em plus
  0.5em minus 0.4em\relax Springer, 1997, pp. 181--188.

\bibitem{Wilde:17b}
M.~M. Wilde, \emph{Quantum information theory}, 2nd~ed.\hskip 1em plus 0.5em
  minus 0.4em\relax Cambridge University Press, 2017.

\bibitem{GyongyosiImreNguyen:18p}
L.~{Gyongyosi}, S.~{Imre}, and H.~V. {Nguyen}, ``A survey on quantum channel
  capacities,'' \emph{IEEE Commun. Surveys Tutorials}, vol.~20, no.~2, pp.
  1149--1205, 2018.

\bibitem{SmithYard:08p}
G.~Smith and J.~Yard, ``Quantum communication with zero-capacity channels,''
  \emph{Science}, vol. 321, no. 5897, pp. 1812--1815, 2008.

\bibitem{Holevo:98p}
A.~S. {Holevo}, ``The capacity of the quantum channel with general signal
  states,'' \emph{IEEE Trans. Inf. Theory}, vol.~44, no.~1, pp. 269--273, Jan
  1998.

\bibitem{SchumacherWestmoreland:97p}
B.~Schumacher and M.~D. Westmoreland, ``Sending classical information via noisy
  quantum channels,'' \emph{Phys. Rev. A}, vol.~56, no.~1, p. 131, July 1997.

\bibitem{Holevo:12b}
A.~S. Holevo, \emph{Quantum systems, channels, information: a mathematical
  introduction}.\hskip 1em plus 0.5em minus 0.4em\relax Walter de Gruyter,
  2012, vol.~16.

\bibitem{BarnumNielsenSchumacher:98p}
H.~Barnum, M.~A. Nielsen, and B.~Schumacher, ``Information transmission through
  a noisy quantum channel,'' \emph{Phys. Rev. A}, vol.~57, no.~6, p. 4153, June
  1998.

\bibitem{Loyd:97p}
S.~Lloyd, ``Capacity of the noisy quantum channel,'' \emph{Phys. Rev. A},
  vol.~55, no.~3, p. 1613, March 1997.

\bibitem{Shor:02l}
P.~W. Shor, ``The quantum channel capacity and coherent information,'' in
  \emph{Lecture notes, MSRI Workshop Quant. Comput.}, 2002.

\bibitem{Devetak:05p}
I.~Devetak, ``The private classical capacity and quantum capacity of a quantum
  channel,'' \emph{IEEE Trans. Inf. Theory}, vol.~51, no.~1, pp. 44--55, 2005.

\bibitem{DevetakShor:05p}
I.~Devetak and P.~W. Shor, ``The capacity of a quantum channel for simultaneous
  transmission of classical and quantum information,'' \emph{Commun. in Math.
  Phys.}, vol. 256, no.~2, pp. 287--303, June 2005.

\bibitem{NielsenChuang:02b}
M.~A. Nielsen and I.~Chuang, ``Quantum computation and quantum information,''
  2002.

\bibitem{BocheJanssenKaltenstadler:17p}
H.~Boche, G.~Jan{\ss}en, and S.~Kaltenstadler, ``Entanglement-assisted
  classical capacities of compound and arbitrarily varying quantum channels,''
  \emph{Quantum Information Processing}, vol.~16, no.~4, p.~88, Feb 2017.

\bibitem{ChitambarGour:19p}
E.~Chitambar and G.~Gour, ``Quantum resource theories,'' \emph{Rev. Modern
  Phys.}, vol.~91, no.~2, p. 025001, April 2019.

\bibitem{BennetWiesner:92p}
C.~H. Bennett and S.~J. Wiesner, ``Communication via one-and two-particle
  operators on einstein-podolsky-rosen states,'' \emph{Phys. Rev. Lett.},
  vol.~69, no.~20, p. 2881, Nov 1992.

\bibitem{BennettShorSmolin:99p}
C.~H. Bennett, P.~W. Shor, J.~A. Smolin, and A.~V. Thapliyal,
  ``Entanglement-assisted classical capacity of noisy quantum channels,''
  \emph{Phys. Rev. Lett.}, vol.~83, no.~15, p. 3081, Oct 1999.

\bibitem{BennettShorSmolin:02p}
C.~H. {Bennett}, P.~W. {Shor}, J.~A. {Smolin}, and A.~V. {Thapliyal},
  ``Entanglement-assisted capacity of a quantum channel and the reverse shannon
  theorem,'' \emph{IEEE Trans. Inf. Theory}, vol.~48, no.~10, pp. 2637--2655,
  Oct 2002.

\bibitem{Swingle:10a}
B.~Swingle, ``Mutual information and the structure of entanglement in quantum
  field theory,'' \emph{\textup{\texttt{arXiv:1010.4038}}}, 2010.

\bibitem{PanJing:08p}
Q.~Pan and J.~Jing, ``Degradation of nonmaximal entanglement of scalar and
  dirac fields in noninertial frames,'' \emph{Phys. Rev. A}, vol.~77, no.~2, p.
  024302, Feb 2008.

\bibitem{CasiniHuertaMyersYale:15p}
H.~Casini, M.~Huerta, R.~C. Myers, and A.~Yale, ``Mutual information and the
  f-theorem,'' \emph{J. High Energy Phys.}, vol. 2015, no.~10, p.~3, Oct 2015.

\bibitem{AgonGaulkner:16p}
C.~A. Ag{\'o}n and T.~Faulkner, ``Quantum corrections to holographic mutual
  information,'' \emph{J. High Energy Phys.}, vol. 2016, no.~8, p. 118, Aug
  2016.

\bibitem{Shannon:48p}
C.~Shannon, ``A mathematical theory of communication,'' \emph{Bell Syst. Tech.
  J}, vol.~27, pp. 379--423, 623--656, Jul 1948.

\bibitem{LMMOR:12p}
D.~Leung, L.~Mancinska, W.~Matthews, M.~Ozols, and A.~Roy, ``Entanglement can
  increase asymptotic rates of zero-error classical communication over
  classical channels,'' \emph{Communications in Mathematical Physics}, vol.
  311, no.~1, pp. 97--111, March 2012.

\bibitem{LeditzkyAlhejjiLevinSmith:20p}
F.~Leditzky, M.~A. Alhejji, J.~Levin, and G.~Smith, ``Playing games with
  multiple access channels,'' \emph{Nature communications}, vol.~11, no.~1, pp.
  1--5, 2020.

\bibitem{CHSH:69p}
J.~F. Clauser, M.~A. Horne, A.~Shimony, and R.~A. Holt, ``Proposed experiment
  to test local hidden-variable theories,'' \emph{Physical review letters},
  vol.~23, no.~15, p. 880, Oct 1969.

\bibitem{PappaChaillouxWehnerDiamantiKerenidis:12p}
A.~Pappa, A.~Chailloux, S.~Wehner, E.~Diamanti, and I.~Kerenidis,
  ``Multipartite entanglement verification resistant against dishonest
  parties,'' \emph{Phys. Rev. Lett.}, vol. 108, no.~26, p. 260502, June 2012.

\bibitem{VaziraniVidick:14p}
U.~Vazirani and T.~Vidick, ``Fully device-independent quantum key
  distribution,'' \emph{Phys. Rev. Lett.}, vol. 113, no.~14, p. 140501, Sep
  2014.

\bibitem{JaiWeiWuGuo:20a}
Z.~A. Jia, L.~Wei, Y.~C. Wu, and G.~C. Guo, ``Quantum advantages of
  communication complexity from bell nonlocality,''
  \emph{\textup{\texttt{arXiv:2004.05098}}}, 2020.

\bibitem{JiNatarajanVidickWrightYuen:20a}
Z.~Ji, A.~Natarajan, T.~Vidick, J.~Wright, and H.~Yuen, ``Mip*= re,''
  \emph{\textup{\texttt{arXiv:2001.04383}}}, 2020.

\bibitem{BennettBrassardJozsaPeres:93p}
C.~H. Bennett, G.~Brassard, C.~Cr{\'e}peau, R.~Jozsa, A.~Peres, and W.~K.
  Wootters, ``Teleporting an unknown quantum state via dual classical and
  einstein-podolsky-rosen channels,'' \emph{Physical review letters}, vol.~70,
  no.~13, p. 1895, 1993.

\bibitem{Shor:04p}
P.~W. Shor, ``The classical capacity achievable by a quantum channel assisted
  by a limited entanglement,'' \emph{Quantum Inf. Comp.}, vol.~4, no.~6, pp.
  537--545, Dec 2004.

\bibitem{DevetakHarrowWinter:08p}
I.~{Devetak}, A.~W. {Harrow}, and A.~J. {Winter}, ``A resource framework for
  quantum shannon theory,'' \emph{IEEE Trans. Inf. Theory}, vol.~54, no.~10,
  pp. 4587--4618, Oct 2008.

\bibitem{HsiehWilde:10p}
M.~{Hsieh} and M.~M. {Wilde}, ``Entanglement-assisted communication of
  classical and quantum information,'' \emph{IEEE Trans. Inf. Theory}, vol.~56,
  no.~9, pp. 4682--4704, Sep. 2010.

\bibitem{HsiehWilde:10p1}
------, ``Trading classical communication, quantum communication, and
  entanglement in quantum shannon theory,'' \emph{IEEE Trans. Inf. Theory},
  vol.~56, no.~9, pp. 4705--4730, Sep 2010.

\bibitem{WildeHsieh:12p1}
M.~M. Wilde and M.~H. Hsieh, ``The quantum dynamic capacity formula of a
  quantum channel,'' \emph{Quantum Inf. Proc.}, vol.~11, no.~6, pp. 1431--1463,
  Sep 2012.

\bibitem{WangHayashi:20c}
K.~{Wang} and M.~{Hayashi}, ``Permutation enhances classical communication
  assisted by entangled states,'' in \emph{2020 IEEE International Symposium on
  Information Theory (ISIT)}, Los Angeles, CA, USA, 2020, pp. 1840--1845.

\bibitem{DevetakHarrowWinter:04p}
I.~Devetak, A.~W. Harrow, and A.~Winter, ``A family of quantum protocols,''
  \emph{Phys. Rev. Lett.}, vol.~93, no.~23, p. 230504, Dec 2004.

\bibitem{Devetak:06p}
I.~Devetak, ``Triangle of dualities between quantum communication protocols,''
  \emph{Phys. Rev. Lett.}, vol.~97, no.~14, p. 140503, Oct 2006.

\bibitem{HorodeckiOppenheim:07p}
M.~Horodecki, J.~Oppenheim, and A.~Winter, ``Quantum state merging and negative
  information,'' \emph{Comm. Math. Phys.}, vol. 269, no.~1, pp. 107--136, 2007.

\bibitem{AbeyesingheDevetakHaydenWinter:09p}
A.~Abeyesinghe, I.~Devetak, P.~Hayden, and A.~Winter, ``The mother of all
  protocols: Restructuring quantum information’s family tree,'' \emph{Proc.
  Royal Society A: Math., Phys. and Engin. Sciences}, vol. 465, no. 2108, pp.
  2537--2563, Aug 2009.

\bibitem{DupuisHaydenLi:10p}
F.~{Dupuis}, P.~{Hayden}, and K.~{Li}, ``A father protocol for quantum
  broadcast channels,'' \emph{IEEE Trans. Inf. Theory}, vol.~56, no.~6, pp.
  2946--2956, June 2010.

\bibitem{HaydenHorodeckiWinterYard:08p}
P.~Hayden, M.~Horodecki, A.~Winter, and J.~Yard, ``A decoupling approach to the
  quantum capacity,'' \emph{Open Sys. Inf. Dynamics}, vol.~15, no.~01, pp.
  7--19, 2008.

\bibitem{Dupuis:08a}
F.~Dupuis, ``Coding for quantum channels with side information at the
  transmitter,'' \emph{arXiv preprint arXiv:0805.3352}, 2008.

\bibitem{Dupuis:10z}
------, ``The decoupling approach to quantum information theory,'' Ph.D.
  dissertation, Universit\'e de Montr\'eal, 2010.

\bibitem{Holevo:02p}
A.~S. Holevo, ``On entanglement-assisted classical capacity,'' \emph{J. Math.
  Phys.}, vol.~43, no.~9, pp. 4326--4333, 2002.

\bibitem{HsiehDevetakWinter:08p}
M.~{Hsieh}, I.~{Devetak}, and A.~{Winter}, ``Entanglement-assisted capacity of
  quantum multiple-access channels,'' \emph{IEEE Trans. Inf. Theory}, vol.~54,
  no.~7, pp. 3078--3090, July 2008.

\bibitem{Shirokov:12p}
M.~E. Shirokov, ``Conditions for coincidence of the classical capacity and
  entanglement-assisted capacity of a quantum channel,'' \emph{Problems. Inf.
  Transm.}, vol.~48, no.~2, pp. 85--101, 2012.

\bibitem{DattaHsieh:13p}
N.~{Datta} and M.~{Hsieh}, ``One-shot entanglement-assisted quantum and
  classical communication,'' \emph{IEEE Trans. Inf. Theory}, vol.~59, no.~3,
  pp. 1929--1939, March 2013.

\bibitem{WildeHsiehBabar:14p}
M.~M. {Wilde}, M.~{Hsieh}, and Z.~{Babar}, ``Entanglement-assisted quantum
  turbo codes,'' \emph{IEEE Trans. Inf. Theory}, vol.~60, no.~2, pp.
  1203--1222, Feb 2014.

\bibitem{QianZhan:18p}
J.~Qian and L.~Zhang, ``On mds linear complementary dual codes and
  entanglement-assisted quantum codes,'' \emph{Designs, Codes and
  Cryptography}, vol.~86, no.~7, pp. 1565--1572, 2018.

\bibitem{AnshuJainWarsi:17a}
A.~Anshu, R.~Jain, and N.~A. Warsi, ``One shot entanglement assisted classical
  and quantum communication over noisy quantum channels: A hypothesis testing
  and convex split approach,'' \emph{\textup{\texttt{arXiv:1702.01940}}}, 2017.

\bibitem{BertaGharibyanWalter:17p}
M.~{Berta}, H.~{Gharibyan}, and M.~{Walter}, ``Entanglement-assisted capacities
  of compound quantum channels,'' \emph{IEEE Transactions on Information
  Theory}, vol.~63, no.~5, pp. 3306--3321, Feb 2017.

\bibitem{CCVH:19a}
A.~S. Cacciapuoti, M.~Caleffi, R.~Van~Meter, and L.~Hanzo, ``When entanglement
  meets classical communications: Quantum teleportation for the quantum
  internet,'' \emph{\textup{\texttt{arXiv:1907.06197}}}, 2019.

\bibitem{AnshuJainWarsi:19p}
A.~Anshu, R.~Jain, and N.~A. Warsi, ``On the near-optimality of one-shot
  classical communication over quantum channels,'' \emph{J. Math. Phys.},
  vol.~60, no.~1, p. 012204, 2019.

\bibitem{BocheCaiNotzel:16p}
H.~Boche, N.~Cai, and J.~N{\"o}tzel, ``The classical-quantum channel with
  random state parameters known to the sender,'' \emph{J. Physics A: Math. and
  Theor.}, vol.~49, no.~19, p. 195302, April 2016.

\bibitem{Pereg:20c1}
U.~Pereg, ``Communication over quantum channels with parameter estimation,'' in
  \emph{Proc. IEEE Int. Symp. Inf. Theory (ISIT'2020)}, June 2020.

\bibitem{Pereg:19a3}
\BIBentryALTinterwordspacing
------, ``Communication over quantum channels with parameter estimation,''
  \emph{\textup{\texttt{arXiv:2001.00836}}}, Jan 2020. [Online]. Available:
  \url{https://arxiv.org/pdf/2001.00836.pdf}
\BIBentrySTDinterwordspacing

\bibitem{WarsiCoon:17p}
N.~A. {Warsi} and J.~P. {Coon}, ``Coding for classical-quantum channels with
  rate limited side information at the encoder: information-spectrum
  approach,'' \emph{IEEE Trans. Inf. Theory}, vol.~63, no.~5, pp. 3322--3331,
  May 2017.

\bibitem{Dupuis:09c}
F.~{Dupuis}, ``The capacity of quantum channels with side information at the
  transmitter,'' in \emph{Proc. IEEE Int. Symp. Inf. Theory (ISIT'2009)}, June
  2009, pp. 948--952.

\bibitem{Pereg:19c3}
U.~Pereg, ``Entanglement-assisted capacity of quantum channels with side
  information,'' in \emph{Int. Z\"urich Seminar Inf. Commun. (IZS'2020)},
  Z\"urich, Switzerland, Feb 2020.

\bibitem{Pereg:19a}
\BIBentryALTinterwordspacing
------, ``Entanglement-assisted capacity of quantum channels with side
  information,'' \emph{\textup{\texttt{arXiv:1909.09992}}}, Sep 2019. [Online].
  Available: \url{https://arxiv.org/pdf/1909.09992.pdf}
\BIBentrySTDinterwordspacing

\bibitem{LuoDevetak:09p}
Z.~{Luo} and I.~{Devetak}, ``Channel simulation with quantum side
  information,'' \emph{IEEE Trans. Inf. Theory}, vol.~55, no.~3, pp.
  1331--1342, March 2009.

\bibitem{WynerZiv:76p}
A.~Wyner and J.~Ziv, ``The rate-distortion function for source coding with side
  information at the decoder,'' \emph{IEEE Trans. Inf. Theory}, vol.~22, no.~1,
  pp. 1--10, Jan 1976.

\bibitem{DevetakWinter:03p}
I.~Devetak and A.~Winter, ``Classical data compression with quantum side
  information,'' \emph{Phys. Rev. A}, vol.~68, p. 042301, Oct 2003.

\bibitem{YardDevetak:09p}
J.~T. {Yard} and I.~{Devetak}, ``Optimal quantum source coding with quantum
  side information at the encoder and decoder,'' \emph{IEEE Trans. Inf.
  Theory}, vol.~55, no.~11, pp. 5339--5351, Nov 2009.

\bibitem{HsiehWatanabe:16p}
M.~{Hsieh} and S.~{Watanabe}, ``Channel simulation and coded source
  compression,'' \emph{IEEE Trans. Inf. Theory}, vol.~62, no.~11, pp.
  6609--6619, Nov 2016.

\bibitem{DattaHircheWinter:19c}
N.~Datta, C.~Hirche, and A.~Winter, ``Convexity and operational interpretation
  of the quantum information bottleneck function,'' in \emph{Proc. IEEE Int.
  Symp. Inf. Theory (ISIT'2019)}, Paris, France, July 2019, pp. 1157--1161.

\bibitem{DattaHircheWinter:18a}
------, ``Convexity and operational interpretation of the quantum information
  bottleneck function,'' \emph{\textup{\texttt{arXiv:1810.03644}}}, 2018.

\bibitem{CHDH:19c}
H.~C. Cheng, E.~P. Hanson, N.~Datta, and M.~H. Hsieh, ``Duality between source
  coding with quantum side information and cq channel coding,'' in \emph{Proc.
  IEEE Int. Symp. Inf. Theory (ISIT'2019)}, Paris, France, July 2019, pp.
  1142--1146.

\bibitem{CHDH:18a}
------, ``Duality between source coding with quantum side information and cq
  channel coding,'' \emph{\textup{\texttt{arXiv:1809.11143}}}, 2018.

\bibitem{KhanianWinter:19c2}
Z.~B. Khanian and A.~Winter, ``Distributed compression of correlated
  classical-quantum sources or: the price of ignorance,'' in \emph{Proc. IEEE
  Int. Symp. Inf. Theory (ISIT'2019)}, Paris, France, July 2019, pp.
  1152--1156.

\bibitem{KhanianWinter:18a}
------, ``Distributed compression of correlated classical-quantum sources or:
  the price of ignorance,'' \emph{\textup{\texttt{arXiv:1811.09177}}}, 2018.

\bibitem{KhanianWinter:19c}
------, ``Entanglement-assisted quantum data compression,'' in \emph{Proc. IEEE
  Int. Symp. Inf. Theory (ISIT'2019)}, Paris, France, July 2019, pp.
  1147--1151.

\bibitem{KhanianWinter:19a}
------, ``Entanglement-assisted quantum data compression,''
  \emph{\textup{\texttt{arXiv:1901.06346}}}, 2019.

\bibitem{CaiWinterYeung:04p}
N.~Cai, A.~Winter, and R.~W. Yeung, ``Quantum privacy and quantum wiretap
  channels,'' \emph{Probl. Info. Transm.}, vol.~40, no.~4, pp. 318--336, 2004.

\bibitem{DevetakWinter:05p}
I.~Devetak and A.~Winter, ``Distillation of secret key and entanglement from
  quantum states,'' \emph{Proc. Royal Society A: Math., Phys. and Engin.
  Sciences}, vol. 461, no. 2053, pp. 207--235, 2005.

\bibitem{HsiehLuoBrun:08p}
M.~H. Hsieh, Z.~Luo, and T.~Brun, ``Secret-key-assisted private classical
  communication capacity over quantum channels,'' \emph{Physical Review A},
  vol.~78, no.~4, p. 042306, 2008.

\bibitem{LiWinterZouGuo:09}
K.~Li, A.~Winter, X.~Zou, and G.~Guo, ``Private capacity of quantum channels is
  not additive,'' \emph{Physical Review Letters}, vol. 103, no.~12, p. 120501,
  2009.

\bibitem{Wilde:11p}
M.~M. Wilde, ``Comment on “secret-key-assisted private classical
  communication capacity over quantum channels”,'' \emph{Phys. Rev. A},
  vol.~83, no.~4, p. 046303, 2011.

\bibitem{Watanabe:12p}
S.~Watanabe, ``Private and quantum capacities of more capable and less noisy
  quantum channels,'' \emph{Phys. Rev. A}, vol.~85, no.~1, p. 012326, 2012.

\bibitem{ElkoussStrelchuk:15p}
D.~Elkouss and S.~Strelchuk, ``Superadditivity of private information for any
  number of uses of the channel,'' \emph{Phys. Rev. Lett.}, vol. 115, no.~4, p.
  040501, 2015.

\bibitem{AnshuHayashiWarsi:18c}
A.~{Anshu}, M.~{Hayashi}, and N.~A. {Warsi}, ``Secure communication over fully
  quantum gel'fand-pinsker wiretap channel,'' in \emph{Proc. IEEE Int. Symp.
  Inf. Theory (ISIT'2018)}, Vail, CO, USA, June 2018, pp. 2679--2683.

\bibitem{QiSharmaWilde:18p}
H.~Qi, K.~Sharma, and M.~M. Wilde, ``Entanglement-assisted private
  communication over quantum broadcast channels,'' \emph{J. Phys. A: Math. and
  Theo.}, vol.~51, no.~37, p. 374001, 2018.

\bibitem{SharmaWakauwaWilde:17a}
K.~Sharma, E.~Wakakuwa, and M.~M. Wilde, ``Conditional quantum one-time pad,''
  \emph{\textup{\texttt{arXiv:1703.02903}}}, 2017.

\bibitem{BochCaiDeppeNotzel:17p}
H.~Boche, M.~Cai, C.~Deppe, and J.~N{\"o}tzel, ``Classical-quantum arbitrarily
  varying wiretap channel: Common randomness assisted code and continuity,''
  \emph{Quantum Info. Proc.}, vol.~16, no.~1, p.~35, 2017.

\bibitem{WildeHsieh:12p}
M.~M. Wilde and M.~H. Hsieh, ``Public and private resource trade-offs for a
  quantum channel,'' \emph{Quantum Information Processing}, vol.~11, no.~6, pp.
  1465--1501, 2012.

\bibitem{KonigRennerBariskaMaurer:07p}
R.~K{\"o}nig, R.~Renner, A.~Bariska, and U.~Maurer, ``Small accessible quantum
  information does not imply security,'' \emph{Physical Review Letters},
  vol.~98, no.~14, p. 140502, 2007.

\bibitem{GHKLLSTW:14p}
S.~Guha, P.~Hayden, H.~Krovi, S.~Lloyd, C.~Lupo, J.~H. Shapiro, M.~Takeoka, and
  M.~M. Wilde, ``Quantum enigma machines and the locking capacity of a quantum
  channel,'' \emph{Physical Review X}, vol.~4, no.~1, p. 011016, 2014.

\bibitem{LupoWildeLloyd:16p}
C.~{Lupo}, M.~M. {Wilde}, and S.~{Lloyd}, ``Quantum data hiding in the presence
  of noise,'' \emph{IEEE Trans. Inf. Theory}, vol.~62, no.~6, pp. 3745--3756,
  June 2016.

\bibitem{SalekHsiehFonollosa:19a}
F.~Salek, M.~H. Hsieh, and J.~R. Fonollosa, ``Publicness, privacy and
  confidentiality in the single-serving quantum broadcast channel,''
  \emph{\textup{\texttt{arXiv:1903.04463}}}, 2019.

\bibitem{SalekHsiehFonollosa:19c}
F.~{Salek}, M.~{Hsieh}, and J.~R. {Fonollosa}, ``Publicness, privacy and
  confidentiality in the single-serving quantum broadcast channel,'' in
  \emph{Proc. IEEE Int. Symp. Inf. Theory (ISIT'2019)}, Paris, France, July
  2019, pp. 1712--1716.

\bibitem{AghaeeAkhbari:19c}
H.~Aghaee and B.~Akhbari, ``Classical-quantum multiple access wiretap
  channel,'' in \emph{Int'l ISC Conf. Info. Secur. Crypt. (ISCISC'2019)},
  Mashhad, Iran, November 2019.

\bibitem{BochJanssenSaeedianaeeni:20p}
H.~Boche, G.~Jan{\ss}en, and S.~Saeedinaeeni, ``Universal superposition codes:
  Capacity regions of compound quantum broadcast channel with confidential
  messages,'' \emph{J. Math. Phys.}, vol.~61, no.~4, p. 042204, April 2020.

\bibitem{ModiPatiSenSen:18p}
K.~Modi, A.~K. Pati, A.~Sen, and U.~Sen, ``Masking quantum information is
  impossible,'' \emph{Phys. Rev. Lett.}, vol. 120, no.~23, p. 230501, 2018.

\bibitem{LieJeong:19a}
S.~H. Lie and H.~Jeong, ``Randomness cost of masking quantum information and
  the information conservation law,''
  \emph{\textup{\texttt{arXiv:1908.07426}}}, 2019.

\bibitem{LieKwonKimJeong:19a}
S.~H. Lie, H.~Kwon, M.~S. Kim, and H.~Jeong, ``Unconditionally secure qubit
  commitment scheme using quantum maskers,'' \emph{arXiv preprint
  arXiv:1903.12304}, 2019.

\bibitem{AlickiFannes:04p}
R.~Alicki and M.~Fannes, ``Continuity of quantum conditional information,''
  \emph{J. Phys. A: Math. General}, vol.~37, no.~5, pp. L55--L57, Jan 2004.

\bibitem{Winter:16p}
A.~Winter, ``Tight uniform continuity bounds for quantum entropies: conditional
  entropy, relative entropy distance and energy constraints,'' \emph{Commun. in
  Math. Phys.}, vol. 347, no.~1, pp. 291--313, 2016.

\bibitem{BocheCaiCaiDeppe:14p}
H.~Boche, M.~Cai, N.~Cai, and C.~Deppe, ``Secrecy capacities of compound
  quantum wiretap channels and applications,'' \emph{Phys. Rev. A}, vol.~89,
  no.~5, p. 052320, 2014.

\bibitem{YardHaydenDevetak:11p}
J.~{Yard}, P.~{Hayden}, and I.~{Devetak}, ``Quantum broadcast channels,''
  \emph{IEEE Trans. Inf. Theory}, vol.~57, no.~10, pp. 7147--7162, Oct 2011.

\bibitem{KingRuskaiNathanson:07p}
C.~King, K.~Matsumoto, M.~B. Ruskai, and M.~B. Nathanson, ``Properties of
  conjugate channels with applications to additivity and multiplicativity,''
  \emph{Markov Processes and Related Fields}, vol.~13, no.~2, pp. 391--423,
  2007.

\bibitem{BradlerHaydenTouchetteWilde:10p}
K.~Br{\'a}dler, P.~Hayden, D.~Touchette, and M.~M. Wilde, ``Trade-off
  capacities of the quantum hadamard channels,'' \emph{Phys. Rev. A}, vol.~81,
  no.~6, p. 062312, June 2010.

\bibitem{BocheNotzel:14p1}
H.~Boche and J.~N{\"o}tzel, ``Positivity, discontinuity, finite resources, and
  nonzero error for arbitrarily varying quantum channels,'' \emph{J. Math.
  Phys.}, vol.~55, no.~12, p. 122201, 2014.

\bibitem{BennettDiVincenzoSmolin:97p}
C.~H. Bennett, D.~P. DiVincenzo, and J.~A. Smolin, ``Capacities of quantum
  erasure channels,'' \emph{Phys. Rev. Lett.}, vol.~78, no.~16, p. 3217, 1997.

\bibitem{Bennett:19t}
C.~H. Bennett, ``Quantum information's birth, growth, and impact on fundamental
  questions,'' July 2019, $\,$Special session in IEEE Int. Symp. Inf. Theory
  (ISIT'2019) and personal communication.

\bibitem{AroaraSinghRandahawa:19p}
K.~Arora, J.~Singh, and Y.~S. Randhawa, ``A survey on channel coding techniques
  for 5g wireless networks,'' \emph{Telecommun. Syst.}, pp. 1--27, 2019.

\bibitem{CostelloForney:07c}
D.~J. Costello and G.~D. Forney, ``Channel coding: The road to channel
  capacity,'' \emph{Proc. IEEE}, vol.~95, no.~6, pp. 1150--1177, 2007.

\bibitem{NisiotiThomos:20a}
E.~Nisioti and N.~Thomos, ``Design of capacity-approaching low-density
  parity-check codes using recurrent neural networks,''
  \emph{\textup{\texttt{arXiv:2001.01249}}}, 2020.

\bibitem{RichardsonKudekar:18p}
T.~Richardson and S.~Kudekar, ``Design of low-density parity check codes for 5g
  new radio,'' \emph{IEEE Commun. Mag.}, vol.~56, no.~3, pp. 28--34, 2018.

\bibitem{Ahlswede:06t}
R.~Ahlswede, ``Towards a general theory of information transfer,'' July 2006,
  $\,$Shannon lecture in IEEE Int. Symp. Inf. Theory (ISIT'2006) and personal
  communication.

\bibitem{Uhlmann:76p}
A.~Uhlmann, ``The ``transition probability" in the state space of a*-algebra,''
  \emph{Reports Math. Phys.}, vol.~9, no.~2, pp. 273--279, 1976.

\bibitem{LiWinter:14p}
K.~Li and A.~Winter, ``Relative entropy and squashed entanglement,''
  \emph{Commun. Math. Phys.}, vol. 326, no.~1, pp. 63--80, Jan 2014.

\bibitem{Renner:08a}
R.~Renner, ``Security of quantum key distribution,'' \emph{Int'l J. Quantum
  Info.}, vol.~6, no.~01, pp. 1--127, 2008.

\bibitem{TomamichelColbeckRenner:09p}
M.~{Tomamichel}, R.~{Colbeck}, and R.~{Renner}, ``A fully quantum asymptotic
  equipartition property,'' \emph{IEEE Trans. Inf. Theory}, vol.~55, no.~12,
  pp. 5840--5847, Dec 2009.

\end{thebibliography}

\end{document}